\documentclass[pdflatex,sn-mathphys-num]{sn-jnl}
\usepackage{graphicx} 
\usepackage{bm}
\usepackage{braket}
\usepackage{amsmath,amssymb}
\usepackage{xcolor}
\usepackage{comment}
\usepackage{ulem}
\usepackage{cancel}
\makeatletter

\newcommand{\e}{\epsilon}
\newcommand{\al}{\alpha}
\newcommand{\dL}{\Delta_L}
\newcommand{\dR}{\Delta_R}
\newcommand{\dLo}{\Delta_{L,0}}
\newcommand{\dRo}{\Delta_{R,0}}
\newcommand{\dosL}{N_L}
\newcommand{\dosR}{N_R}
\newcommand{\tL}{T_L}
\newcommand{\tR}{T_R}
\newcommand{\tcL}{T_{c,L}}
\newcommand{\tcR}{T_{c,R}}
\newcommand{\SAl}{S_\mathrm{Al}}
\newcommand{\SAlCu}{S_\mathrm{Al/Cu}}
\newcommand{\dAl}{\Delta_\mathrm{Al}}
\newcommand{\dAlCu}{\Delta_\mathrm{Al/Cu}}
\newcommand{\dAlo}{\Delta_\mathrm{Al,0}}
\newcommand{\dAlCuo}{\Delta_\mathrm{Al/Cu,0}}
\newcommand{\tAl}{T_\mathrm{Al}}
\newcommand{\tAlCu}{T_\mathrm{Al/Cu}}
\newcommand{\tcAl}{T_{c,\mathrm{Al}}}
\newcommand{\tcAlCu}{T_{c,\mathrm{Al/Cu}}}

\usepackage{chngcntr}
\counterwithin{equation}{section}
\counterwithin{figure}{section}
\DeclareUnicodeCharacter{2009}{\,}
    \title{Bipolar Thermoelectric Superconducting Quantum Devices}

\author*[1]{\fnm{F.} \sur{Antola}}\email{filippo.antola@sns.it}
\author[2]{\fnm{G.} \sur{Marchegiani}}
\author[3,4]{\fnm{A.} \sur{Braggio}}
\author[3]{\fnm{F.} \sur{Giazotto}}
\affil[1]{\orgname{NEST Istituto Nanoscienze-CNR and Scuola Normale Superiore},
  \orgaddress{\city{Pisa}, \postcode{I-56127}, \country{Italy}}}
\affil[2]{\orgdiv{Quantum Research Center}, \orgname{Technology Innovation Institute},
  \orgaddress{\city{Abu Dhabi}, \postcode{9639}, \country{UAE}}}
\affil[3]{\orgname{Istituto Nanoscienze – CNR, NEST-SNS},
  \orgaddress{\city{Piazza San Silvestro 12, Pisa}, \country{Italy}}}
\affil[4]{\orgdiv{Institute for Quantum Studies}, \orgname{Chapman University},
  \orgaddress{\city{Orange}, \state{CA}, \postcode{92866}, \country{USA}}}
\abstract{
Quantum technologies increasingly require accurate modeling of their hardware components and of the non-equilibrium regimes in which they operate, where managing heat and energy flow becomes a central challenge. Thermoelectric effects, the direct conversion of a thermal gradient into electrical signals, offer one such route to this control.
In this review, we present an overview of the bipolar thermoelectric effect, a recent development for thermoelectric conversion in reciprocal systems, where linear effects are forbidden by symmetry. This symmetry yields a bipolar thermoelectric signal, in which the generated voltage can exhibit both polarities at a fixed temperature gradient. This represents a non-trivial novelty relative to conventional thermoelectric effects, in which carrier dominance determines the sign of the thermoelectric signal.
We summarize the underlying physical principles, showing how thermoelectricity emerges as a strong violation of detailed balance. Concrete physical conditions for obtaining bipolar thermoelectricity are then outlined, of which a tunnel junction between two superconductors with unequal energy gaps and suppressed Josephson coupling is the paradigmatic example. Afterward, we discuss the experimental observation of the effect to date and related proposals for different applications, including volatile memories and radiation detection. Finally, we briefly survey recent developments and outlooks, ranging from extensions to new platforms to a proposal for a novel quantum thermoelectric effect.}
\enlargethispage{\baselineskip}

\begin{document}

\maketitle
\newpage

\section{Introduction}
\subsection{\bf From superconducting tunneling to thermoelectricity}
Tunneling through a classically forbidden region is among the most distinctive manifestations of quantum mechanics arising from the wave nature of particles. This fundamental phenomenon has far-reaching consequences across different areas of physics, from nuclear processes (e.g., $\alpha$-decay) to condensed-matter systems. In particular, since the 1960s, tunneling has played a crucial role in understanding the superconducting state, characterized by the Fermi surface 
instability with a gapped excitation spectrum and, intriguingly, dissipation-free current flow. In 1960, Giaever measured the superconducting energy gap from the charge current flowing through a thin oxide barrier between a superconductor and a normal 
metal and soon extended the technique to junctions between two superconductors~\cite{giaever1960energy,giaever1960electron}.  Only two years later, Josephson predicted that in such a junction a current can flow even in the absence of a voltage bias, driven only by the phase difference between the two condensates~\cite{josephson1962possible}. This prediction was experimentally confirmed soon afterward~\cite{anderson1963probable}, despite the initial skepticism of Bardeen himself~\cite{Bardeen_Tunneling_1962}.

It is hard to overstate the technological impact of superconducting tunnel junctions. The interplay between electrons (or, more precisely, quasiparticles) and Cooper pairs in transport makes these devices a key element of today's superconducting quantum technologies. Their phase sensitivity underlies the SQUID magnetometers~\cite{jaklevic1964quantum,clarke2004squid,fagaly2006superconducting}, while the synchronization of Josephson phase oscillations with an external microwave drive gives rise to Shapiro steps, enabling a direct and highly accurate frequency-to-voltage conversion for metrological voltage standards~\cite{shapiro1963josephson,hamilton2000josephson}. 
The persistent interest in tunneling phenomena is testified by the 2025 Nobel Prize in Physics, awarded to J.~Clarke, M.~H.~Devoret, and J.~M.~Martinis for macroscopic quantum tunneling and energy-level quantization in a Josephson circuit~\cite{nobel2025physics}, in the International Year of Quantum Science and Technology. These pioneering works set the basis for the ongoing development of superconducting quantum processors, of which the Josephson junctions form the backbone ~\cite{krantz2019quantum,blais2021circuit}.

Transport through the tunnel barrier is not limited to charges. Quasiparticles carry both charge and energy, and, therefore, a heat current can flow across the junction in the presence of a thermal gradient. Theoretical works on heat transport closely followed those on charge transport. In 1965, Maki and Griffin predicted that the heat current through a Josephson junction is affected by the superconducting phase difference~\cite{Maki_entropy_1965}, as in the case for the charge current.  A clear experimental identification of this modulation was confirmed only nearly 50 years later in a Josephson heat interferometer~\cite{martinez2012josephson}, which led to the development of the phase-coherent caloritronics field~\cite{fornieri_towards_2017}. Since quasiparticles contribute to both the charge and heat currents, it is natural to investigate the coupling between these two flows, which includes thermoelectric phenomena. While on-chip refrigeration under voltage bias has been investigated for over three decades~\cite{giazotto_opportunities_2006,Muhonen2012}, thermoelectric generation of power in superconductors has been characterized only recently, and forms one of the main subjects of this review.  

\subsection{\bf Basic concepts on thermoelectricity}
\label{subsec:linearTE}
Thermoelectricity denotes the coupling between charge and heat transport. In nonequilibrium statistical mechanics, the charge ($I$) and heat ($\dot{Q}$) currents are the fluxes driven by the thermodynamic forces (or affinities) $\delta\mu/(e T$) and $\delta T/T^2$, where $\delta\mu$ and $\delta T$ are the chemical potential and temperature differences with $e$ the elementary charge. In the linear response regime, $\delta\mu/k_B,\delta T \ll T$, where $k_B$ is the Boltzmann constant, the fluxes are related to the forces by the Onsager matrix~\cite{onsager1931reciprocalI,benenti_fundamental_2017}
\begin{equation}
\begin{pmatrix} I \\ \dot{Q} \end{pmatrix}
=
\begin{pmatrix} L_{11} & L_{12} \\ L_{21} & L_{22} \end{pmatrix}
\begin{pmatrix} \delta \mu/(eT) \\ \delta T/T^2 \end{pmatrix},
\label{eq:Onsager}
\end{equation}
whose elements are denoted as Onsager coefficients. The diagonal elements $L_{11}$ and $L_{22}$ (which must be non-negative according to the thermodynamic laws) are related to the charge and heat conductance, respectively, while the off-diagonal coefficients account for the thermoelectric response. Specifically, $L_{12}$ is proportional to the current driven by a thermal gradient in $\delta\mu=0$, while $L_{21}$ is proportional to the heat current associated with a voltage bias in thermal equilibrium. These coefficients are related by the Onsager reciprocal relations $L_{12}=L_{21}$~\cite{onsager1931reciprocalI,benenti_fundamental_2017} (assuming for simplicity a time-reversal symmetric Hamiltonian), and therefore can be expressed in terms of a single parameter 
$S=-\delta\mu/(e\,\delta T)|_{I=0}$ (Seebeck coefficient or thermopower). 

Within the transport theory for fermionic systems, one finds the Mott relation~\cite{MOTT} for the thermopower $S=-\frac{\pi^2 k_B^2 T}{3e} d[\ln G(\mu)]/d\mu$, where $G=L_{11}/T$ is the charge conductance. Then, an electronic system in which a large change in the chemical potential modifies the conductance is expected to exhibit a linear thermoelectric effect. 
Using the scattering approach to quantum transport, the Seebeck coefficient can be shown to be proportional to the following integral~\cite{benenti_fundamental_2017}
\begin{equation}
S\propto \int_{-\infty}^{\infty} d\e\ \frac{(\e-\mu)}{4 k_B T\cosh[(\e-\mu)/2k_BT]^2}\ \mathcal{T}(\e)
\label{eq:OnsagerElement}
\end{equation}
where $\mathcal{T}(\e)$ is the transmission function that characterizes the scattering region.  Since the fraction in the integrand of Eq.~\eqref{eq:OnsagerElement} is odd in $(\e-\mu)$, the Seebeck coefficient vanishes when $\mathcal{T}(\e-\mu)=\mathcal{T}(\mu-\e)$, i.e., for a transmission function symmetric around the chemical potential. This property indicates that when electrons ($\e>\mu$) and holes ($\e<\mu$) contribute equally to the transport, the thermoelectric effect is suppressed. Normal metals sit close to this symmetric limit: transport properties vary on the scale of the Fermi energy $E_F$, but only a narrow window $k_B T\ll E_F$ around the Fermi level carries the current, so the electron-hole asymmetry seen by transport is only of order $k_B T/E_F$. The Seebeck coefficient is correspondingly small, $S\sim (k_B/e)(k_B T/E_F)$, a few $\mu$V/K at room temperatures. Research on thermoelectricity~\cite{goldsmid2010introduction}, therefore, focused on materials with strong particle-hole asymmetry. Semiconductors are an example: band engineering and doping produce responses up to hundreds of $\mu$V/K at room temperatures. There, this asymmetry usually fixes which carrier dominates: the material is $n$-type ($p$-type) when the dominant carriers are electrons (holes).
This review addresses a unique nonlinear thermoelectric effect, in which a thermoelectric response coexists with a reciprocal IV characteristic; the system shows both $n$-type and $p$-type character, but thermoelectricity emerges via a nonequilibrium spontaneous symmetry breaking.

\subsection{\textbf{Thermoelectricity in superconductors}}
Thermoelectric effects (TEs) in superconductors have been elusive for a long time. An early measurement by Meissner in 1927~\cite{Meissner1927} led to questions about the existence of TEs in superconductors. Ginzburg later pointed out in 1944 that a temperature gradient can drive a quasiparticle (QP) charge current inside a superconductor when the QP spectrum retains the weak electron-hole asymmetry of a real metal~\cite{ginzburg1944thermoelectric,ginzburg1978thermoelectric}. However, in a homogeneous bulk superconductor, this current is compensated by a counterflowing supercurrent, leaving no net charge signal in an open circuit. Hence, TEs could be traced by measuring a thermally induced phase winding. In other words, TEs are not absent in superconductors but often characterized by coherent effects~\cite{Claughton_Thermoelectric_1996,Kalenkov_Large_2017,Kalenkov_Phase_2021}, charge imbalance phenomena~\cite{Schon_Thermoelectric_1981,VANHARLINGEN19821710}, or magnetic impurities~\cite{Kalenkov_Theory_2012}.  In a bimetallic superconducting loop, the mismatch between the compensating currents in the two arms produces a measurable magnetic flux~\cite{van1980experimental}, whose long-standing disagreement with theory was resolved only recently~\cite{shelly2016resolving}. 

In Josephson-junction-based devices, the supercurrent can shunt and obscure thermoelectric transport, as in bulk superconductors. Even when the Josephson coupling is absent, the thermoelectric response of a superconducting tunnel junction is expected to be extremely small, because the tunneling transmission $\mathcal{T}(\e)$ inherits the electron-hole symmetry of the Bardeen–Cooper–Schrieffer (BCS) density of states (DoS)~\cite{BCS}. By the argument of the previous subsection, this symmetry ensures that the linear Seebeck coefficient will vanish. 

One way to obtain a small but finite response is to give the barrier itself an energy-dependent transmission, so that QPs above and below the Fermi level contribute unequally to the current~\cite{smith1980new}. Higher responses were reported nearly 20 years later on Andreev interferometers, in which a superconductor and a normal metal are in good electrical contact~\cite{eom1998phase,Chandrasekhar_Thermal_2009}; there, however, a sizeable thermoelectric response requires a nonlocal multi-terminal interferometer~\cite{Galaktionov_PRB85_2012,Kalenkov_Large_2017,Dolgirev_current_2018,Kirsanov_Heat_2019,Hussein_nonlocal_2019,blasi_nonlocal_2020_PRL,Kalenkov_Phase_2021,Tan2021}.

Superconductors are, in principle, good materials for thermoelectricity. The sharp variation of the superconducting DoS at the gap edge~\cite{tinkham_introduction_2004} makes the transmission strongly energy-dependent through the electrode spectrum itself, a desirable but not sufficient property according to the Mott relation, since a finite response also requires lifting the electron-hole symmetry of the electrodes. 

This feature was recently proposed to enhance the thermoelectric responses by exploiting spin-active interfaces
~\cite{machon2013nonlocal,Ozaeta,Kalenkov_Electron_2014}. For instance, in junctions comprising superconductors and ferromagnetic elements, the spin-up and spin-down components of the DoSs are split either by an external magnetic field or by the exchange interaction with a nearby ferromagnetic layer. A spin-filtering element can then convert this splitting into a large linear Seebeck coefficient, enabling the experimental observation shortly after~\cite{kolenda2016observation,kolenda2017thermoelectric}.
Thermoelectricity associated with hybrid ferromagnetic/superconducting elements has also been investigated in superconductors with non-trivial order parameters, such as odd-frequency~\cite{Dutta_PRB102,Savander_PRR2} or spin triplet~\cite{Sonar2026GapAnisotropy,Pierattelli2026}. 
Finally, strong linear thermoelectricity has been predicted at the core of a fluxon of a type-II superconductor in the quantum limit~\cite{Singh_PRL133}. These results suggest that interfaces and hybrid structures are promising candidates for generating and detecting thermoelectricity~\cite{Arrachea2025}.
However, the effect discussed in this review works differently: the particle-hole symmetry breaking arises spontaneously,  triggered by a large (nonlinear) thermal bias, and is not engineered as in the previous cases.

\subsection{\textbf{Bipolar thermoelectricity}}
Superconducting-ferromagnetic junctions provide a suitable platform for cryogenic thermoelectricity and hold promise for applications, such as radiation detection. A few reviews have already been devoted to this topic~\cite{bergeret2018colloquium,HEIKKILA2019100540,Beckmann_AnnPhys}. This review focuses on a thermoelectric effect that finds its standard realization in a closely related structure, i.e., gap-asymmetric superconductor-insulator-superconductor junctions (SIS$^\prime$)~\cite{marchegiani_nonlinear_2020,marchegiani_superconducting_2020}, but is based on a crucially different mechanism.

The standard premise for thermoelectricity is that charge carriers with different signs (electrons and holes) should contribute unequally to the transport. This review covers works that have been stimulated by a fundamental question: can reciprocal systems display a thermoelectric response? This question is meaningful because no thermodynamic law forbids such a phenomenon. However, the reciprocal symmetry, characterized by an charge current that is odd in the voltage bias, rules out any linear TE at first glance; it is straightforward to prove that reciprocal systems have zero off-diagonal linear Onsager coefficients. 

This apparent incompatibility breaks down in the nonlinear response regime, as we originally proved for a specific class of tunnel junctions~\cite{marchegiani_nonlinear_2020}: driven far enough from equilibrium, the reciprocal system develops a finite thermoelectric response, and, as anticipated above, its sign is not fixed but can switch between the two polarities at a given temperature difference. We call this a \emph{bipolar thermoelectric effect}.
We stress that this bipolar thermoelectricity is unrelated to the so-called "bipolar effect" of conventional thermoelectrics~\cite{goldsmid2010introduction}, where both carriers contribute to conduction in a semiconductor or semimetal at high temperature, typically reducing the response. Here, instead, the term denotes that the thermoelectric response can have either electron or hole character at a fixed temperature gradient, a feature that is strictly impossible in the linear regime.

The effect has since been experimentally confirmed and explored across hybrid platforms and in various setups. More recently, it has even been extended to the development of a purely quantum TE. This review presents these developments and the underlying physics comprehensively.

\subsection{\bf Organization of the present review}
This review is organized as follows. Section~\ref{Sec: Th} develops the theoretical framework, from the tunneling rates to the conditions on the DoS and the thermodynamics of the conversion. Section~\ref{sec:Experiments} describes the first experimental observation of the effect, the bipolar thermoelectric Josephson engine. 
Section~\ref{Sec: Applications} turns to applications: detection, thermal diodes and pipes, and circuit elements such as memories, oscillators, and amplifiers. Section~\ref{Sec: hybrids} extends the effect to other platforms, where charging effects, electrostatic gating, and spin splitting add further control and enrich the mechanism. 
Section~\ref{Sec: Env} addresses an environment-induced variant driven by a genuinely quantum mechanism, namely the imbalance between emission and absorption of energy quanta exchanged with a cold electromagnetic environment. Section~\ref{sec:conclusions} closes with an outlook.

\newpage
\section{Theoretical Framework for Tunneling in Superconducting Junctions}\label{Sec: Th}
\subsection{\bf Tunneling Hamiltonian}
\label{subsec:Tunneling}
We consider a two-terminal system in which two electrodes, hereafter denoted left ($L$) and right ($R$), are weakly coupled through a thin (nanometer-thick) insulating tunneling barrier. 
The total Hamiltonian of the system is of the form $\hat{H}=\hat{H}_L+\hat{H}_R+\hat{H}_T$ with $\hat{H}_{\alpha}$ ($\alpha=R,L$) denoting the Hamiltonian of the lead $\alpha$. When the lead is superconducting, we model it using the minimal mean-field description of the BCS theory \cite{tinkham_introduction_2004} 
\begin{equation}
\hat{H}_{\alpha}=\sum_{{\bf k}\sigma} \epsilon_{\alpha,{\bf k}\sigma}\hat{c}^\dagger_{\alpha,{\bf k}\sigma}\hat{c}_{\alpha,{\bf k}\sigma}+
\sum_{\bf k} \Delta_\alpha \hat{c}^\dagger_{\alpha,{\bf k}\uparrow}\hat{c}^\dagger_{\alpha,-{\bf k}\downarrow}+\Delta_\alpha^* \hat{c}_{\alpha,-{\bf k}\downarrow}\hat{c}_{\alpha,{\bf k}\uparrow}\, ,
\label{eq:Hlead}
\end{equation}
where $\hat{c}^\dagger_{\alpha,{\bf k}\sigma}(\hat{c}_{\alpha,{\bf k}\sigma})$ are the creation (annihilation) operators for an electron with wave-vector ${\bf k}$, spin $\sigma$ in the lead $\alpha$, and
$\Delta_\alpha$ is the mean-field superconducting order parameter, which we can assume independent of ${\bf k}$ for our purposes. Finally, the tunneling Hamiltonian can be expressed  
as~\cite{ingold_charge_1992}
\begin{equation}
\hat{H}_T=\sum_{{\bf kq}\sigma} t_{{\bf kq}} \hat{c}^\dagger_{L,\bf{q}\sigma}\hat{c}_{R,{\bf k}\sigma} e^{-i\varphi}+t_{{\bf kq}}^* \hat{c}^\dagger_{R,{\bf k}\sigma}\hat{c}_{L,\bf{q}\sigma} e^{i\varphi}\,,
\label{eq:Htun}
\end{equation}
where we introduced the circuit phase difference  
$\varphi(t)=(e/\hbar)\int_{-\infty}^t dt' V(t')$, where $e$ is the elementary charge, $\hbar$ is the reduced Planck's constant, and $V$ is the voltage bias across the junction. For our goals, the phase is treated as a classical variable (i.e., expectation value of the corresponding phase operator) in this section; a full quantum treatment is required if one needs to investigate the coupling of the system to an electromagnetic environment \cite{ingold_charge_1992,nazarov_quantum_2009,Vool_Devoret} (see also Sec.~\ref{Sec: Env}). For a constant voltage bias $\mu_L-\mu_R=eV$, it is convenient to make a time-dependent unitary transformation on the lead operators  $\hat{U}_\alpha=\prod_{k\sigma}\exp(i\mu_\alpha t\, \hat{c}^\dagger_{\alpha,\bf{k}\sigma}\hat{c}_{\alpha,\bf{k}\sigma}/\hbar)$ which corresponds to moving to a rotating frame. In the new frame, 
the phase difference $\tilde\varphi=\varphi-e V t/\hbar$ becomes time-independent while 
the energies of the leads in Eq.~\eqref{eq:Hlead} shift as $\epsilon_{\alpha,\bf{k}\sigma}\to \epsilon_{\alpha,\bf{k}\sigma}-\mu_\al$. This procedure is convenient for computing the tunneling 
rates for a single barrier, which 
usually depend only on the applied bias $V$, and not separately on the chemical potential values $\mu_\alpha$, at least if gating effects (see, e.g. Ref.~\cite{Datta2005AtomToTransistor}) can be neglected as usually done with metals.

\subsection{\bf Tunnelling rates}
\label{subsec:tunnelingRates}
In this review, we primarily address charge and heat transport in tunnel-coupled electrodes using the general Hamiltonian of Eqs.~\eqref{eq:Hlead} and \eqref{eq:Htun}. In the tunneling regime, we can use perturbation theory to compute the tunneling rates. We can evaluate the transition rate for the tunneling of a QP initially located in one lead (with initial energy $\e_i$) into the other lead (with final energy $\e_f$) using the standard result for time-dependent perturbation theory popularized by Fermi as ``Golden Rule''~\cite{Fermi1950}, but earlier derived by Dirac~\cite{Dirac1927,VisserAJP},
\begin{equation}
m_{if}=\frac{2\pi}{\hbar}|\braket{i|\hat{H}_T|f}|^2\delta(\e_i-\e_f)\,.
\label{eq:Goldenrulerate}
\end{equation}
The total rate is then obtained by integrating over all initial states, weighted by their initial occupation probabilities, and over all available final states. After some algebra, and assuming that the tunneling coefficient $t_{\bf kq}$ of Eq.~\eqref{eq:Htun} is momentum-independent in the range of interest, i.e., $t_{\bf kq}\simeq t$, we can express the tunneling rate from the $\alpha$-electrode to the $\bar{\alpha}$-electrode as ($\al=R,L$)
\begin{equation}
\label{eq:RateTunneling}
\Gamma_{\al\bar\al}=\frac{G_T}{e^2 }\int_{-\infty}^{+\infty} d\e  N_\al(\e_\al)N_{\bar{\al}}(\e_{\bar{\al}}) f_\al(\e_\al) [1-f_{\bar{\al}}(\e_{\bar{\al}})]\, ,
\end{equation}
where $\e_\alpha=\e-\mu_\alpha$, $\bar{L}=R$ and vice versa. Expressions of this kind were originally derived by Bardeen~\cite{BardeenPRL6} and were subsequently used to analyze charge transport in superconducting tunnel junctions~\cite{Giaever_Study_1961}. The prefactor $G_T=4\pi e^2\nu_{0L}\nu_{0R}|t|^2/\hbar^3$, where $\nu_{0\alpha}$ is the DoS in the normal state at the Fermi energy in electrode $\alpha$, gives the junction conductance when both electrodes are in the normal state (see the discussion below).
In Eq.~\eqref{eq:RateTunneling}, $N_\al(\e)$ represents the energy-dependent tunneling DoS, normalized with $\nu_{0\alpha}$, for the $\al$-lead measured with respect to the chemical potential $\mu_\al$ and the Fermi functions are $f_\al(\e)=f_0(\e,T_\alpha)=(e^{\beta_\al \e}+1)^{-1}$ with $\beta_\al=1/k_BT_\al$ and $T_\al$ the electronic temperature of the $\al$-lead.
Under a constant DC voltage bias, only the mismatch between the chemical potentials, $\delta\mu_\al=\mu_\al-\mu_{\bar \al}$, matters.\footnote{This result holds only when we can neglect gating effects; this approximation typically holds for metals and superconductors, as mainly considered in this review, but not necessarily for other materials such as semiconductors or 2D materials.} So the rate is invariant under energy shift of the dummy variable $\e$ since the integral runs on the full real axis; below we will occasionally use this invariance to express the transport quantity in a more convenient way and derive some crucial results.

We first summarize the general properties of the tunneling rates. When the temperature of the two electrodes is the same 
$T_\al=T$, we find
\begin{equation}
\Gamma_{\al\bar\al}(-\delta \mu_\al)=e^{-\delta\mu_\al/k_BT}\Gamma_{\al\bar{\al}}(\delta\mu_\al)\,,
\label{eq:detail}
\end{equation} 
an identity which holds generally at thermal equilibrium, known as detailed balance (see, e.g., ~\cite{ingold_charge_1992}).
This condition shows that for $\delta\mu_\al>0$ the tunneling rate $\Gamma_{\al\bar\al}$ corresponds to a transition flowing in the preferred direction of the electrical bias (forward). In particular, this establishes that, at thermal equilibrium, the transition probabilities in the direction favored by the bias (forward) are always greater than the same tunneling process done against the bias (backward), which is activated only by the thermal energy, and it clearly happens more rarely. 

When the DoS satisfies the energy symmetry,
\begin{equation}
\label{eq:PHSdos}
N_{\al}(\e_{\al})=N_{\al}(-\e_{\al}),
\end{equation}
a symmetry that reflects the particle-hole symmetry (PHS) of the lead Hamiltonian, the forward and the backward rates are also related by the following property
\begin{equation}
\Gamma_{\al\bar\al}(\delta \mu_\alpha)=\Gamma_{\bar\al\al}(-\delta \mu_\alpha)\,,
\label{eq:symmetryRates}
\end{equation}
which holds for general energy dependence of the lead DoSs (that still satisfy the energy symmetry separately) and for arbitrary values of the lead electronic temperatures $T_\al$.\footnote{To derive this result, the following property of the Fermi function $f_\al(-\e)=1-f_\al(\e)$ is explicitly used.} 

The energy symmetry is trivially satisfied for a constant DoS, which is usually used to model normal metals 
for energy scales $\epsilon_\alpha\ll E_F$ with $E_F$ the Fermi energy \footnote{Assuming a constant DoS for non-superconducting normal metals, in the temperature range of interest, is an approximation which holds for $\e_{\alpha}\ll E_F$ then neglecting corrections in the small parameters $k_BT/E_F, \delta\mu/E_F\ll 1$.}. However,
this energy symmetry, expressed with respect to the chemical potential (the Fermi energy), also characterizes superconducting materials (see below).

We note that Eq.~\eqref{eq:RateTunneling}, obtained assuming the energy (and spin) independent tunneling coefficient $t$, implies the energy inversion symmetry EIS~\cite{Arrachea2025} of the transport coefficients when the DoS satisfies the energy symmetry. This point has a strong impact on the thermoelectric response, as
linear TEs disappear when the EIS is satisfied~\cite{benenti_fundamental_2017,Arrachea2025}. In the literature, the EIS of the transport coefficients is occasionally associated with the PHS, even though the latter is more general (holds by definition, for instance, in the Bogoliubov-De Gennes description of the superconducting state~\cite{gennes_superconductivity_2018}). 
We remark, indeed, that the PHS implies EIS only in specific cases, such as spin-rotationally invariant Hamiltonians.
An example where PHS is satisfied, while EIS is strongly broken, has been recently investigated in Ref.~\cite{Singh_PRL133}: the vortex-bound states arising from the winding of the order parameter around the vortex and angular momentum quantization strongly break the energy symmetry of the DoS, and, in the quantum limit, lead to a strong thermoelectric response.
In this work, unless explicitly stated otherwise, we consider spin-degenerate fermionic systems and sometimes use both notions interchangeably.

\subsection{\bf Transport quantities} 
\label{subsec:transport}
The tunneling current through a single barrier can be shown to follow bidirectional Poissonian statistics~\cite{LevitovReznikov,nazarov_quantum_2009}.
The cumulant generating function $S(\lambda,t)$ defined as $e^{S(\lambda,t)}=\sum_N P(N,t) e^{i\lambda N}$ where $P(N,t)$ is the probability distribution that $N$ QPs tunneled through the junction in a time $t$. For a bidirectional Poissonian,
$S(\lambda,t)=t[(e^{i\lambda}-1)\Gamma_{LR}+(e^{-i\lambda}-1)\Gamma_{RL}]$, and so all the cumulants of the charge transport for a finite measurement time $t$ are known. Specifically, the QP current $k$th-cumulants are~\cite{LevitovReznikov,nazarov_quantum_2009,ferraro}
\begin{equation}
\langle\langle \mathcal{I}\rangle\rangle_k=\frac{d}{dt}\frac{\partial^m S(\lambda,t)}{\partial^m (i\lambda)}\Biggl|_{\lambda\to0,t\to\infty }= \begin{cases} \Gamma_{LR}-\Gamma_{RL}\  \textrm{for}\  k\ \textrm{odd}\\ \Gamma_{LR}+\Gamma_{RL}\  \textrm{for}\  k\ \textrm{even}\end{cases}
\label{eq:cumulants}
\end{equation}
calculated in the stationary limit as a statistical average over $P(N,t)$. Equation~\eqref{eq:cumulants} implies that the odd QP current cumulants can be written in terms of the stationary QP current $\mathcal{I}\equiv\langle\langle \mathcal{I}\rangle\rangle_1$ [related to the charge current $I=-e\mathcal{I}$ where we assumed $e>0$], while the even cumulants can be written in terms of the zero frequency current noise $S_{II}=e^2 \langle\langle \mathcal{I}\rangle\rangle_2$.

When the condition of Eq.~\eqref{eq:symmetryRates} is valid [for instance, for EIS DoSs], the QP current, obtained by setting $k=1$ in Eq.~\eqref{eq:cumulants}, satisfies the reciprocity condition 
\begin{equation}
\mathcal{I}(\delta \mu)=- \mathcal{I}(-\delta \mu)
\label{eq:currentReciprocity}
\end{equation}
irrespective of the lead temperatures. 
In particular, when current reciprocity is satisfied, linear QP thermoelectric effects are necessarily absent 
$L_{12}\propto d\mathcal{I}/dT|_{\delta\mu=0}=0$ (see Sec.~\ref{subsec:linearTE})
since $\mathcal{I}\to 0$ as $\delta \mu\to 0$ 
independently of the lead temperatures. 
Furthermore, when
Eq.~\eqref{eq:currentReciprocity} holds, and the junction is in thermal equilibrium, one finds  
\begin{equation}
\mathcal{I}(\delta \mu)=\Gamma_{LR}(\delta\mu)(1-e^{-\delta\mu/k_B T})
\label{eq:Currentdetail}
\end{equation}
applying the detailed balance relation Eq.~\eqref{eq:detail}.
It follows that the QP current is necessarily positive (negative) for $\delta\mu>0$ ($\delta\mu<0$) since the tunneling rates are transition probabilities and therefore not negative.   
Physically, this result expresses the fact that, in thermal equilibrium, the QP current always flows in the direction of the chemical potential gradient and that the junction displays a dissipative response, characterized by positive power $\dot{W}=\mathcal{I}(\delta\mu) \delta\mu>0$.

In closing this subsection, we note that the current noise [$k=2$ Eq.~\eqref{eq:cumulants}] at thermal equilibrium reads 
\begin{equation}
\langle\langle \mathcal{I}
(\delta \mu)\rangle\rangle_2
=  \mathcal{I}(\delta \mu)  \coth\left(\frac{\delta\mu}{2k_BT}\right)\,,
\end{equation}
a result also found using the detailed balance, which expresses the standard nonequilibrium fluctuation-dissipation theorem (see, e.g., Ref.~\cite{Scalapino}). In the limit $\delta\mu\to 0$ one recovers the standard Johnson-Nyquist (charge current) noise $S_{II}\approx2 k_B T G$ where the junction linear electrical conductance is $G=\delta I/\delta V|_{V\to0}$ while, for $\delta\mu\gg k_BT$, one gets the shot noise limit $S_{II}\approx e I$.

\subsection{\bf Thermally biased tunnel junctions}\label{Sec: Th SIS'} 
We now discuss transport when a thermal bias is established between the two leads, i.e., $\tL\neq \tR$. 
First, we address the linear response regime, where the thermal gradient is small, i.e., $\delta T=\tL-\tR\ll T=(\tL+\tR)/2$.
We are interested in determining the conditions for a finite QP current in the absence of an electrochemical bias, i.e., $\delta\mu=0$. 
By linearly expanding the Fermi functions $f_{L/R}(\e)=f_0(\e,T)\mp(\e/2T) f_0'(\e,T) \delta T$, with $f_0'(\e,T)=\partial f_0/\partial\e=
-[4k_B T \cosh^2(\e/2k_B T)]^{-1}$, the thermoelectric current response is expressed by the gradient~\cite{Ozaeta,LinderPRB93} 
\begin{equation}
\label{eq:dIoverdT}
\frac{\delta \mathcal{I}}{\delta T}\Big|_{\delta\mu\to 0}=
\frac{G_T}{e^2 }\int_{-\infty}^{+\infty} d\e \dosL(\e) \dosR(\e)\frac{\e}{4k_BT^2\cosh^2(\e/2k_BT)}\,.
\end{equation}
If both DoSs are energy symmetric, i.e. $N_\al(\e)=N_\al(-\e)$, no thermocurrent is expected due to the odd-parity in energy of the integrand of Eq.~\eqref{eq:dIoverdT}. As mentioned in Sec.~\ref{subsec:tunnelingRates}, this situation
applies to normal metals, for which the DoS is approximately constant, and to superconductors (see also Sec.~\ref{SubSec:asymmetricSIS}). This result is clearly consistent with the findings of Eqs.~\eqref{eq:symmetryRates} and~\eqref{eq:currentReciprocity};  
in fact, the reciprocity of the current   
imposes $(\delta\mathcal{I} \mu\to0)=0$ for arbitrary values of the lead temperatures.
In the linear response regime, this result necessarily coincides with the one obtained by applying 
the scattering theory~\cite{benenti_fundamental_2017} to opaque barriers. In the scattering approach, the even symmetry in the energy of the scattering probability $\mathcal{T}_{RL}(\e)$, measured with respect to the Fermi energy, equivalently forbids any TE.

We note that the bipolar thermoelectricity discussed in this review inherits its main features from the fact that the thermoelectricity is generated in a system that satisfies the reciprocal condition in Eq.~\eqref{eq:currentReciprocity}. As shown above, this symmetry forbids a thermoelectric current at zero electrochemical bias in the linear response regime. Therefore, thermoelectric power generation can occur only beyond the linear response~\cite{sanchezNonlinear}, in the presence of a finite temperature difference or electrochemical bias $\delta\mu$. When both $\mathcal{I}$ and $\delta\mu$ are finite, power is either generated, i.e., $\mathcal{I}(\delta\mu) \delta\mu<0$, or dissipated, i.e., $\mathcal{I}(\delta\mu)\delta\mu>0$. According to the discussion following Eq.~\eqref{eq:Currentdetail}, power generation is possible only when a temperature difference is established across the junction. In this regime, the current $\mathcal{I}$ flows against the applied bias $\delta\mu$, a condition that can usually be referred to as negative absolute mobility or as absolute negative conductance (ANC) with charge systems~\cite{Ros2005NegativeMobility,PhysRevLett.122.070602,BenentiPRL124}.

Before proceeding, we note that the equivalence between the scattering and the tunneling (Green's function) approaches is not guaranteed in the nonlinear regime when interaction effects become significant; the Green's-function formalism provides a suitable framework for treating these effects~\cite{Caroli_1971,Caroli_1971_2,Meir_Landauer_1992,HaugJauho2008}. Then, in the following, we focus only on the tunneling approach, introduced in Sec.~\ref{subsec:Tunneling}. 

The reciprocity [Eq.~\eqref{eq:currentReciprocity}] for arbitrary lead temperatures is a key property for the bipolar TE, and follows from the energy-symmetry of Eq.~\eqref{eq:PHSdos} (see Sec.~\ref{subsec:transport}).
The latter symmetry implies that the DoS depends only on the energy relative to the chemical potential (i.e., the Fermi energy for our purposes), a feature that is typically ensured by interaction effects.

The relation between interaction and energy symmetry holds for BCS superconductors -- the main example we consider in this review -- where the DoS gap, determined by the superconducting phase transition, opens at the Fermi surface~\cite{BCS}. 
However, the energy symmetry of Eq.~\eqref{eq:PHSdos} also approximately holds in other systems:  single-impurity Kondo systems~\cite{Hewson_1993,PhysRevB.36.675}, magnetic atoms on metals~\cite{Yu2014KondoEffect}, quantum dots in the Kondo valley~\cite{Cronenwett_1998,Sasaki2000,Jeong2001}, Kondo lattice models at half filling~\cite{Sykora2013,Nakamura2023}, Kondo insulators~\cite{PhysRevB.58.15483}, heavy-fermions~\cite{Vano2021}, and twisted bilayer graphene in the flat-bands regime~\cite{10.1063/5.0303858,Andrei2020}.

\subsection{\bf Conditions for bipolar thermoelectrics}
\label{SubSec:Necessary}
Tunnel junctions between electrodes with energy-symmetric DoS can display a
nonlinear thermoelectric response in the QP current only under specific conditions~\cite{sanchezNonlinear}. In the following, we specifically investigate the requirements for achieving
bipolar thermoelectricity. Since the rates are expressed in terms of energy integrals of the convolution of four functions [see Eq.~\eqref{eq:RateTunneling}], finding general necessary and sufficient conditions for thermoelectricity is technically hard and likely not physically insightful. For this reason, we will focus on determining a minimal set of conditions to generate the bipolar TE, highlighting the qualitative features required in the excitation spectrum of each lead. These conditions, in turn, impose constraints at the junction design to achieve the aforementioned effect.

First, we show that the two DoSs must be unequal for thermoelectricity to occur. 
Using the energy symmetry of the DoS and the property $f_\alpha(-\e)=1-f_\alpha(\e)$ of the Fermi function, we can rewrite the rate of Eq.~\eqref{eq:RateTunneling}, restricting the integration only on the positive energies (defined with respect to the average chemical potential) as follows,
\begin{equation}
\Gamma_{\al\bar\al}(\delta\mu_\al)=\frac{G_T}{e^2 }\int_{0}^{+\infty}\!\!\!\!\!\!\!\!\!\!\! d\e  N_\al(\e-\delta\mu_\al/2)N_{\bar{\al}}(\e+\delta\mu_\al/2) f_\al(\e-\delta\mu_\al/2) [1-f_{\bar{\al}}(\e+\delta\mu_\al/2)]\, +\alpha\leftrightarrow\bar\alpha\,.
\label{eq:rateSymmetrized}
\end{equation}
where the symbol $\alpha\leftrightarrow\bar\alpha$ denote the contribution obtained by exchanging $\alpha$ and $\bar{\alpha}$ in the first term.
Using the reciprocity relation for the rates in Eq.~\eqref{eq:symmetryRates}, the QP current can then also be written as 
\begin{equation}
\mathcal{I}(\delta\mu)=
\Gamma_{LR}(\delta\mu)-\Gamma_{LR}(-\delta \mu)\,,
 \label{eq:currentmu}
\end{equation}
where we recall $\delta\mu=\delta\mu_L$. 
Then we can demonstrate that for identical DoSs, i.e. $N_\al(\e)=N(\e)$ independently of $\al$, transport can be only dissipative.
Inserting Eq.~\eqref{eq:rateSymmetrized} into Eq.~\eqref{eq:currentmu}, one finds that the terms containing the product of two Fermi functions cancel out, and that the particle current can be expressed just as
\begin{align}
 \mathcal{I}(\delta\mu)=\frac{G_T}{e^2}\int_{0}^{+\infty} d\e  N(\e-\delta\mu/2)N(\e+\delta\mu/2)[f_L(\e-\delta\mu/2)-f_L(\e+\delta\mu/2) \nonumber\\
+f_R(\e-\delta\mu/2)-f_R(\e+\delta\mu/2)]\, .
\label{eq:symmetrizedqpcurrent}
\end{align}
Since the DoSs are always positive and the Fermi distribution is monotonically decreasing, this expression is manifestly positive (negative) for $\delta\mu>0$ ($\delta\mu<0$). 
Then, with identical lead DoSs, electrical power is always positive, i.e. $\mathcal{I}(\delta\mu) \delta\mu >0$, meaning that the junction can only be dissipative independently of the lead temperatures.

Equation~\eqref{eq:currentmu} shows that, for a positive electrochemical bias $\delta\mu>0$, thermoelectric generation requires a negative current $\mathcal{I}<0$, i.e. $\mathcal{I}\delta\mu<0$. 
This can occur only if there exist at least bias values $\delta\mu$, for which the following inequality is satisfied:
\begin{equation}
\Gamma_{LR}(\delta \mu)<\Gamma_{LR}(-\delta \mu)\,,
\label{eq:strong}
\end{equation}
a condition previously referred to by some of the authors of this review as a strong detailed-balance violation~\cite{battisti2024bipolar}
\footnote{Clearly, current reciprocity implies the opposite sign in the inequality for $\delta\mu<0$.}. 
The name for this condition is inspired by the fact that when the detailed balance is satisfied, at thermal equilibrium, [c.f. Eq.~\eqref{eq:detail}] 
we always have that $\Gamma_{LR}(\delta \mu)>\Gamma_{LR}(-\delta \mu)$ for $\delta \mu>0$ at any finite temperature~\cite{Crooks_PhysRevE60,Seifert_2012}. In conclusion, to achieve a strong detailed-balance violation, one needs to consider at least different DoS in the leads and apply a finite thermal bias to the junction.   

If we consider two different energy-symmetric DoS, i.e., $\dosL(\e)\neq \dosR(\e)$ using Eqs.~\eqref{eq:RateTunneling} one easily finds the transition probability rate
\begin{equation}
\Gamma_{LR}(\delta\mu)=\frac{G_T}{e^2}\int_{-\infty}^{\infty}d\e \dosL(\e-\delta\mu) f_L(\e-\delta\mu) \dosR(\e) [1-f_R(\e)]\,.
\label{eq:RateSimplified}
\end{equation} 
where we observe that the rate integrand is always a positive quantity, since $N_\al(\e)>0$ and $0\leq f_\al(\e)\leq 1$.
\footnote{Notably, due to the particle-hole symmetry, one can analogously write the same rate in terms of the bias shifted only on the right side, leaving the left side unaffected instead.}
First, we consider the case in which one of the electrodes ($L$, with no loss of generality) has an energy-independent DoS, i.e., $\dosL(\e)=1$, which physically corresponds to considering a normal metal on that side. To satisfy the strong detail balance violation given in Eq.~\eqref{eq:strong}, the integrand of Eq.~\eqref{eq:RateSimplified} must be smaller than the expression obtained by changing the sign of $\delta\mu$ in it, at least for a given value of $\e$, i.e.
\begin{equation}
f_L(\e-\delta\mu) \dosR(\e) [1-f_R(\e)]<f_L(\e+\delta\mu) \dosR(\e) [1-f_R(\e)]\,,
\end{equation}
a condition which is impossible since the Fermi distribution is monotonically decreasing.
Alternatively, the impossibility of the strong detailed balance violation can be obtained showing that $\Gamma_{LR}(\delta\mu)$, is increasing monotonically for all $\delta\mu$ values, since $\partial_{\delta\mu}\Gamma_{LR}(\delta\mu)>0$.
In summary, no thermoelectricity is obtained if either of the DoS values is constant, as in the case of normal metals. 
We now assume that both DoSs have a nontrivial energy dependence and that the temperatures of the two leads differ, as necessary to violate detailed balance. 
Without loss of generality, we analyze the case $T_L> T_R$ again. To derive explicit conditions for thermoelectricity, we consider the case $T_R\to 0$, where the thermoelectric response is expected to be maximized. In this limit, we approximate $1-f_R(\e)\simeq \theta(\e)$ in Eq.~\eqref{eq:RateSimplified}, then in the integrand only $\e>0$ matters, and to get the strong violation of the detailed balance, one requires  
\begin{equation}
\dosL(\e-\delta\mu)f_L(\e-\delta\mu)<\dosL(\e+\delta\mu)f_L(\e+\delta\mu)\,,
\label{eq:necessaryNew}
\end{equation}
which implies that the DoS of the hot (left) lead must exhibit a local growth sufficiently fast to compensate for the decrease in the Fermi function.
In particular, this condition can be satisfied, for example, by a gapped DoS, i.e., $N_L(\e)=0$ for $|\e|<\Delta_L$ in the energy interval, $\dL-\delta\mu<\e<\dL+\delta\mu$ provided that $\delta\mu<\Delta_L$. A further condition can be derived by shifting the bias to the right-contact terms of the integrand in Eq.~\eqref{eq:strong}, i.e., performing the variable change $\e-\delta\mu\to \e$. Then, in such a case, the strong violation of the detailed balance requires at least, for some energy ($\e>0$),
\begin{equation}
\dosR(\e+\delta\mu)
<\dosR(\e-\delta\mu)
\,,
\label{eq:necessary}
\end{equation}
which implies that the DoS in the right electrode is, at least locally, monotonically decreasing.\footnote{Reader should note that a similar conclusion can be obtained even assuming a finite temperature for the cold (right) side if $k_BT_R\ll\Delta_L$.}
 
In particular, combining the conditions Eq.~\eqref{eq:necessaryNew} and ~\eqref{eq:necessary}, with more restrictive requirements as we detail below, one finds a set of sufficient conditions for bipolar thermoelectricity. This can be clearly seen by writing the particle current as:
\begin{equation}
\mathcal{I}(\delta\mu)=\frac{G_T}{e^2}\int_{-\infty}^{\infty}\!\!\!\!\!\!\!\!d\e \dosL(\e)f_L(\e)\{\dosR(\e+\delta\mu)[1-f_R(\e+\delta\mu)]-\dosR(\e-\delta\mu)[1-f_R(\e-\delta\mu)]\},
\label{eq:currentSecondExpression}
\end{equation}
where we still take the hot side to be on the left, i.e. $T_L>T_R$. Given the previous discussion, we need to assume energy-dependent DoSs. To satisfy Eq.~\eqref{eq:necessaryNew}, we start by considering a gapped DoS for the left contact $\dosL(\e)=\theta(|\e|-\Delta_L)$. 
Furthermore, we consider $T_R\ll T_L,\Delta_L/k_B$, to replace $1-f_R(\e\pm\delta\mu)\simeq \theta(\e\pm\delta\mu)$ -- up to exponentially small corrections of the type $\exp[-(\Delta_L\pm \delta\mu)/k_BT_R]$. 
With these approximations and for $\delta\mu<\Delta_L$, the integrand of Eq.~\eqref{eq:currentSecondExpression} is nonzero only for $\e>0$, and the junction becomes thermoelectric [$\mathcal{I}(\delta\mu)\delta\mu<0$] if the inequality of Eq.~\eqref{eq:necessary} is satisfied for $\e>\dL-\delta\mu$.
\footnote{For negative energies $\e<0$, one can follow very similar arguments in which the DoS should increase monotonically with energies, as the energy symmetry implies.} 

\begin{figure}[t]
    \centering
    \includegraphics[width=0.35\linewidth]{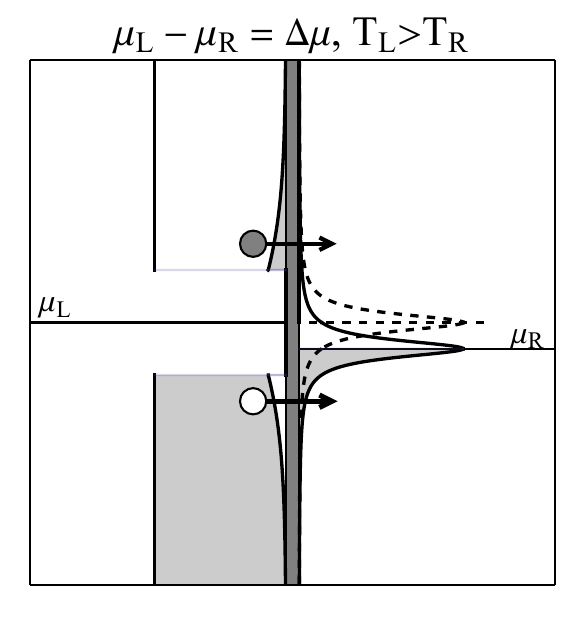}
   \caption{Band diagram schematic illustrating the sufficient conditions for bipolar thermoelectricity. A gapped electrode (left) is tunnel coupled to a lead with a resonance-like DoS (right). At zero bias $\mu_L=\mu_R$ (dashed curves), the electron current (filled circles) is exactly compensated by the hole current (empty circles). When $\mu_L>\mu_R$, the electron current decreases while the hole current increases, due to the monotonically decreasing DoS on the right lead. As a result, a net particle current flows in the opposite direction of the chemical potential gradient.  
}
    \label{fig:bandDiagramSmLorentzian}
\end{figure}

This result was first presented by some of the authors of this review in Ref.~\cite{marchegiani_nonlinear_2020}. The sufficient conditions for bipolar thermoelectricity are generically illustrated in the band diagram plot of Fig.~\ref{fig:bandDiagramSmLorentzian}, which represents a purely hypothetical example. It is important to stress that, in this argument, the DoS of the cold terminal, here the $R$ lead, necessarily shifts with the chemical potential of that lead. In other words, the DoS is pinned to the chemical potential. Otherwise, one cannot formulate the conditions as done above, leading to the inequality of Eq.~\eqref{eq:necessary}. 
This property follows from the PHS, defined relative to the local lead chemical potential (Fermi energy), and it is required to guarantee the IV reciprocity and, consequently, the bipolar nature of the effect. 
Hereafter, we mainly consider systems comprising BCS superconductors. However, as anticipated before, other interacting systems, even gapless (e.g. Kondo~\cite{Hewson_1993} systems) may also satisfy the mentioned sufficient conditions.\footnote{The particle-hole symmetry is preserved in interacting systems if the self-energy $\Sigma(\e)$ near the Fermi energy ($\e=0$) has a even- (odd-)in energy imaginary (real) part, respectively.} In the mean-field treatment of the BCS Hamiltonian [c.f. Eq.~\eqref{eq:Hlead}], the BCS gap opens around the Fermi energy, fixed by the chemical potential. 
This example unveils, to some extent, the key role of interaction effects in bipolar thermoelectric phenomena.
Nevertheless, we stress that this requirement may be relaxed in a two-terminal junction. For instance, in Sec.~\ref{subsec:grapheneSuper} we briefly discuss a concrete case in which only one of the two leads is interacting. 

\subsection{\bf Toy models}
\label{subsec:toy_model}
It is instructive to first consider a toy model to convince the reader that bipolar thermoelectricity is a general concept, realizable across a wide range of cases. We assume that DoS in each electrode is strongly localized at energy $\e_\al\simeq \Delta_\alpha$. This energy profile mimics the superconducting gap discussed below. We take $\Delta_{\rm L}\geq \Delta_{\rm R}$ to realize different DoSs between the two terminals. In order to derive elementary closed form expressions, we investigate the extreme case where the two normalized DoSs are Dirac-delta functions, namely $N_{\alpha}(E)=n_{\alpha}\delta(|E|-\Delta_\alpha)$ with $n_{\alpha}>0$ film-dependent constants (with the dimension of an energy). Inserting the DoSs into Eq.~\eqref{eq:currentSecondExpression}, we immediately find:
\begin{equation}
\mathcal{I}(\delta\mu)=-\frac{G_{\rm T}n_L n_R}{e^2}\delta(\Delta_{\rm L}-\Delta_{\rm R}-|\delta\mu|)[f_{\rm L}(\Delta_{\rm L})-f_{\rm R}(\Delta_{\rm R})]
\operatorname{sign}(\delta\mu).
\label{Eq.ToyModel_Currents}
\end{equation}
Despite its simplicity, this toy model captures many of the general features of the bipolar thermoelectricity that we will discuss later: 
\begin{itemize}
    \item since both DoSs have the symmetry of Eq.~\eqref{eq:PHSdos}, the current $\mathcal{I}(\delta\mu)$ is reciprocal, as in Eq.~\eqref{eq:currentReciprocity}. Therefore, whenever thermoelectricity is produced, it is necessarily \emph{bipolar}.
    \item the maximum thermoelectric power occurs for a chemical potential equal to the gap difference, corresponding to the matching-peak condition discussed below.
    \item thermoelectricity occurs only for a given sign of the thermal bias. For $\delta\mu>0$, the sign of $\mathcal{I}$ depends on the difference of Fermi functions in the square brackets of Eq.~\eqref{Eq.ToyModel_Currents}. For $T_{\rm L}\leq T_{\rm R}$, we have $\mathcal{I}(\delta\mu)\geq0$ since $\Delta_{\rm L}\geq \Delta_{\rm R}$. Hence, there is no thermoelectricity in this specific temperature configuration.
    \item thermoelectricity requires a minimum temperature difference. In the thermal bias configuration where $T_{\rm L}> T_{\rm R}$, we then have $\mathcal{I}(\delta\mu)\delta\mu<0$ when $f_{\rm R}(\Delta_{\rm R})-f_{\rm L}(\Delta_{\rm L})<0$, which reads
\begin{equation}
\frac{\Delta_{\rm R}}{T_{\rm R}}>\frac{\Delta_{\rm L}}{T_{\rm L}},
\label{eq:constraint}
\end{equation}
and therefore the minimum thermal bias for thermoelectricity is
\begin{equation}
\Delta T=T_{\rm L}-T_{\rm R}>T_{\rm R}\frac{\Delta_{\rm L}-\Delta_{\rm R}}{\Delta_{\rm R}}.
\end{equation}
\end{itemize}
These quite general characteristics of the toy model are shared by a set of different realizations that we discuss later in this review. To some extent, these features can be regarded as hallmarks of the nonlinear bipolar TE. However, we stress that, due to the nonlinear nature of the effect, it is not restricted to the conditions discussed before. 

\subsection{\bf Asymmetric gap superconductor-insulator-superconductor junctions}
\label{SubSec:asymmetricSIS}
We now present a concrete system in which the conditions for thermoelectricity are satisfied. That is, we consider an SIS$^\prime$ junction between two superconductors with unequal superconducting gaps to ensure different DoSs in the two leads. For simplicity, we assume that the non-dissipative current associated with the Josephson effect can be neglected. In Sec.~\ref{sec:Experiments}, we comment on the impact of Josephson coupling and on practical approaches to suppress this channel.
The DoS of a BCS superconductor reads
\begin{equation}
N^{\rm BCS}_\al(\e,\Delta_\al)=\frac{|\e|\theta(\e^2-\Delta_\al^2)}{\sqrt{\e^2-\Delta_\al^2}}\,,
\label{eq:BCSdos}
\end{equation}
where $\Delta_\al$ is the superconducting gap of lead $\alpha$. For BCS superconductors, the gap in the DoS usually corresponds to the modulus of the superconducting order parameter. In the following analysis, when we discuss the gap, we always refer to the superconducting gap appearing in the lead DoSs.
In particular, the superconducting gap depends on the electronic temperature in the leads, and its temperature dependence can be calculated independently~\cite{tinkham_introduction_2004}. 
However, in simplified treatments, an interpolation formula is often used whose relative error is only a few percent with respect to the numerical complete self-consistent computation~\cite{Gross1986},
\begin{equation}
\Delta_\al(T_\al)\approx\Delta_{\al,0}\tanh[1.74\sqrt{T_{c,\al}/T_\al-1}]\,,
\label{eq:gapInterpolation}
\end{equation}
where $\Delta_{\al,0}\equiv\Delta_\al(T_\al\to0)$ and $T_{c,\al}$ are the zero-temperature gap and the critical temperature, respectively. In the BCS model, these two quantities are related by $\Delta_{\al,0}\simeq1.76 k_BT_{c,\al}$, but their ratio can deviate from this value for strong-coupling superconductors (see e.g. Ref.~\cite{Mitrovic_Ratio_1984}) or in proximity systems (see, e.g. Ref.~\cite{McMillanPR175}).
The temperature dependence of the gap in the BCS limit is plotted in the bottom 
inset of Fig.~\ref{fig:PRL1}.  

We remark that the temperature dependence of the gap is not a necessary condition for thermoelectricity; rather, it is a peculiarity of superconductors that actually limits the maximum temperature range in which such effects can possibly be observed. In fact, as in the discussion of the toy model, for $\Delta_L>\Delta_R$, thermoelectricity can only be observed when $T_L>T_R$. This temperature profile decreases $\Delta_L$, ultimately reducing the junction's gap asymmetry. As a result, thermoelectricity can only be observed in a finite temperature range, which ends when $\Delta_L(T_L)=\Delta_R(T_R)$. 

Due to the square-root singularity of Eq.~\eqref{eq:BCSdos}, the tunneling current of Eq.~\eqref{eq:currentmu} diverges when the bias equals the gap-difference, i.e., $\delta\mu=\pm eV=\pm|\Delta_L-\Delta_R|$, commonly called the matching-peak singularity. 
However, under suitable conditions, these divergences at specific bias values are usually truncated by an energy-resolution cutoff. In particular, in a real system, the DoS's square-root divergences are usually regularized since the real DoSs are finite even for $\e_\al=\Delta_\al$. In this review, we adopt the standard phenomenological correction by introducing a Dynes parameter $\gamma$~\cite{DynesPRL41,DynesPRL53}. For the standard case, 
\begin{equation}
    N_{\rm Dynes}(\e,\Delta,\gamma)=\left|\Re\left[\frac{\e+i\gamma}{\sqrt{(\e+i\gamma)^2-\Delta^2}}\right]\right|\, ,
    \label{eq:DynesDOS}
\end{equation} 
it is straightforward to verify that the corresponding DoS satisfies the energy symmetry. This phenomenological parameter can be associated with QP-lifetime effects and can incorporate deviations in the junction transport from BCS predictions occurring at low temperatures or bias~\cite{Jung_potential_1980,Nahum_Ultrasensitive_1993,Martinis_Energy_2009}; 
it has been shown that Dynes-like DoS can emerge
by coupling the junction with electromagnetic environmental modes~\cite{PekolaDynes} or including other possible mechanisms, such as coupling to magnetic impurity potentials~\cite{Hlubina_PRB84} and proximity effects~\cite{Hosseinkhani_Proximity_2018}.
In standard SIS junctions with identical gaps, this parameter can usually be estimated from the subgap conductance and serves as a characteristic of junction quality. However, for SIS$^\prime$ junctions, QP lifetime can differ on the two sides, so one could also assume a lead-dependent $\gamma_\al$. Nevertheless, these quantities must be treated as energy-independent phenomenological parameters that capture complex physical processes at the junction and its interaction with the environment~\cite{marchegiani_proximity_2026}.

The Dynes parameter quantitatively affects the current 
at the matching peak voltage bias, and consequently the electrical power dissipated (or generated) in the junction. We comment more extensively on this point in Sec.~\ref{SubSec:PowerSIS}.
Typical values for $\gamma$ in superconducting aluminum junctions range from $10^{-7}{\Delta}-10^{-3}{\Delta}$~\cite{Feshchenko2015,SairaPRB85,PekolaDynes,Esat2023,SethiPRApplied24}.

\subsection{\bf General behavior of IV characteristics}
\label{subsec:generalIV}
\begin{figure}[t]
    \centering
    \includegraphics[
    ]{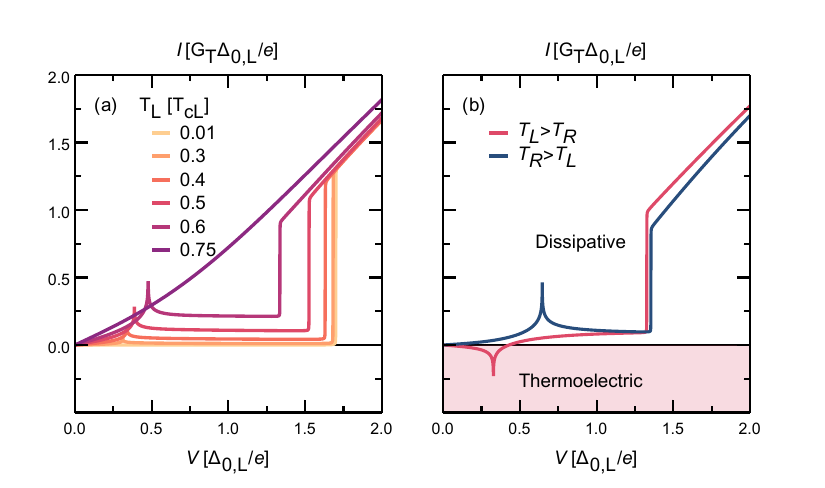}
   \caption{Quasiparticle IV characteristic of a tunnel junction between two gap-asymmetric superconductors (SIS$'$ junction). (a) Thermal equilibrium, (b) thermally biased junctions, with $\tL=0.7\tcL$ and $\tR=0.1\tcL$ (red), $\tR=0.4\tcL$ and $\tL=0.1\tcL$ (blue). Parameters: $\gamma_\alpha=10^{-4}\Delta_{\al,0}$, and $r=0.7$ and $r=0.5$ in panels (a) and (b), respectively. 
   Panel (b) adapted from~\cite{marchegiani_nonlinear_2020}.
}
    \label{fig:IvOriginal}
\end{figure}

It is now interesting to discuss the general behavior of the IV characteristics of the asymmetric junction. Although the thermal equilibrium case has been extensively reviewed in the literature~\cite{barone1982physics,tinkham_introduction_2004}, the case with different temperatures between the electrodes has been less explored. In all cases, the IVs are reciprocal, independently of the lead temperatures, because the superconducting DoSs satisfy the energy-inversion symmetry. We therefore restrict the discussion in Fig.~\ref{fig:IvOriginal} to the $V>0$ branch. 

We briefly review charge transport at thermal equilibrium $\tL=\tR=T$. The IV characteristics exhibit subgap matching peaks at $eV_p=\pm|\Delta_L(T)-\Delta_R(T)|$, together with an abrupt increase in current at the edges of the gap, $eV=\pm[\Delta_L(T)+\Delta_R(T)]$. At thermal equilibrium [Fig.~\ref{fig:IvOriginal}(a)], the matching peaks grow in height and width with temperature, as more QPs are thermally activated, while thermal broadening smoothens the curves. Since $\tcR<\tcL$, the lower gap is more strongly suppressed as the temperature increases [c.f. Eq.~\eqref{eq:gapInterpolation}], hence the position of the matching peak shifts toward higher bias.\footnote{The BCS superconducting gap is practically unaffected by the temperature if, roughly, $T\leq 0.4\,T_{c}$.} 

In Fig.~\ref{fig:IvOriginal}(b), we show the IV characteristics for $\tL\neq\tR$. When the smaller-gap electrode is hotter, $T_R>T_L$, the matching peak appears on the dissipative side, $I(V)V>0$ (blue curve). In contrast, when the larger-gap electrode is hotter, $T_L>T_R$, and provided that $\Delta_L(T_L)>\Delta_R(T_R)$, the current can flow against the bias over a finite voltage range, $I(V)V<0$ (red curve), signaling subgap thermoelectric generation. However, in this temperature configuration, the matching peak shifts toward lower biases due to the different temperature profile.
Therefore, the nonlinear IV characteristic, together with the temperature dependence of the matching-peak position, can provide information on the electronic temperatures of the two electrodes. This feature is crucial in experiments where the temperature profile across the junction cannot always be measured independently.

\begin{figure}[t]
    \centering
    \includegraphics[width=0.9\linewidth]{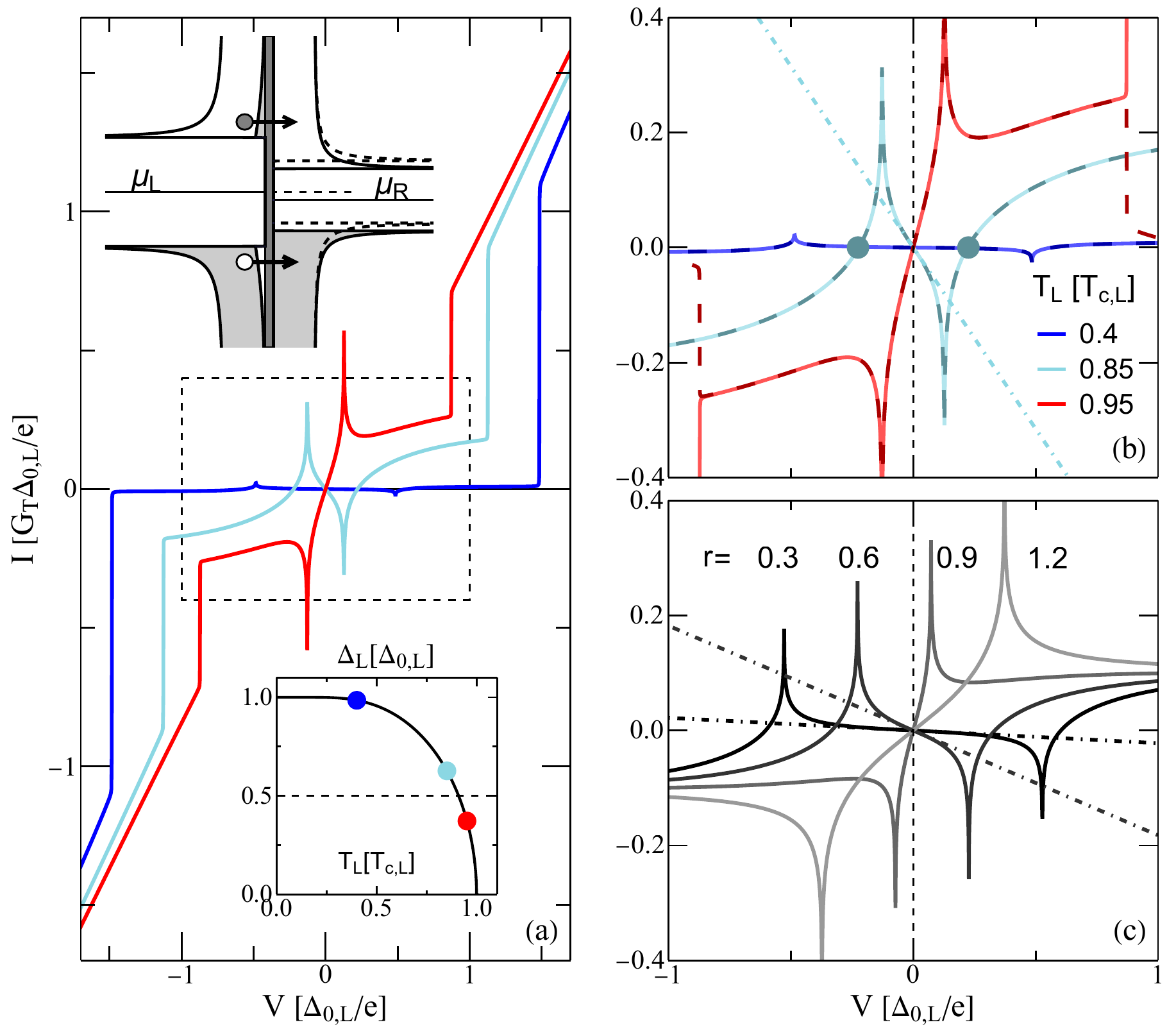}
   \caption{Bipolar thermoelectricity in SIS$^\prime$ junctions.
   (a) Quasiparticle IV characteristic of a thermally biased SIS$^\prime$ junction for $\tR = 0.01\tcL$, $r = 0.5$ and different values of $\tL > \tR$. The curves display NAC and thermoelectric power $\dot{W} = -IV > 0$ at small voltage bias if $\dL(\tL) > \dR(\tR)$. Top inset: energy band diagram of the SIS junction. The combination of the gap in the hot electrode (left) and the monotonically decreasing DoS above the gap of the cold electrode (right) produces a particle current which flows in the opposite direction to the chemical potential gradient. Bottom inset: temperature dependence of the superconducting gap $\dL$. Colored points mark the values of $\dL(\tL)$ for the curves displayed in panel (a). The horizontal dashed line intercepts the $\dL$ curve at the point where $\dL(\tL) =\dR(\tR)\approx\dRo$, i.e., the maximum temperature for the existence of the ANC in panels (a) and (b). (b) Enlargement of the subgap transport in panel (a) (dashed rectangle). Dashed curves give the term in the first line of Eq.~\eqref{eq:IV-mod}. The light-blue dots give the values of the Seebeck voltage $V_S$. (c) Subgap IV characteristics for $\tL = 0.7\tcL$, $\tR = 0.01\tcL$, and different values of $r$. The slopes of the dash-dotted lines in panels (b) and (c) give the values of the ANC at $V\simeq0$, as expressed by Eq.~ \eqref{eq:g0BCS}. Adapted from~\cite{marchegiani_nonlinear_2020}.
}
    \label{fig:PRL1}
\end{figure}

We now specialize the discussion to the SIS$^\prime$ junction sketched in the top inset of Fig.~\ref{fig:PRL1}(a), where the right electrode is kept cold, $\tR\ll\tcL$. Indeed, as in the toy model, we do not expect thermoelectric generation for the opposite configuration, $\tL<\tR$, as long as $\dL(\tL)>\dR(\tR)$.
In this configuration, we vary the temperature of the hot left electrode, $\tL$, and analyze the resulting evolution of the IV characteristics.
The gap asymmetry is defined as $r=\Delta_{R,0}/\Delta_{L,0}=\tcR/\tcL$ and, for the moment, is fixed to $r=0.5$.\footnote{It is important to note that $r$ defines gap ratio only for $T_\alpha\ll T_{c,\alpha}$, while the effective gap ratio $\dL/\dR$ depends on the leads temperature due to temperature evolution of the gaps.}
In Fig.~\ref{fig:PRL1}(a), we consider three different values for $T_L$ identified by the different color lines and show the full IV characteristics. Notably, we see that for blue and cyan, the system produces thermopower around the matching peaks, whereas for the higher temperature (red line), the matching peaks are again purely dissipative. This nonmonotonic evolution is essentially determined by the fact that as $T_L$ increases, the larger superconducting gap $\Delta_L(T_L)$ progressively closes, as shown by the colored points in the bottom inset of the figure. When $\Delta_L(T_L)<\Delta_R(T_R)$, the effective gap asymmetry in the junction is inverted and the thermoelectricity disappears. This peculiar result reveals the intrinsic nonlinearity of the TE: in this system, an increase in the temperature difference does not necessarily yield a stronger thermoelectric effect. 
This counterintuitive behavior of the bipolar  can also be used to distinguish it from the usual thermoelectric behavior.

When the side with the smaller gap is very cold, it is very convenient to express the current  as~\cite{marchegiani_nonlinear_2020}  
\begin{align}
 I=&\frac{G_{\rm T}}{e}\int_{0}^{\infty}d\e \dosL(\e)f_{\rm L}(\e)[\dosR(\e_+)-\dosR(\e_-)]+\nonumber\\
&\frac{G_{\rm T}}{e}\int_{0}^{\infty}d\e \dosL(\e)[\dosR(\e_-)f_{\rm R}(\e_-)-\dosR(\e_+)f_{\rm R}(\e_+)] 
\label{eq:IV-mod}
\end{align}
with $\e_{\pm}=\e\pm eV$, where we used the energy symmetry of the DoSs and of the Fermi functions to restrict the integral to positive energies and collect them in two terms. 
In the limit $eV,k_BT_R\ll\Delta_L$, the second term is 
exponentially suppressed $\exp[-\Delta_L/k_B T_R]\ll 1$, and can be neglected. The current is then dominated by the term in the first line of Eq.~\eqref{eq:IV-mod}, which involves the combination 
$\dosR(\e_+)-\dosR(\e_-)$. For a positive (negative) bias $eV > 0$ ($eV < 0$), this difference is negative (positive) wherever $\dosR(\e)$ is locally decreasing, i.e., for $\e > \Delta_R(T_R)$ in the BCS case. Since $k_BT_L \lesssim \Delta_L$ ensures that $f_L(\e)$ still provides a non-negligible contribution in this energy range, the integral can be negative (or positive for $eV<0$), providing a thermoelectric response  [$I(V)V < 0$]. 

It is interesting to zoom in on the subgap bias regime [see dashed box of panel (a)] as shown in Fig.~\ref{fig:PRL1}(b). The IV calculated from the full formula is compared with the approximation retaining only the first term of Eq.~\eqref{eq:IV-mod} (dashed lines). As can be seen, this fully captures the tunnel-junction behavior in the limit $T_R\to 0$ for almost all values of $T_L$ and biases considered. 

We also indicate, with blue circles for the cyan curve, the Seebeck voltages $\pm V_S$ at which the external voltage compensates for thermoelectric generation, resulting in $I=0$. Equivalently, these are the maximum voltages that would develop in an infinite resistive load connected to the junction. It is intriguing to note that the bipolarity of the effect implies two possible equivalent Seebeck voltages $\pm V_S$ for the \emph{same} sign of the temperature difference. This contrasts strongly with the usual unipolar s, in which the sign of the Seebeck voltage indicates whether the transport is dominated by particles or holes. In other words, the junction essentially behaves as both n-type and p-type material, depending on the settled voltage bias configuration.  

Finally, in panel (c), we show how the IV characteristic changes for different values of the gap asymmetry $r$, while keeping the temperatures $T_L=0.7 T_{c,L}$ and the cold temperature $T_R\approx 0$ fixed. At these temperatures, thermoelectricity completely disappears for $r\gtrsim 0.9$, when the gap asymmetry becomes too small to 
sustain the effect at the given temperature difference. These observations show that to optimally observe the bipolar , one has to tune the thermal gradient and the junction asymmetry accordingly. In panels (b) and (c), we also show, in dashed lines, the slope for the differential conductance $g_0=dI/dV|_{V\to0}$, which can be negative. This result shows that the bipolar thermoelectric behavior can also be present at very low biases if the residual Josephson current can be neglected (see Sec.~\ref{Sec: Exp}). 

\begin{figure}[t]
    \centering
    \includegraphics[width=1\linewidth]{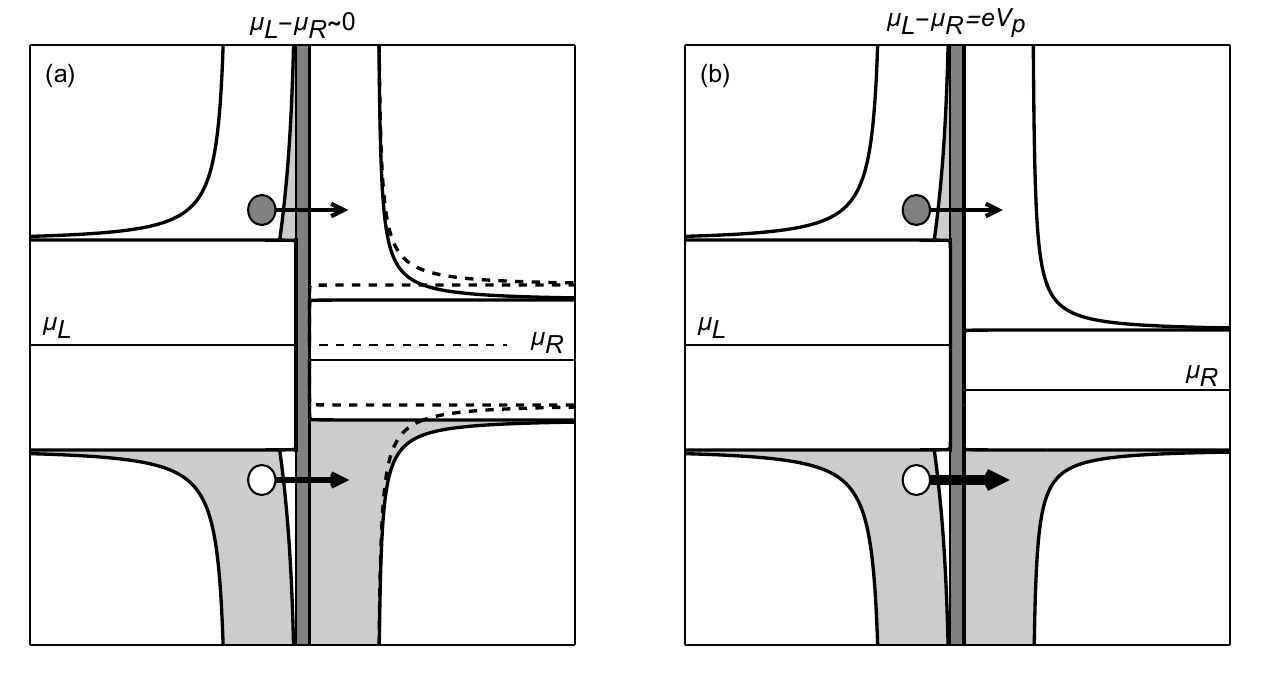}
   \caption{Energy band schematics of different thermoelectric regimes for nonlinear thermoelectricity in SIS$^\prime$ junction. (a) Linear-in-bias regime: thermoelectricity arises since the hole current (hollow circles) is larger than the particle current (filled circles), due to the DoS in the cold electrode (on the right) being locally monotonically decreasing. (b) Nonlinear-in-bias regime: the thermoelectric current gets enhanced when the bias aligns the singularity peaks of the superconducting DoS. Adapted from Ref.~\cite{marchegiani_nonlinear_2020}.
   \label{fig:diagramSIS} 
}
\end{figure}

In Fig.~\ref{fig:diagramSIS}, we illustrate the two regimes of bipolar thermoelectricity in the SIS$^\prime$ junction. The linear-in-bias regime is sketched in Fig.~\ref{fig:diagramSIS}(a). When a small positive bias is applied, the chemical potential of the cold electrode shifts relative to the zero-bias case (dashed), so that the available DoS is reduced (increased) for quasiparticle (quasihole) states at positive (negative) energies. 
This imbalance, which also determines the strong detail-balance violation, favors the quasihole flux over the quasiparticle flux, driving a net current opposite to the bias and giving rise to an absolute negative conductance. 
In the nonlinear-in-bias regime, shown in Fig.~\ref{fig:diagramSIS}(b), thermoelectric current is further enhanced when the bias aligns the BCS singularities of the two electrodes at the matching peak, where the thermoelectric power reaches its maximum. In what follows, we discuss the two regimes in more detail. 

It is important to note that for SIS$^\prime$ a photoelectric effect has also been envisioned in Ref.~\cite{aronov1975photoeffect}, which in spirit is very similar to the discussed mechanism; however, in such a case, the involvement of microwave irradiation and the competition with photon-assisted tunneling make the experimental verification much more complex~\cite{gershenzon1986absolute,gershenzon1988absolute,gijsbertsen1996quasiparticle}.

\subsubsection{\textbf{Linear-in-bias thermoelectricity}}
\label{SubSec:linear-in-bias}
\begin{figure}[t]
    \centering
    \includegraphics[width=1\linewidth]{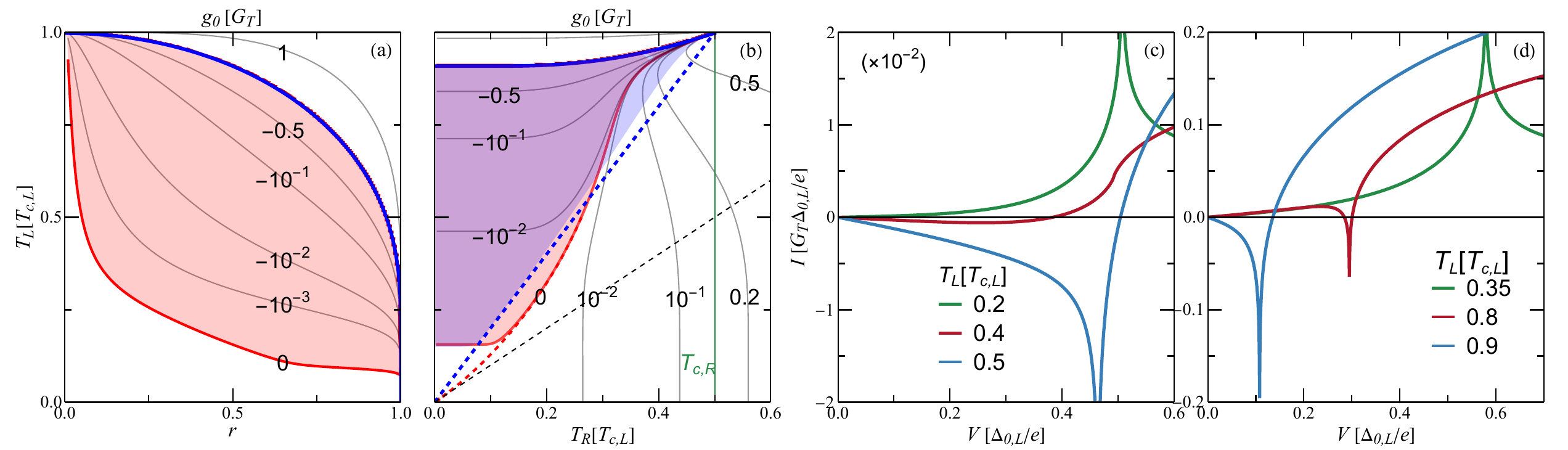}
   \caption{
   Linear-in-bias and nonlinear-in-bias thermoelectric regimes in SIS$'$ junctions.
   (a) Contour plot of $g_0$ vs $\tL$ and $r$ for $\tR = 0.001T_{c,L}$. The linear-in-bias thermoelectric contribution is represented by the red area. (b) Contour plot of $g_0$ vs $\tL$ and $\tR$ for $r=0.5$. The red area and the blue area denote the linear-in-bias thermoelectric region and the nonlinear-in-bias thermoelectric region, respectively. The dashed lines give the zero temperature difference contour $\tL = \tR$ (black), the contour $\tL = \tR/r$ (blue), and the boundary from negative (left to the curve) to positive (right to the curve) $g_0$ for BCS DoSs (red). (c)-(d) Onset of the thermoelectricity by raising the temperature of the left electrode for $r = 0.5$ and (c) $\tR = 0.2\tcL$ (first linear-in-bias then nonlinear-in-bias) or (d) $\tR = 0.35\tcL$ (first nonlinear-in-bias then linear-in-bias). The Dynes parameters are set to $\gamma_j=10^{-4}\Delta_{0,j}$. Adapted from~\cite{marchegiani_superconducting_2020}.
}
    \label{fig:PRB3}
\end{figure}

The distinctive signature of thermoelectricity is a current flowing against the applied voltage bias, that is, $I(V)V<0$. Although for EIS DoSs no thermoelectricity is possible in the linear-response regime, a TE response is still possible, linearly in the voltage bias, but for a \emph{nonlinear} temperature difference. 
In this regime, the relevant quantity to analyze is the zero-bias conductance, defined as the derivative of the current with respect to the voltage bias at $V=0$. 
For DoSs satisfying EIS, one obtains
\begin{equation}
g_0=2G_T\int_{0}^\infty d\e \dosL(\e)\left\{[f_L(\e)-f_R(\e)]\frac{d\dosR(\e)}{d\e}-\dosR(\e)\frac{df_R(\e)}{d\e}\right\}\,.
\label{eq:g0}
\end{equation}
This expression follows from Eq.~\eqref{eq:RateSimplified} by expanding the current to the first order in the bias, with the voltage shift assigned to the right contact. This equation reduces to the expression given in Ref.~\cite{marchegiani_superconducting_2020} [Eq. (3) in the reference], taken in the limits $T_L,\dL/k_B\gg T_R$ and $\dL(\tL)>\dR(\tR)$, where the last term of the previous equation can be neglected.
Indeed, the linear-in-bias TE response is signaled by a negative value of $g_0$.
Then, for a superconducting junction in the limit of vanishing Dynes parameters, i.e. $\gamma_i\to0$, this gives

\begin{align}
g_0&=2G_T\int_{0}^\infty d\e \dosL(\e)f_L(\e)\frac{d\dosR(\e)}{d\e}\\
&=-2G_T\dR^2\int_{\dL}^\infty d\e \frac{\e f_L(\e)}{\sqrt{\e^2-\dL^2}}\frac{1}{(\e^2-\dR^2)^{3/2}}\,,
\label{eq:g0BCS}
\end{align}
that is indeed always a negative expression.
Figure~\ref{fig:PRB3}(a) displays the contour plot of $g_0$, with the corresponding values reported on the gray lines, as a function of the hot lead temperature and the gap ratio in the zero temperature limit of the cold electrode $T_R\to 0$. The shaded red region indicates the parameters for which linear-in-bias thermoelectricity occurs, i.e., $g_0<0$. When the temperature of the hot lead is too large, such that the condition $\dL(\tL)>\dR(\tR)$ (solid blue curve gives the equality) is no longer satisfied, the junctions display a dissipative response. In the figure, the minimum temperature required for negative differential conductance is also shown (red line). 

A complementary analysis is shown in Fig.~\ref{fig:PRB3}(b), where the contour plot of $g_0$ is shown as a function of both lead temperatures 
at a fixed gap asymmetry ratio $r=0.5$. The red shaded area identifies the region where $g_0<0$, corresponding to linear-in-bias TE generation (see also Sec.~\ref{SubSec:linear-in-bias}). This region does not necessarily coincide with the blue shaded area, which marks the operating conditions for nonlinear-in-bias TE generation. Several boundaries delimit the TE regions. First, TE generation is not possible when the temperature of the larger-gap electrode is too high, since in that case $\dL(\tL)<\dR(\tR)$ and the effective gap asymmetry is inverted, as indicated by the upper blue line. Second, a minimum temperature difference is required and, notably, the equilibrium line $\tL=\tR$ (black dashed line) does not intersect the red or blue area. Third, TE generation is also suppressed when the smaller-gap electrode enters the normal state, $\tR>\tcR$, as marked by the green vertical line. The onset of nonlinear-in-bias TE generation is well approximated by $\tL\simeq \tR/r$ (blue dashed line). Finally, for the finite Dynes parameter $\gamma$ used in the plot, the minimum value of $\tL$ required for linear-in-bias TE generation saturates as $\tR\to0$. This saturation decreases as $\gamma$ is reduced; the red dashed line shows the corresponding boundary in the ideal limit $\gamma=0$.

Concerning the different shapes of the IV characteristics when the linear-in-bias and nonlinear-in-bias are not present at the same time, we show a selection of the different possible cases in Figs.~\ref{fig:PRB3} (c) and (d), which reflect the rich nonlinear physics of the SIS$^\prime$ junction out of equilibrium.

\subsubsection{\textbf{Nonlinear in bias thermoelectricity}}
\label{SubSec:PowerSIS}
\begin{figure}[t]
    \centering
    \includegraphics[
    ]{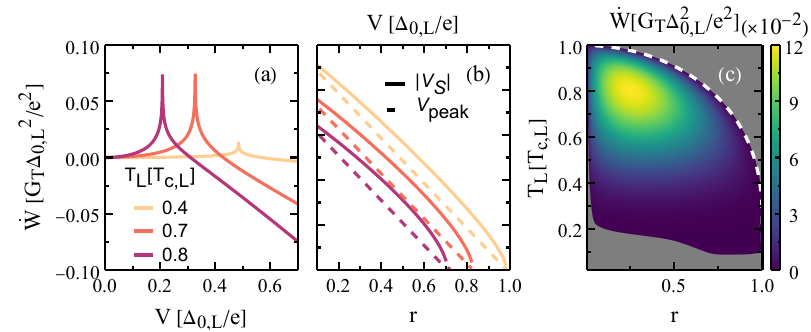}
   \caption{
   Nonlinear-in-bias thermoelectricity. (a) Thermoelectric power $\dot{W}$ vs voltage bias $V$ for $r=0.5$, $\tR = 0.01\tcL$ and a few values of $\tL$.  (b) Seebeck voltages $V_S$ vs $r$ for $\tR = 0.01T_{c,L}$ and the same values of the temperature as in panel (a) (solid lines). The dashed curves show the corresponding singularity-matching peak value $V_p$. (c) Density plot of the thermoelectric power $\dot{W}$ as a function of $r$ and $\tL$ for $\tR = 0.01\tcL$. The dashed line gives the condition $\dL(\tL)=\dR(\tR)$. Adapted from~\cite{marchegiani_superconducting_2020}. 
}
    \label{fig:PRL2}
\end{figure}

At finite bias $V\neq 0$, power is generated ($\dot{W}>0$) or dissipated ($\dot{W}<0$) in the junction, where $\dot{W}=-IV$. Although bipolar thermoelectricity can be probed arbitrarily close to a zero voltage bias, the thermoelectric power is vanishingly small for $V\to 0$. In this section, we discuss, at finite voltages, the dependence of the bipolar TE and the resulting thermoelectric power. 
Figure~\ref{fig:PRL2}(a) displays the voltage dependence of the thermoelectric power $\dot{W}=-IV$ as a function of the applied voltage for $r=0.5$, $\tR=0.01\tcL$, and different values of the temperature of the hot lead. We only show positive values of $V$, taking advantage of the even symmetry of thermoelectric power, which is again the direct consequence of the reciprocity of the current discussed before. The power is non-monotonic with the voltage bias, being zero at $V=0$ (as expected) and at the Seebeck voltage $V_S$ (where $I=0$ as well). Between these values, the maximum power is reached very near the matching-peak singularity $V=V_p$. For higher voltages, $V\geq V_S$, the power is negative, reflecting the dissipative behavior of the junction. It is also interesting to note that increasing the hot temperature, and consequently the temperature difference, does not significantly affect the maximum thermoelectric power. This is in stark contrast to linear thermoelectricity, where, in general, as the temperature difference $\Delta T$ increases, the maximum thermoelectric power is quadratic in the thermal gradient $\dot{W}_{\rm max}=GS^2 \Delta T^2/4$, where $G=(dI/dV)|_{V,\Delta T=0}$ is the charge conductance and $S=-(dV/dT)|_{I=0}$ the Seebeck coefficient~\cite{benenti_fundamental_2017}. This result further evidences the nonlinear nature of the effect discussed in this review. 

In panel (b), we show how Seebeck voltages $V_S$ (solid) and matching-peak voltages $V_p$ (dashed) evolve when we change the junction asymmetry $r$ or the temperature of the larger-gap lead (lines of different colors). Clearly, Seebeck voltages are always higher than the voltage at which the maximum thermopower is generated, $V\approx V_p$.   

Finally, to explore the thermoelectric power as a function of the system parameters, we can compute its value at the singularity of the matching peak $V=V_p$, which well represents the maximum value for sufficiently low Dynes parameters.
In panel (c), we show a density plot of $\dot{W}$ as a function of the hot-lead temperature and the gap ratio, with the cold electrode temperature fixed at $\tR=0.01\tcL$. The thermoelectric power is maximum in a parameter region (yellow area) located around the hot-temperature lead $\tL=0.8\tcL$ and $r\sim 0.35$. Clearly, no thermoelectric response is possible for temperature values for which $\dL(\tL)\leq\dR(\tR)$ (see dashed white line), and at low values of the temperature, with the latter feature depending on the specific Dynes parameter used ($\gamma_\al=10^{-4}\Delta_\al$) as we also discussed before. 

In this respect, we note that the maximum thermoelectric power depends explicitly on $\gamma_\al$ (diverging in the limit $\gamma_\al\to 0$). This hypothetical singular behavior requires an infinite heat flux to be externally supplied to the hot lead and removed from the cold one. However, as the thermoelectric power grows, it becomes progressively 
more challenging to keep the superconducting cold terminal at the desired temperature, and eventually the thermoelectric power becomes limited by overheating effects in the cold terminal~\cite{Maasilta_PRB_85,lucchesi2025out}. Nevertheless, even ignoring these practical considerations, the small but finite proximity effects between the leads~\cite{marchegiani_proximity_2026} or effects due to the electromagnetic environment~\cite{PekolaDynes} are expected to regularize the divergences in the DoS, determining a finite maximum power.
Finally, we note that the optimal temperature bias and gap ratio for power generation are almost independent of the Dynes parameter, as shown in the Appendix~\ref{app:optimalPowerGamma}.

\subsection{Heat Current} 
To investigate the thermodynamic efficiency of the , we compute the heat currents $\dot Q_{\alpha}$ in the junction and compare them to the thermoelectric power generated by the junction. We can generalize the expression for the charge and heat currents coming out from the $\alpha$-lead $I_{\alpha}$ using the compact notation~\cite{marchegiani_nonlinear_2020} 
\begin{equation}
\begin{pmatrix}
I_{\alpha}   \\   
\dot Q_{\alpha}
\end{pmatrix}
=\frac{G_{\rm T}}{e^2}\int_{-\infty}^{+\infty}d\epsilon 
\begin{pmatrix}
-e \\   
\epsilon_\alpha
\end{pmatrix} N_{\alpha}(\epsilon_{\alpha})N_{\bar\alpha}(\epsilon_{\bar\alpha})[f_\alpha(\epsilon_\alpha)-f_{\bar\alpha}(\epsilon_{\bar\alpha})]
\label{eq:IVandQ}
\end{equation}
where $\epsilon_\alpha=\epsilon-\mu_\alpha$ and it is easy to verify that $I\equiv I_L=-I_R$ is connected to the equations discussed previously. Indeed, the general expressions for the heat fluxes can also be derived using the Fermi Golden rule, as for the charge current. The main difference is that the transition probability at each energy of Eq.~\eqref{eq:Goldenrulerate} 
needs to be multiplied by the QP energy measured with respect to the lead chemical potential $\mu_\alpha$.
Although the current $I$ is reciprocal due to the symmetry of~\eqref{eq:PHSdos}, the heat fluxes are instead even in bias, i.e., $\dot Q_{\alpha}(-V,\tL,\tR)=\dot Q_{\alpha}(V,\tL,\tR)$, independently of $\tL$ and $\tR$.  

\subsubsection{\textbf{Energy Conservation}} 
\begin{figure}[t]
    \centering   \includegraphics[width=1\linewidth]{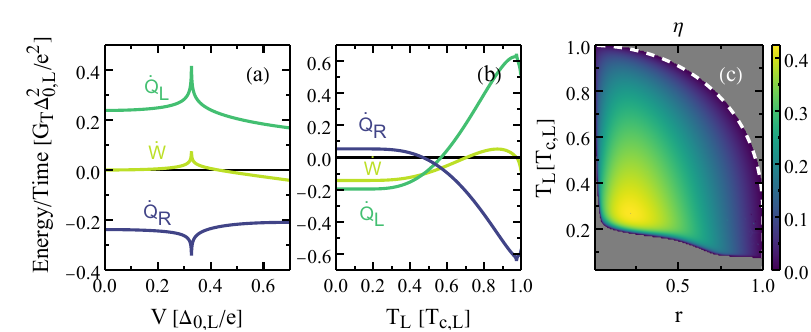}
   \caption{
   Energy conservation and thermodynamic efficiency in bipolar thermoelectricity. Power generated, and heat current flowing out of the two leads vs 
   (a) $V$ for $\tL = 0.7\tcL$, $\tR = 0.01\tcL$, $r=0.5$ and  (b) $\tL$ for $\tR = 0.2\tcL$, $r=0.3$ at the singularity matching bias. (c) Density plot of efficiency vs $\tL$ and $r$ for $\tR = 0.01\tcL$. In all the panels, we set $\gamma_j=10^{-4}\Delta_{0,j}$.
}
    \label{fig:energyCons}
\end{figure}

Using Eq.~\eqref{eq:IVandQ}, one can show that the tunneling currents satisfy 
the first law of thermodynamics (energy conservation)~\cite{WhitneyPRB87,benenti_fundamental_2017}. Energy conservation can be written as 
\begin{equation}
\label{eq:energyconservation}
\dot{Q}_L+\dot{Q}_R-\dot{W}=0\,.
\end{equation}
where $\dot{W}=-IV$ is the thermoelectric power introduced before. 
Clearly, this relationship shows the deep connection between the heat transfer within the junction and the power generated/dissipated by the junction itself.
For completeness, in Fig.~\ref{fig:energyCons}(a)-(b) we show the typical behavior of the heat fluxes $\dot{Q}_i$ and their relation to electrical power $\dot{W}$ for the bipolar . In particular, in panel (a), we show the bias dependence, with the maximum thermoelectric power observed at the matching peak, as anticipated earlier. We see that this corresponds exactly to the condition where the heat transfer from the hot lead to the cold one is maximum. In panel (b), we discuss the dependence on the hot temperature $T_L$, keeping the bias $V=V_p$ fixed at the matching peak and assuming the cold temperature to be very low. We see that if the hot temperature is not sufficiently high, the junction is dissipative $\dot{W}<0$. In particular, in such a biased regime, we see that the heat flows from the cold $\dot{Q}_R>0$ to the hot terminal $\dot{Q}_L<0$. The junction operates as a refrigerator for the cold terminal (see below). When 
$T_L\gtrsim 0.6 T_{c,L}$, the junction starts to produce thermoelectric power: as in any other thermal engine, the hot terminal transfers heat to the cold terminal, part of which is converted into electrical work.  

Similarly, one can show that the second law of thermodynamics is satisfied. In particular, recalling that the entropy production rate of the reservoirs at the stationary state satisfies~\cite{benenti_fundamental_2017}
\begin{equation}
\label{eq:entropy1}
\dot{S}=-\frac{\dot{Q}_L}{T_L}-\frac{\dot{Q}_R}{T_R}\,, 
\end{equation}
where, indeed, on the right side, we included only the entropy production rate of the contacts.
One immediately finds that no spontaneous ($\dot S\geq 0$) thermoelectric power generation can occur at thermal equilibrium $\tL=\tR=T$, since $\dot{W}=\dot{Q}_L+\dot{Q}_R=-T\dot S\leq 0$, then implying only a dissipative behavior for the junction. More generally, by combining Eqs.~\eqref{eq:energyconservation} and \eqref{eq:entropy1}, it is also straightforward to show that, for $\tL>\tR$, the efficiency $\dot{W}/\dot{Q}_L$ cannot be greater than the Carnot limit $\eta_C=1-\tR/\tL$. 

\subsubsection{\textbf{Thermodynamic Efficiency}}\label{SubSec:Efficiency}
We now discuss the thermodynamic efficiency of bipolar thermoelectric conversion.
It is instructive first to explicitly show how the toy model introduced in Sec.~\ref{subsec:toy_model} obeys the second law of thermodynamics. Inserting the DoS $N_\alpha=n_\alpha\delta(\e-\e_\alpha)$ into Eq.~\eqref{eq:IVandQ}, one finds for the efficiency  
\begin{equation}
\eta=\frac{\dot W}{\dot Q_{\rm L}}=\frac{\Delta_{\rm L}-\Delta_{\rm R}}{\Delta_{\rm L}}=1-\frac{\Delta_{\rm R}}{\Delta_{\rm L}}.
\label{eq:eff2delta}
\end{equation}
Recalling that a thermoelectric response only occurs when the inequality of Eq.~\eqref{eq:constraint}  is satisfied, the efficiency is necessarily smaller than Carnot limit $\eta_{\rm C}$. Nevertheless, for nonlinear temperature differences $T_R\ll T_L$, the Carnot efficiency is very close to the theoretical bound dictated by energy conservation, i.e., $\eta_C\approx 1$.  

Next, we consider the thermodynamic efficiency of the SIS$^\prime$ junction. Figure~\ref{fig:energyCons}(c) shows a density plot of efficiency as a function of asymmetry $r$ and the temperature of the hot-lead $T_L$. 
The maximum efficiency, $\eta_{\rm max}\simeq0.4$, is obtained for relatively low hot-lead temperatures, around $\tL\simeq0.3\tcL$, and for gap asymmetries close to $r\simeq0.3$. This maximum lies near the boundary between the TE-generator regime and the refrigeration regime discussed below. However, we report that an \emph{absolute} thermodynamic efficiency of $40\%$ is remarkable since the best absolute thermoelectric performances measured are of the order of $14-15$\%~\cite{Bu2021GeTeEfficiency,Li_Adma_2023}.
It is also fruitful to compare this behavior with the TE power of Fig.~\ref{fig:PRL2}. We note that even if the conditions to have maximum efficiency and maximum power do not coincide in terms of the optimal hot-lead temperature, they are instead almost the same for the asymmetry, showing that $r\approx 0.3$ may be optimal for the SIS$^\prime$ junctions to reach the best performances.

\begin{figure}[t]
    \centering   \includegraphics[width=1\linewidth]{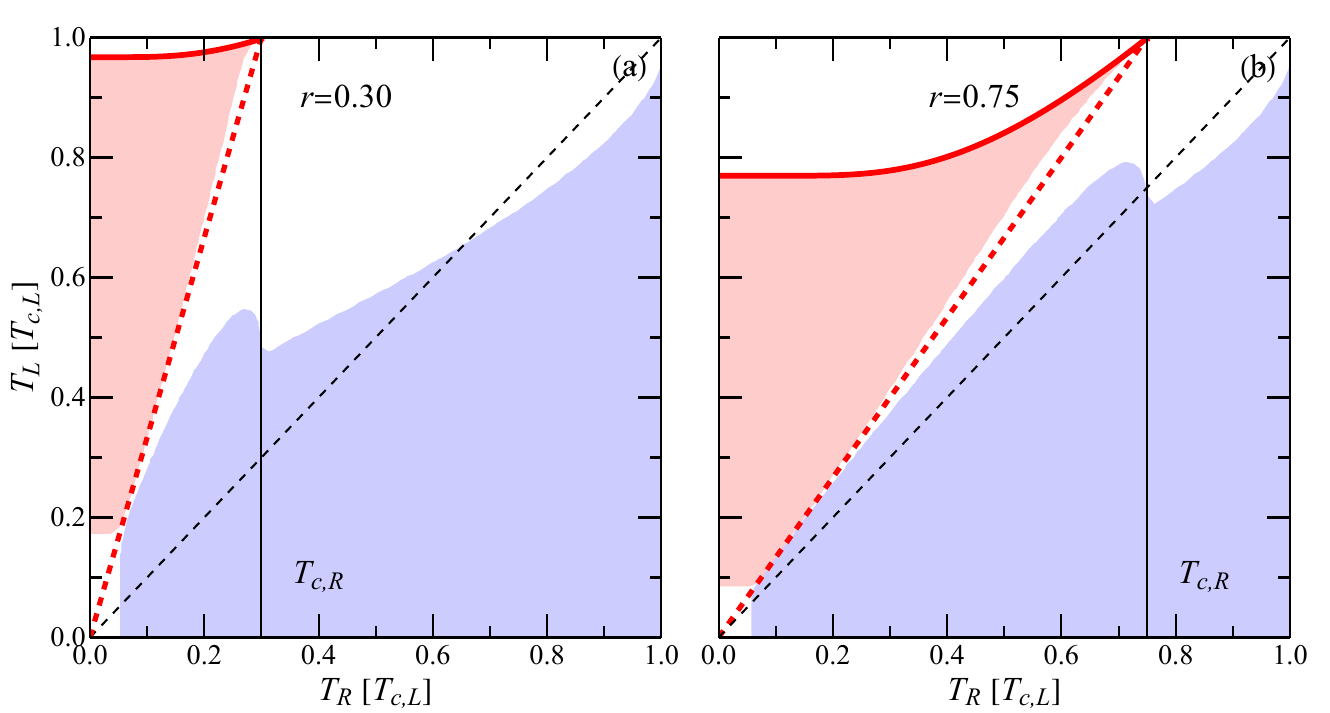}
   \caption{
   Cooling vs engine regimes for a thermally biased SIS$^\prime$ junction for (a) $r=0.3$ and (b) $r=0.75$. The red area denotes the heat engine region ($\dot W>0$), while the blue area denotes the region where heat is extracted from the low-gap electrode ($\dot {Q} _ {R}>0$). The dashed black line is the equal temperature line, while the red dashed line corresponds to $T_L=T_R/r$. The upper solid red curve ($\dL(\tL)=\dR(\tR)$) also separates the heat engine region from the dissipative region (white) as determined by the inversion of the gap asymmetry. The Dynes parameters are $\gamma_j=10^{-4}\Delta_{0,j}$. Adapted from~\cite{marchegiani_superconducting_2020}. 
}
    \label{fig:PRB4}
\end{figure}

\subsubsection{\textbf{Operating regimes for SIS$^\prime$  junction}} The TE physically signals the intrinsic coupling between thermal and electrical transport. While a finite voltage/current can be generated under a temperature bias (Seebeck effect), a temperature gradient can be generated from a current flowing through a thermoelectric element (Peltier effect). In the linear regime, the two phenomena are usually interconnected by the Onsager symmetry principle~\cite{benenti_fundamental_2017} (see also Sec.~\ref{subsec:linearTE}). However, since the bipolar thermoelectricity is a nonlinear effect, it is interesting to explore the relationship between the two cases.

We previously mentioned that electrical currents can be used to cool one side of a thermoelectric element; we note that this effect is fundamentally different from Joule heating which concerns dissipation, and also holds for non-thermoelectric materials. While in the previous subsection we addressed only thermoelectric power generation, we present here a more extensive thermodynamic analysis, as reported in~\cite{marchegiani_superconducting_2020}. In particular, this Peltier-like cooling corresponds to evaporative cooling in superconducting tunnel junctions~\cite{giazotto_opportunities_2006,Muhonen2012}, established both theoretically and experimentally almost thirty years before the clear identification of the bipolar TE. It is interesting to note that one application of SIS$^\prime$ tunnel junctions in thermal management is as a cooler~\cite{FRANK1997281,ManninenAPL74,Muhonen2012}. The large superconducting gap, indeed, acts as an energy filter for QPs; then a suitable voltage can provide cooling of the electronic temperature of (a) the lower-gap superconductor in an asymmetric-gap SIS$^\prime$ junction ($r\neq 1$) and of (b) a normal metal island in a normal metal-insulator-superconductor (NIS) junction (known as NIS cooling). 

First, we show that thermoelectric power generation and Peltier cooling are mutually exclusive, with a general thermodynamic argument.  We recall that the first principle of thermodynamics (energy conservation) imposes the relation $\dot{Q}_R = \dot{W} -\dot{Q}_L$. In a thermoelectric generator, we have $\dot{W},\dot{Q}_L>0$; therefore, the condition for cooling, i.e., $\dot{Q}_R>0$, would require $\dot{W} > \dot{Q}_L > 0$ in clear violation of the second law of thermodynamics.

We now discuss the thermodynamic behavior (cooling vs thermoelectric generation) of a voltage-biased asymmetric junction (at $V=V_p$) between two superconductors $(r\neq 1)$, in relation to the temperatures of the two electrodes $\tL, \tR$. The interplay between thermoelectric power generation (red areas, $\dot W>0$) and Peltier-like cooling (blue areas, $\dot{Q}_R>0$) is shown in Fig.~\ref{fig:PRB4}, for $r = 0.3$ (panel a) and for $r = 0.75$ (panel b). The diagonal dashed lines set the equal temperature contours $\tL = \tR$, and the vertical solid lines
the thresholds $\tR = \tcR = r\tcL$. For $\tL\leq \tR$ we do not see a nonlinear thermoelectric response, as already discussed in Sec.~\ref{subsec:generalIV}.  
The region $\tL\geq \tR$ is richer from a thermodynamic perspective. For small temperature differences, the low-gap electrode cools, a phenomenon that can persist even for $\tR>\tcR$ (thereby becoming NIS cooling). As the temperature difference increases, the junction is dissipative at first (i.e., $\dot{Q}_R<0$) and when it becomes large enough (roughly given by $\tL > \tR/r$, see red dashed lines), it displays thermoelectric power generation.  Finally, TE disappears for values of $\tL$ where $\dL(\tL) < \dR(\tR)$ [region bounded by solid red curves in Figs.~\ref{fig:PRB4}(a)-(b)].

This progression from thermoelectricity towards cooling 
may remind one of the typical behavior of a linear thermoelectric engine~\cite{BenentiPRL124}.  In the linear regime, the parameter that controls the transition from cooling to heating is given by the change in bias around the Seebeck voltage. However, in our case, this is determined solely by the temperature difference. In contrast with the linear response regime, where the two operating regimes are separated by a bias proportional to the small temperature gradient, here the two regimes are well separated in temperature; again, this is a consequence of the purely nonlinear nature of the effect.

The plots also show that thermoelectric power generation in the superconducting junction occurs over a narrower parameter range than evaporative cooling.  
In fact, for $\tR \geq \tcR$, i.e., when the lower-gap superconductor is in the normal phase, evaporative cooling is still achievable [see $\tR > 0.3\tL$ in Fig.~\ref{fig:PRB4} (a)], while the TE necessarily requires energy-dependent DoSs in both leads (see Sec.~\ref{SubSec:Necessary}). This requirement is guaranteed in our system only when the right electrode remains superconducting.

\subsection{\textbf{Out-of-equilibrium spontaneous symmetry breaking.}}
\label{SubSec:symmetryBreaking}
\begin{figure}[t]
    \centering   \includegraphics[
    ]{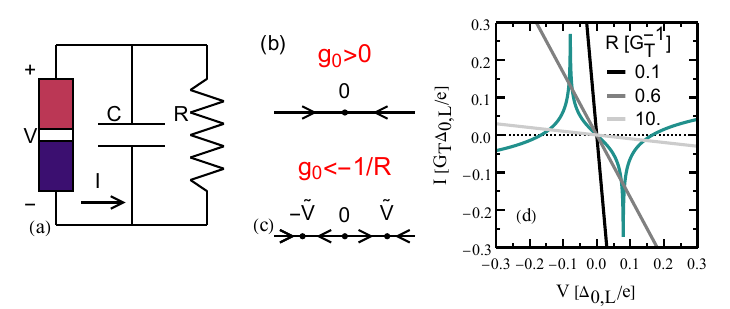}
   \caption{
   Bistability of the thermoelectric voltage. (a) Circuit scheme: a thermally biased Josephson junction is connected in parallel with a load with resistance $R$ and capacitance $C$. (b)-(c) Stability diagram. For a dissipative junction $g_0>0$, the zero-current solution is stable. For thermoelectric systems, two metastable solutions (with opposite voltages) are possible for sufficiently large loads. (d) Graphical solution of Eq.~\eqref{eq:currConsRC0}. The solutions are given by the intersections of the $I(V)$ characteristic (green curve) and the load line $-V/R$ (grayscale). Depending on $R$, we can have up to 5 different solutions. Adapted from~\cite{marchegiani_nonlinear_2020}. 
}
    \label{fig:PRLEH}
\end{figure}

In Sec.~\ref{subsec:generalIV}, we discussed the current-voltage characteristic of a thermally biased tunnel junction between asymmetric-gap superconductors (assuming no Josephson coupling). Here, we show how the occurrence of an absolute negative conductance affects the junction's electrical stability when placed in an external circuit. The electrical nonlinearity of the junction, which exhibits absolute negative conductance (power production) as well as negative differential conductance (electrical instability), will crucially determine its electrical behavior.

For concreteness, we focus on the minimal circuit of Fig.~\ref{fig:PRLEH}(a): the junction is modeled as a nonlinear element (colored) with characteristic $I(V, \tL, \tR)$, connected in parallel with a capacitance $C$ (accounting for the junction's own capacitance) and a load resistance $R$. Current conservation in the circuit imposes the equation:
\begin{equation}
I(V, \tL, \tR) = -C\frac{dV}{dt} -\frac{V}{R}\,.
\label{eq:currConsRC0}
\end{equation} 
Hereafter, we limit ourselves to stationary DC solutions, i.e. $V (t) = \tilde{V} $, obtained by setting $\dot{V} = 0$ in Eq.~\eqref{eq:currConsRC0}; the possible values of  $\tilde{V}$ are then given by the implicit equation 
$RI(\tilde V, \tL, \tR) +\tilde{V}=0$.
\footnote{In Sec.~\ref{subsec:TEcircuit}, we also discuss possible time-dependent solutions, which explicitly depend on the AC response of the external circuitry.}
Due to IV reciprocity, $I(V, \tL, \tR) = -I(-V, \tL, \tR)$, the implicit equation has an odd
number of solutions, and $\tilde{V} = 0$ is always a solution, irrespective of $R, \tL, \tR$. Examples with one, three, and five solutions are shown in Fig.~\ref{fig:PRLEH}(d). Changing the load resistance, indeed, one can find different solutions due to the nonlinearity of the 
IV characteristic.

The stability of these solutions can be investigated by linearizing Eq.~\eqref{eq:currConsRC0} around $ v = V - \tilde{V}$, which reads
\begin{equation}
\frac{dv}{dt} = -\frac{1}{C}
\left[g(\tilde{V}) + \frac{1}{R}\right]v\,,
\label{eq:linearizedRC}
\end{equation} 
where
$g(\tilde{V}) = dI/dV |_{V =\tilde{V}}$. 
The solution is dynamically stable if the term in the square bracket of Eq.~\eqref{eq:linearizedRC} is positive and unstable otherwise.

For a dissipative system, $IV \geq 0$ and $g_0\geq0$, thus $\tilde{V} = 0$ is the unique solution of Eq.~\eqref{eq:currConsRC0}, and it is
stable as shown in Fig.~\ref{fig:PRLEH}(b). In this scheme, we show that the positive differential conductance around the stable solution naturally stabilizes the circuit toward the stable fixed point $V=0$.  
The situation is richer when the junction displays a thermoelectric response. When the temperature of the cold superconductor is kept low $\tR\ll \tL,\Delta_R/k_B$ (see the discussion in Secs.~\ref{SubSec:linear-in-bias} and~\ref{SubSec:PowerSIS}), the junction may display linear-in-bias $g_0<0$ and/or nonlinear-in-bias thermoelectricity $I(V_{p})V_p<0$ depending on the different temperature or bias regimes. For a resistive load, roughly $R \gg -1/G_T$, the implicit equation has a total of three solutions to  $V=0,\pm \tilde V$. 
However, it follows that the zero-voltage solution is dynamically unstable, i.e., an arbitrarily small deviation drives the junction into one of the two finite voltage states $\pm\tilde{V}$~\cite{marchegiani_nonlinear_2020}, depending on the details of the initial condition [see Fig.~\ref{fig:PRLEH}(c)]. Then $\pm\tilde{V}$ represents stable solutions for the junction SIS$^\prime$ connected in parallel to a resistor when bipolar thermoelectricity is generated. \footnote{This result holds when the Josephson effect in the junction is fully suppressed, since for a finite Josephson coupling, the zero bias state can be metastable (see Sec.~\ref{sec:Experiments}).} 

The thermal bias, therefore, induces a \emph{spontaneous} breaking of the electron-hole symmetry: Although the characteristic is perfectly odd, the system selects one of two equivalent finite-voltage branches of opposite polarity, much as a ferromagnet selects one of two magnetization directions below its Curie temperature. In the absence of an external bias generator, the state symmetry is, in principle, not broken, but the nonequilibrium temperature profile opens that possibility. 
This bistability has direct consequences.  The two finite-voltage states provide the basis for a volatile thermoelectric memory, in which the sign of the voltage plays the role of the magnetization in a ferromagnetic cell~\cite{Giazotto2025ThermoelectricMemoryUS,germanese_phase_2023}, as discussed together with the experiment in Sec.~\ref{Sec. BTJE}. Moreover, when the circuit includes an inductive element, the instability of the zero-voltage state, combined with the net power produced by the junction, can sustain self-oscillatory dynamics~\cite{marchegiani_nonlinear_2020}, examined in Sec.~\ref{subsec:TEcircuit}.

\subsubsection{\textbf{Noise effects}}\label{Sec: Noise}
It is instructive, especially in view of the bipolar thermoelectric superconducting memory application~\cite{Giazotto2025ThermoelectricMemoryUS} (See Sec.~\ref{subsec:TEcircuit}), to investigate the stability of the device with respect to unavoidable noise fluctuations. In a tunneling system, the junction can either generate or dissipate power; in both cases, the current is affected by noise fluctuations~\cite{Scalapino}, which can be thermal (Johnson-Nyquist noise)~\cite{Johnson_PhysRev32,Nyquist_PhysRev32}, induced by voltage difference (shot or Schottky noise)~\cite{Schottky_AnnPhys} or even by strong temperature differences ($\Delta T$-noise)~\cite{SheinLumbroso2018DeltaTNoise,Pierattelli_2025}.
The bipolar thermoelectric element is bistable: the two finite-voltage states $\pm V_L$ are separeted by the unstable symmetric solution $V=0$. This unstable state is intrinsically sensitive to voltage fluctuations, which can can drive stochastic transitions between the two thermoelectric branches, in close analogy to noise-activated switching dynamics studied extensively in classical bistable systems~\cite{kramers1940brownian,hanggi1990reaction} and in current-biased Josephson junctions~\cite{fulton1974lifetime,devoret1985measurements}. A theoretical analysis of these effects was carried out in~\cite{marchegiani_noise_2020}, and we briefly summarize its results here.

To account for fluctuations, a stochastic Langevin current source $I_n(t)$ must be included in the current-conservation equation [c.f. Eq.~\eqref{eq:currConsRC0}]
\begin{equation}
I(V, \tL, \tR) = -C\frac{dV}{dt} -\frac{V}{R}- I_n(t)\,.
\label{eq:Vdyn}
\end{equation}
with $\langle I_n(t)\rangle=0$ and $\langle I_n(t)I_n(t')\rangle = S_I(\tilde V)\,\delta(t-t')$. In the tunnel limit, the spectral density follows from the bidirectional Poissonian statistics of QP
tunneling~\cite{rogovin1974fluctuation,levitov2004counting,golubev2001nonequilibrium} 
[cf.\ Eq.~\eqref{eq:cumulants}],
\begin{equation}
S_I(V) = G_T e^2\bigl[\Gamma_{\alpha\bar\alpha}(V)+\Gamma_{\bar\alpha\alpha}(V)\bigr]\,.
\label{eq:SI}
\end{equation}
The stability of noiseless solutions is governed by the same criterion as above: $G(\tilde V)+1/R>0$. Deterministically, the unstable zero-voltage state would relax to $\pm\tilde V$ according to the sign of the initial perturbation. Noise removes this deterministic selection: fluctuations can steer the system toward either state, and the voltage acquires a root mean square spread $\sigma_V$ about each stable solution. 
Linearizing Eq.~\eqref{eq:Vdyn} around a stable solution $\tilde V$, one finds at the stationary 
state~\cite{risken1989fokker,coffey2004langevin}
\begin{equation}
\sigma_V = \sqrt{\frac{S_I(\tilde V)}{2C\bigl[1/R+G(\tilde V)\bigr]}}\,.
\label{eq:stdV}
\end{equation}
When $\sigma_V$ becomes comparable to $|\tilde V|$, fluctuations bridge the two wells and 
stochastic switching between $\pm\tilde V$ sets in, in close analogy to Kramers escape from a 
double-well potential~\cite{hanggi1990reaction}. The capacitance plays a dual role: a larger 
$C$ suppresses the noise amplitude and stabilizes the two states, whereas a smaller $C$ 
enhances fluctuations and promotes noise-induced switching. In conclusion, the effective electrical capacitance 
of the junction can be adjusted, for instance, with a shunting capacitor~\cite{barone1982physics,MartinisPRB35}, to prevent the noise fluctuations from affecting the stability of the bipolar thermoelectric memory. On the other hand, an increase in junction capacitance degrades performance in terms of switching time and effective device bandwidth.

\newpage
\section{Experimental observations of the bipolar thermoelectric effect}\label{Sec: Exp}
\label{sec:Experiments}
Tunneling experiments on asymmetric-gap SIS$^\prime$ junctions date back to the early days of superconducting tunneling spectroscopy~\cite{giaever1960electron,giaever1960energy,Nicol_Direct_1960,Giaever_Tunneling_1962,Taylor_Excess_1963}. 
These early measurements, conducted at thermal equilibrium, were already able to resolve distinctive features of the quasiparticle $I$--$V$ characteristic, such as the singularity matching peak for $eV=|\dL-\dR|$, a region of negative differential conductance, and a sharp current onset at the gap sum $eV=\dL+\dR$~\cite{Nicol_Direct_1960}. Later, these nonlinearities enabled the design of low-noise SIS quasiparticle mixers operating close to the quantum limit~\cite{Tucker_Quantum_1979,gurvitch1983high}.
Aronov and Spivak~\cite{aronov1975photoeffect} predicted that  SIS$^\prime$ junctions exposed to optical radiation can exhibit absolute negative conductance, an exotic response in which the current flows against the applied bias, discussed as the photoelectric effect of a superconducting tunnel junction. Their prediction was later confirmed in quasiparticle injection experiments by Gershenzon and Falei~\cite{gershenzon1986absolute,gershenzon1988absolute}, and by Gijsbertsen and Flokstra~\cite{gijsbertsen1996quasiparticle}. 
ANC was also observed more recently by Nagel et al. in junctions subjected to microwave driving~\cite{nagel2008observation}. In all these cases, the current reversal was sustained by driving the junction out of equilibrium, but was not directly related to a temperature difference, which defines a thermoelectric response.

The bipolar TE addressed in this section produces the same absolute negative conductance, a current flowing against the bias, but from a 
pure thermal gradient rather than an external drive. Because the electrodes 
retain their particle-hole-symmetric BCS spectra, no built-in symmetry breaking 
is required, and the effect is intrinsically bipolar: the \emph{same} temperature 
difference supports two opposite Seebeck voltages $\pm V_S$. The resulting power 
can be delivered to an external load, turning the junction into a heat engine.

This section is organized as follows. We first discuss the key experimental 
requirements for observing the bipolar TE, such as junction fabrication, suppression of the Josephson contribution, and generation and readout of the thermal bias. Then we review the experiments that demonstrated the defining features of the effect, from the observation of bipolar thermoelectric generation ~\cite{germanese_bipolar_2022} to the phase control of the thermoelectric response~\cite{germanese_phase_2023}.

\subsection{\textbf{Key Experimental Requirements}}~\label{sec:requirements} 
The experimental observation of thermoelectricity in superconducting tunnel junctions relies on specific choices in device design and fabrication, which we briefly discuss hereafter.

\subsubsection{\textbf{Fabrication of superconducting tunnel junctions}}\label{Sec. Fab}
The bipolar TE discussed in Sec.~\ref{Sec: Th} is realized in superconducting tunnel junctions, which can be fabricated through different techniques. Aluminum (Al) is a widely used material for tunnel junctions because it spontaneously forms a thin oxide layer upon exposure to oxygen, enabling the controlled formation of high-quality tunnel barriers.
\begin{figure}[t]
    \centering
    \includegraphics[width=\linewidth]{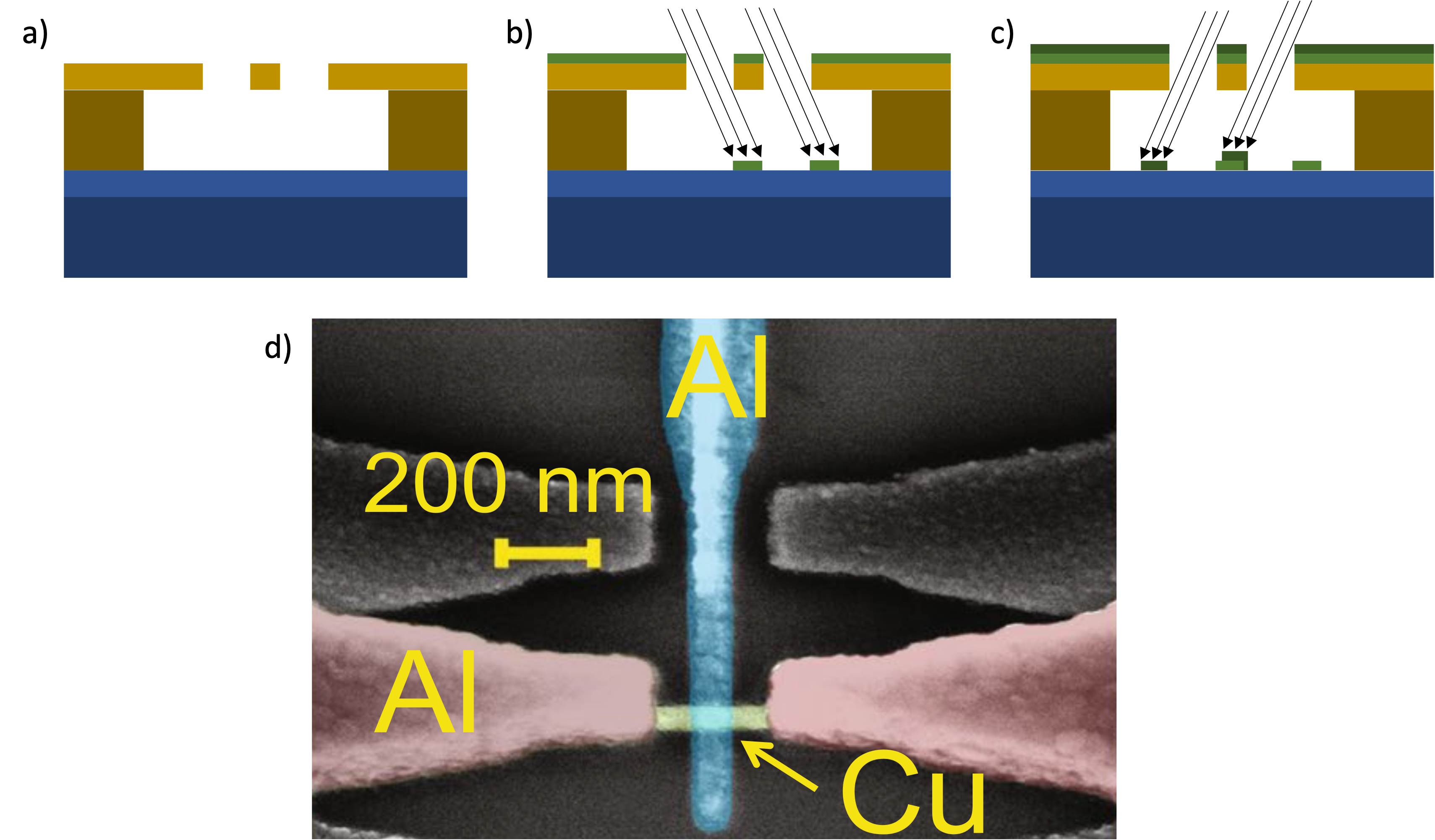}
   \caption{Nanofabrication of metallic junctions. (a)-(c) Schematics of the shadow mask evaporation process. (a) Patterned bilayer resist mask with undercut profile. (b) First angled deposition of the electrode material. 
   (c) Second deposition at the opposite angle, completing the junction. (d) False-color scanning electron microscopy (SEM) image of a device fabricated by three-angle shadow-mask evaporation. Adapted from \cite{jabdaraghi2017noise}.}
    \label{fig:shadow}
\end{figure}
Angle-evaporation techniques are widely used to fabricate small (submicrometer-squared) superconducting tunnel junctions. They require highly directional deposition, typically achieved by electron-beam evaporation, and the tunnel barrier is formed in situ by oxidation without breaking the vacuum. 
The most widely used shadow-mask evaporation techniques are the Dolan-bridge and Manhattan-style geometries. The Dolan bridge technique~\cite{Dolan1977} defines the junction area through a suspended resist bridge, as schematized in Fig.~\ref{fig:shadow}.  
First, a bilayer resist mask (light and dark yellow) is deposited on the substrate (blue) and patterned to form a suspended bridge with an undercut profile [panel (a)]. 
This geometry originates from the different development rates of the two resist layers and defines the regions that can be reached during angled deposition. A first metallic film (light green) is then evaporated at a given angle [panel (b)]. When a tunnel barrier needs to be formed, the deposited layer is subsequently exposed to a controlled \textit{in situ} oxidation step (not shown), 
to form the insulating layer. The second metallic film (dark green) is then deposited at a different angle [panel (c)], so that the two evaporated films overlap only in the junction region. Additional replicas may form on the substrate along the sides of the junction as a consequence of multi-angle deposition. 
When the oxidation step is included, this protocol yields tunnel junctions, as in SIS devices, whereas skipping this step yields clean metallic contacts, as in proximity junctions. The suspended mask can be defined by either electron-beam or optical lithography, depending on the target device dimensions and required resolution. 
Typical e-beam implementations employ bilayer resist stacks such as PMMA and an MMA-MAA copolymer, while analogous shadow-evaporation schemes can also be realized with suitable bilayer photoresists in optical lithography. The mask thickness, 
in combination with the deposition angles, sets the lateral displacement between the two evaporations, and consequently, the effective size of the junction.  
Figure~\ref{fig:shadow}(d) shows a scanning electron microscopy image of a representative device (SQUIPT magnetometer, see Sec.~\ref{Subsec: SQUIPT}) fabricated by three-angle shadow-mask evaporation. First, an Al layer is deposited to form the superconducting probe (blue). Following \textit{in situ} oxidation, a Cu layer is evaporated to form the normal metal wire in the magnetometer (green), whose overlap with the oxidized Al layer defines a NIS tunnel junction (the cross-shaped structure). Finally, a thicker Al layer is deposited to define the superconducting side electrodes (red), which are connected in a loop geometry not shown in the figure. The aluminum ring, interrupted by the Cu bridge, forms a superconductor-normal metal-superconductor (SNS) interferometer, whose transport properties can be controlled through the magnetic flux threading the loop (see discussion in Sec.~\ref{subsec:suppressionIj}).

The Manhattan process~\cite{potts2001novel} is based on the same \textit{in situ} 
oxidation principle, but does not rely on a suspended bridge. A single mask defines two perpendicular trenches, and the two films are evaporated at different angles into them, so that the junction is formed at their right-angle overlap. Being bridgeless, the process avoids the mechanical fragility and resist-height sensitivity of the Dolan bridge, thereby improving wafer-scale uniformity and scalability~\cite{muthusubramanian2024wafer}.

When less directional deposition methods are employed, different fabrication routes are preferred. Trilayer technologies, such as sputtered Nb/Al-AlOx/Nb stacks~\cite{gurvitch1983high,morohashi1986self}, deposit the full superconducting structure first and then define the junction by lithography and etching. Finally, overlap processes~\cite{wu2017overlap,van2024advanced} follow a different scheme. The two electrodes are deposited and patterned in separate steps, with the tunnel barrier formed by oxidation after \textit{in situ} cleaning of the first electrode surface.

\subsubsection{\textbf{Engineering the Gap Asymmetry}}
Bipolar thermoelectricity requires different DoSs between the two electrodes (see Sec.~\ref{SubSec:Necessary}). In a BCS superconductor, the DoS is characterized by the gap [cf. Eq.~\eqref{eq:BCSdos}], so combining two superconductors with different zero-temperature gaps, $\dLo>\dRo$, is necessary and sufficient to meet this condition. The gap ratio $r=\dRo/\dLo$ is a crucial parameter for thermoelectricity: the effect vanishes when $r$ approaches unity, while in the optimal range $r\sim 0.2-0.4$ the generated power is maximized (see Sec.~\ref{SubSec:PowerSIS}).

Electron-beam evaporation constrains the choice of superconducting materials available for gap engineering. Refractory superconductors such as niobium (Nb) or vanadium (V) require high deposition temperatures~\cite{surgers1994effect,delacour2011persistence,Samaddar_2013}, which complicates multilayer processing and may affect previously deposited structures~\cite{alekseevskii1976superconducting,gutsche1994growth}. In contrast, aluminum (Al) is a standard material for shadow-mask evaporation. Its superconducting gap can be tuned through the film thickness, since reducing the thickness enhances the critical temperature from the bulk value $T_c\simeq 1.2$ K, up to values of the order of a few kelvin, with $\Delta_0 \simeq 1.764 k_B T_c$~\cite{ChubovThinAl,CherneyThinAl,meservey1971properties,Court_2007,LiuPRL111,Nho2026}. The maximum gap asymmetry achievable with this strategy is limited by the challenge of growing continuous ultrathin films. Below a certain thickness, which depends on the deposition method and substrate~\cite{Kaiser2002}, Al tends to form disconnected islands rather than a uniform superconducting electrode. For sputter-deposited Al films, this percolation threshold has been characterized in detail~\cite{Kunz1988}.

A flexible route to engineer the superconducting gap, compatible with shadow-mask evaporation, is based on the proximity effect~\cite{DeutscherDeGennes1969,PannetierCourtois2000}. When a superconductor is placed in good electrical contact with a normal metal, superconducting correlations leak into the normal region through Andreev reflection, and the normal metal acquires superconducting-like spectral features over the phase-coherence length.
In an SNS weak link, the proximity effect opens a minigap in the DoS of the normal wire, which depends on the superconducting phase difference and can therefore, in a loop geometry, be tuned by the magnetic flux threading the loop. This principle is exploited by superconducting quantum interference proximity transistors (SQUIPTs)~\cite{giazotto2010superconducting}, where the flux-controlled minigap modulates transport through a tunnel-coupled probe, as discussed in more detail in Sec.~\ref{Subsec: SQUIPT}.

Just as a superconductor induces superconducting correlations in an adjacent normal metal, the proximity of the normal metal, in turn, suppresses the superconductor order parameter~\cite{DeutscherDeGennes1969,McMillanPR175}. This inverse proximity effect provides a viable route for gap engineering in tunnel junctions, using thin SN bilayers rather than phase-biased weak links. When both layers are thinner than the superconducting coherence length, and the SN interface is highly transmissive, i.e., a direct metallic contact with no oxide barrier, the bilayer acts as an effective superconducting electrode with a reduced gap, tunable through the relative layer thicknesses~\cite{FominovPRB63}. Gap engineering based on SN bilayers was used in the experiments reported in Refs.~\cite{germanese_bipolar_2022,germanese_phase_2023} discussed in the following.

\subsubsection{\textbf{Suppression of the Josephson effect}}
\label{subsec:suppressionIj}
The hallmark of bipolar thermoelectricity, i.e., current flowing against the voltage bias, appears in the subgap regime (see Sec.~\ref{Sec: Th}) of a thermally biased Josephson junction, and is due to QP transport. In the same structure, the Josephson effect provides a coherent, reactive channel for Cooper-pair transport that typically dominates the low-bias dynamics and can even prevent detection of the QP thermoelectric response. In a typical experimental setup, the junction is connected to room-temperature electronics through filtered cryogenic lines, whose finite resistance and frequency-dependent impedance constitute an essential part of 
the electromagnetic environment coupled to the junction~\cite{devoret_effect_1990,cleland_influence_1991,
ingold_charge_1992,holst_effect_1994}. The Josephson contribution generally
interacts with these environmental modes, further modifying the low-bias transport and the thermoelectric signal~\cite{marchegiani_phase-tunable_2020,germanese_phase_2023,antola2026quantum} (see the discussion in Sec.~\ref{Sec:Phase_Exp}). Efficient suppression of the 
Josephson coupling is therefore essential for the clear identification of subgap QP current.

The Josephson current in a tunnel junction is described by the sinusoidal current-phase relation~\cite{josephson1962possible}
\begin{equation}
I_J = I_c \sin\phi,
\end{equation}
where $I_c$ is the critical current and $\phi$ is the gauge-invariant superconducting phase difference across the junction~\cite{tinkham_introduction_2004,barone1982physics}. 
Several strategies can be used to reduce $I_c$, as we shall discuss now. 

A standard way to suppress $I_c$ is to exploit phase interference in a superconducting loop interrupted by two Josephson junctions, namely a superconducting quantum interference device (SQUID)~\cite{tinkham_introduction_2004,barone1982physics}. 
Fluxoid quantization, associated with the single-valuedness of the superconducting wave function, puts a constraint on the relative values of the phase drops across the junctions, so that the total supercurrent depends periodically on the magnetic flux threading the loop~\footnote{This periodicity breaks down when the field approaches the critical magnetic field in the out-of-plane junction direction.}. For a two-junction DC SQUID with negligible loop inductance and junction critical currents $I_{c1}$ and $I_{c2}$, the flux-dependent critical current reads~\cite{barone1982physics}
\begin{equation}
I_c(\Phi) = \sqrt{I_{c1}^2 + I_{c2}^2 + 
2I_{c1}I_{c2}\cos\left(2\pi\frac{\Phi}{\Phi_0}\right)},
\end{equation}
where $\Phi$ is the applied magnetic flux and 
$\Phi_0 = h/(2e)$ is the superconducting flux quantum. At half-integer flux, $\Phi = (n+1/2)\Phi_0$, destructive interference minimizes the critical current to
\begin{equation}
I_c^{\min} = |I_{c1} - I_{c2}|.
\end{equation}
However, complete suppression is obtained only for identical junctions, whereas any junction asymmetry leaves a finite residual supercurrent.
\begin{figure}[t]
    \centering
    \includegraphics[width=\linewidth]{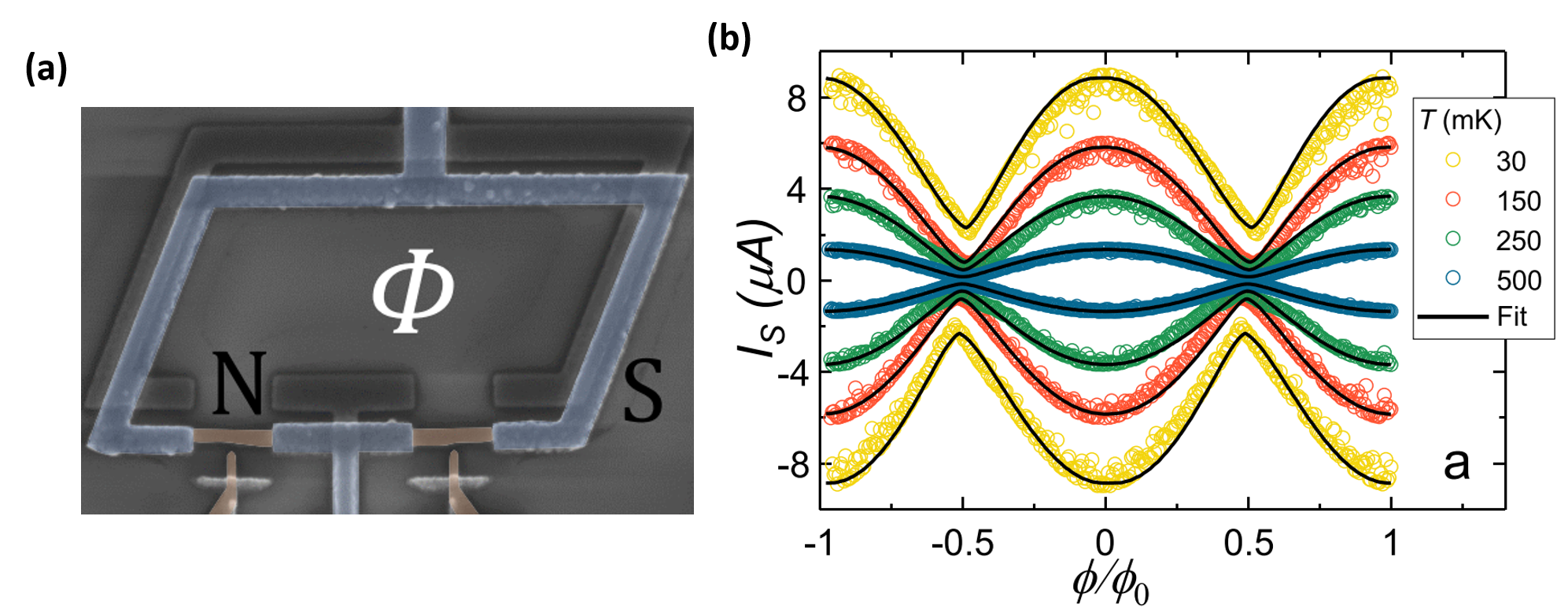}
    \caption{Flux-controlled suppression of the Josephson current. (a) False-color SEM image of a fabricated DC SQUID. (b) Critical current $I_c$ versus magnetic flux $\Phi$ measured at different temperatures. Adapted from~\cite{de2021gate}.}
    \label{fig:SQUID}
\end{figure}
Figure~\ref{fig:SQUID} shows a SQUID consisting of an Al loop interrupted by two SNS junctions with Cu weak links in parallel. The loop geometry enables supercurrent 
suppression by phase interference, controlled by magnetic flux. 
The minimum supercurrent decreases as the temperature increases, since the critical currents of both junctions are reduced as the superconducting gap of the two films is suppressed with temperature~\cite{ambegaokar_tunneling_1963, tinkham_introduction_2004}. 

\begin{figure}[t]
    \centering
    \includegraphics[width=0.9\linewidth]{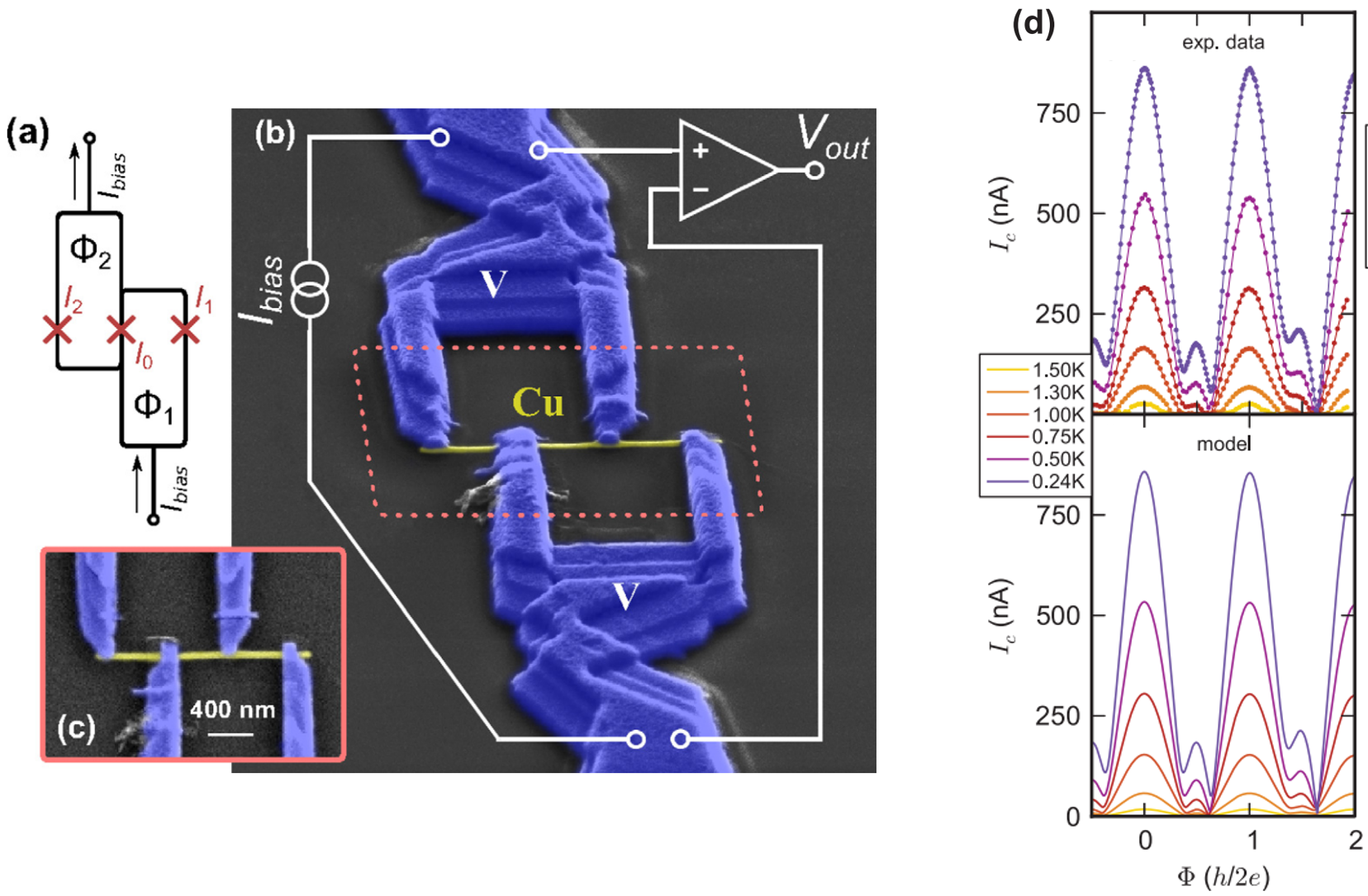}
    \caption{Double-loop interferometer for critical-current suppression. (a) Scheme of the three-junction double-loop geometry. (b,c) Scanning electron micrographs of a realization based on V/Cu SNS weak links. (d) Experimental and simulated critical current $I_c$ as a function of magnetic flux. Adapted from~\cite{ronzani_balanced_2014}}
    \label{fig:Inter}
\end{figure}

The impact of junction asymmetries on the critical-current suppression can be reduced by increasing the number of junctions in the interferometer. 
In a balanced three-junction configuration~\cite{kemppinen_suppression_2008,ronzani_balanced_2014}, two superconducting loops are threaded by reduced magnetic fluxes $\tilde{\Phi}_1 = \Phi_1/\Phi_0$ and $\tilde{\Phi}_2 = \Phi_2/\Phi_0$. Assuming sinusoidal current-phase relations and negligible loop inductance, the critical current reads~\cite{chiarello_optimal_2008,fornieri2016nanoscale}
\begin{equation}
I_c(\tilde\Phi_1,\tilde\Phi_2)
= I_{c2}
\sqrt{
1 + r_1^2 + r_2^2
+ 2r_1 \cos(2\pi\tilde\Phi_1)
+  2r_2 \cos(2\pi\tilde\Phi_2)
+  2r_1 r_2 \cos[2\pi(\tilde\Phi_1+\tilde\Phi_2)]
},
\end{equation}
where $r_1 = I_{c1}/I_{c2}$ and $r_2 = I_{c3}/I_{c2}$. For a spatially homogeneous magnetic field, the fluxes can be written as $\Phi_{1,2}=(1\pm\alpha)\Phi$, where $\Phi=(\Phi_1+\Phi_2)/2$ is the average flux, and $\alpha=(\Phi_1-\Phi_2)/(\Phi_1+\Phi_2)$ quantifies the effective loop-area asymmetry. Using the previous formula, one can show that if $|r_1-r_2|\leq 1$ and $r_1+r_2\geq 1$ then one can always find a value of $\Phi$ corresponding to a completely destructive interference, i.e. $I_c(\tilde\Phi_1,\tilde\Phi_2)=0$~\cite{ronzani_balanced_2014}.
Figure~\ref{fig:Inter} shows the double-loop geometry and its implementation in an SNS device based on vanadium and copper. As shown in panel~(d), the additional interference path makes the suppression more robust against residual junction asymmetries and produces a more structured $I_c(\Phi)$ pattern than in a two-junction DC SQUID.

\begin{figure}[t]
    \centering
    \includegraphics[width=0.9\linewidth]{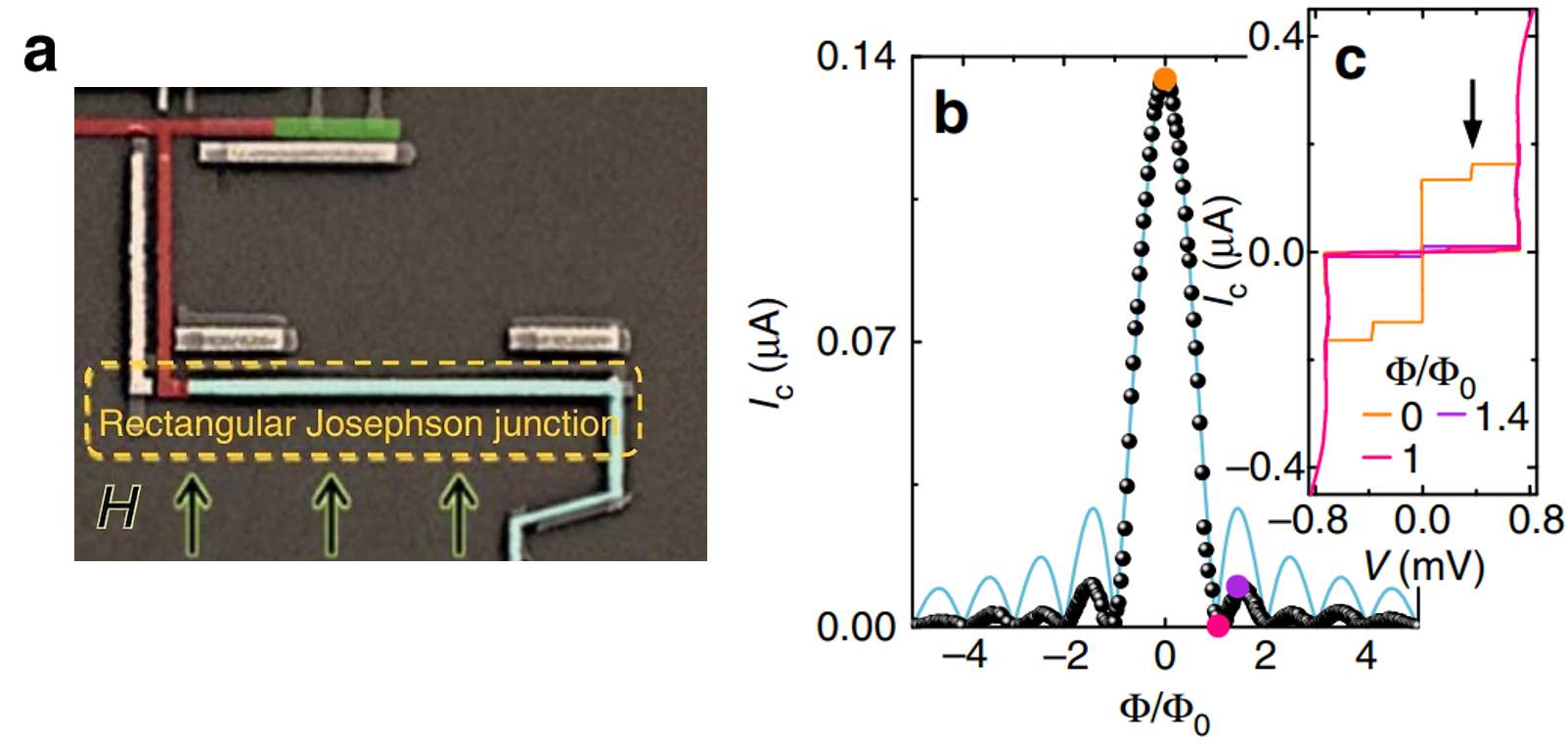}
    \caption{Fraunhofer suppression in an extended SIS junction. (a) False-color SEM of the Al SIS junction, designed for operation under an in-plane magnetic field. (b) Critical current $I_c$ as a function of magnetic field. (c) $I$--$V$ characteristics for selected values of the applied flux. Adapted from~\cite{jose2014quantum}}
    \label{fig:fraunhofer}
\end{figure}
The Josephson coupling can also be suppressed in a single-junction device by applying a uniform magnetic field in the plane of the junction. The flux that threads through the junction area induces a spatially varying phase difference, leading to a Fraunhofer-like modulation of the critical current~\cite{rowell_magnetic_1963,chiodi2012geometry,granata2013spatial,jose2014quantum, trnjanin2025magnetotransport}. For a rectangular junction in the short-junction limit, i.e., dimensions much smaller than the Josephson penetration depth~\cite{barone1982physics}, the critical current reads
\begin{equation}
I_c(\Phi) = I_{c0}\left|\frac{\sin(\pi\Phi/\Phi_0)}
{\pi\Phi/\Phi_0}\right|,
\end{equation}
where $\Phi = BA_\text{eff}$ is the flux that flows through the effective junction area $A_\text{eff}$. The critical current vanishes at integer multiples of $\Phi_0$,
reflecting the destructive interference of the spatially-dependent phase difference across the junction. The junction length can be designed so that the Fraunhofer period falls in a field range of a few mT, which does not significantly suppress the order parameter in the superconducting films. An experimental realization of Fraunhofer-like interference
is shown in Fig.~\ref{fig:fraunhofer} (a) and (b), with the $I$--$V$ characteristics measured for selected values of the magnetic flux in panel (c).

Clearly, the critical current decreases as the transmissivity of the tunneling barrier is reduced, scaling inversely with $R_T$. For example, for a gap-symmetric Josephson junction, the critical current is described by the Ambegaokar--Baratoff relation~\cite{ambegaokar_tunneling_1963,likharev1979superconducting},
\begin{equation}
I_c R_T = \frac{\pi\Delta}{2e} \quad (T=0),
\end{equation}
with an analogous expression for SIS$^\prime$ junctions~\cite{barone1982physics}. 
Since the QP current is also inversely proportional to $R_T$, fabricating junctions with high tunnel resistance is not, in principle, convenient in terms of output thermoelectric power. 
For sufficiently low critical currents, such as the Josephson energy $E_J=\hbar I_c/2e\lesssim k_B T$, thermal fluctuations can induce thermally activated phase slips.
In overdamped junctions, this leads to the phase-diffusion regime (Ambegaokar--Halperin theory~\cite{Ambegaokar_PRL22}) characterized by a finite dissipative subgap response. The critical current in thermally biased SIS$^\prime$ junctions displays a nonmonotonic dependence with the temperature difference, a feature which can further modify the junction behavior~\cite{Guarcello_Nonlinear_2019,Guarcello_Calorimeter_2019}. In summary, the almost complete suppression of the Josephson coupling is, in general, advantageous for the measurement of the bipolar TE.

\subsubsection{\textbf{Thermal Bias Generation}}
The reliable detection of a thermoelectric signal relies on establishing a tunable temperature difference across the junction. In early experiments, such as those of Smith, Tinkham, and Skocpol~\cite{smith1980new}, thermal bias was generated by heating one electrode with laser irradiation.  
In more recent experiments, heaters and electronic thermometers have been directly integrated on the chip and have been used successfully for refrigeration and temperature readout in the past few decades~\cite{nahum1994electronic,leivo1996efficient,giazotto_opportunities_2006,Muhonen2012}.  
These elements enable a controlled increase in the electronic temperature in a specific film while providing a sufficiently precise local readout. 

Heating of the electronic system in a specific metallic film is enabled by 
the weak electron-phonon coupling at sub-kelvin temperatures. When the electron--phonon collision rate becomes sufficiently low, electrons can thermally decouple from the film phonons, the latter being thermalized to the substrate phonons (and so to the cryostat mixing chamber) due to the negligible Kapitza resistance~\cite{giazotto_opportunities_2006}. Provided that the electron--electron scattering rate is much faster than the electron--phonon scattering rate and the injection rate, the electronic system in the metallic film can still be described using a Fermi distribution with effective temperature $T_{qp}$, distinct from phonon temperature $T_{ph}$. This regime is commonly referred to as the quasi-equilibrium regime~\cite{giazotto_opportunities_2006,wellstood1994hot}.

In superconductors, the thermal decoupling of electrons and phonons is enhanced by the presence of the superconducting gap; energy relaxation is mediated by QPs whose number is exponentially suppressed at low temperatures. For $T_{ph}\ll T_{qp}\ll \Delta/k_B$, the electron–phonon power 
exchange reads~\cite{timofeev_recombination-limited_2009}
\begin{equation}
P_{\text{qp-ph}} \simeq \frac{64}{63\,\zeta(5)}\Sigma\, \mathcal{V}\, T_{qp}^5\, 
e^{-\Delta/k_B T_{qp}},
\end{equation}
where $\zeta(z)$ is the Riemann zeta function, $\Sigma$ is the material-dependent electron–phonon coupling constant and $\mathcal{V}$ the film volume. The exponential suppression, absent in the normal-state electron--phonon heat conductance, makes the superconducting electrode an excellent thermal island, enabling the generation of large temperature differences relative to the phonon bath.

In practice, the electronic temperature in the target electrode is raised by Joule heating through voltage-biased on-chip heaters, typically integrated into the device, with additional tunnel junctions coupled to the film. 
Normal metal-insulator-superconductor (or NIS) heaters dissipate at any finite current, while SIS heaters dissipate only when the voltage bias exceeds the gap sum. The heater circuit must be electrically isolated from the measurement circuit, for instance by using floating battery sources~\cite{martinez2012josephson}, so that no spurious net charge current flows through the device under study.

At the steady state, the QP temperature $T_{qp}$ in each element is determined by energy balance. For instance, for a superconducting electrode connected to a heater and tunnel coupled to an additional metallic element (which acts as the cold lead with temperature $T_{qp}'<T_{qp}$), the injected power $P_{\text{in}}$ must equate the energy exchanged with the phonons in the film 
and the heat current $\dot{Q}$ flowing through the junction 
\begin{equation}
P_{\text{in}} = P_{\text{qp-ph}}(T_{qp}, T_{ph}) + 
\dot{Q}(T_{qp}, T_{qp}').
\end{equation} 
However, a similar energy balance equation should be written for $T_{qp}'$. In the ideal case of a well thermalized cold terminal $T_{qp}'$ matches the bath temperature $T_{ph}$. In practice, the QP heat current $\dot{Q}$ flowing through the junction can increase the temperature of the cold terminal if it is also thermally decoupled from the phonons. This unavoidable heating could partially reduce the temperature difference that is required to observe the bipolar thermoelectric response. This potential issue can be mitigated by coupling the cold superconductor to a normal-metal electrode, which acts as a trap for QPs and evacuates the incoming heat, favoring the thermalization with the bath~\cite{rajauria2009quasiparticle,o2012measurement}.
   
Independent readout of the electronic temperature relies on the same tunnel-junction physics: a junction biased below the gap acts as a thermometer, since the temperature dependence of its QP current provides a direct measure of the electronic temperature of the probed electrode~\cite{giazotto_opportunities_2006}. NIS junctions are particularly convenient in this respect, their subgap current being exponentially sensitive to the temperature of the normal side, and are the standard tool for local electronic thermometry in mesoscopic circuits~\cite{nahum1994electronic,Muhonen2012}. 

To measure the temperature of a superconductor, a possible strategy is Josephson thermometry~\cite{giazotto_opportunities_2006,Fornieri_0Pi_2017}, which exploits the electronic temperature dependence of the critical current~\cite{ambegaokar_tunneling_1963}. However, this approach requires careful calibration and a specifically design of the Josephson junction to enable sufficient sensitivity in the desired temperature range.

In the experiments on bipolar thermoelectricity discussed below, the strong nonlinearity of the IV characteristics of SIS$'$ junctions can be exploited to determine the two leads' electronic temperatures after a specific designed calibration procedure. In other words, the nonlinearity of the IV characteristic is a secondary thermometer for the electronic temperatures of both superconductors. 

\subsection{\textbf{Experimental Realization of the Bipolar Thermoelectric Josephson Engine}}
The experiments discussed in this section are designed to observe
the bipolar TE theoretically described in Sec.~\ref{Sec: Th}. A temperature difference is applied across the SIS$^\prime$ junction, in which  
the Josephson coupling has been suppressed. We recall that linear TEs effects in this structure are vanishingly small due to the nearly perfect particle-hole symmetry of the superconductor DoSs. 
Thermoelectric power conversion, signaled by a QP current flowing against the applied voltage, can be observed for nonlinear temperature differences, (see Sec .~\ref {Sec: Th}). 

The defining feature is bipolarity: for a single direction of the temperature difference, the open-circuit thermovoltage can take either of two opposite values $\pm V_S$. This feature distinguishes these experiments from conventional thermoelectrics, in which the dominant carrier type uniquely determines the sign of the Seebeck voltage for a given thermal gradient. 

The realization of this effect requires assembling, in a single structure, the three ingredients identified in Sec.~\ref{sec:requirements}: engineered gap asymmetry, flux-controlled suppression of the Josephson coupling through a double-SQUID geometry, and on-chip thermal biasing. This is achieved in the devices of Refs.~\cite{germanese_bipolar_2022,germanese_phase_2023}, which we refer to in the following as the Bipolar Thermoelectric Josephson Engine (BTJE).

\subsubsection{\textbf{BTJE device}}
The operating principle of the BTJE is sketched in Fig.~\ref{fig:BTJE1}(a). Under specific conditions, the heat current arising from the electronic temperature  difference established across the junction is partially converted into electrical work delivered to an external load, with the bipolar response enabling two opposite output voltages for the same temperature bias.

A false-color SEM image of a representative device is shown in Fig.~\ref{fig:BTJE1}(b). The BTJE features a double-loop DC-SQUID formed by three Al/AlO$_x$/Al-Cu tunnel junctions connected in parallel, a geometry motivated by the interferometric suppression of the Josephson effect of  multi-junction (double-SQUID) setups
(see discussion in Sec.~\ref{sec:requirements}). The higher-gap superconductor is a high-purity Al film ($\SAl$, red), while the lower-gap one is an Al/Cu proximity bilayer ($\SAlCu$, light blue). The thicknesses of the two films in the bilayer are chosen to reduce via inverse proximity effect the gap of $\SAlCu$ to $\dAlCuo \simeq 0.35\,\dAlo$, a value close to the theoretical optimum for power generation (cf.\ Sec.~\ref{SubSec:PowerSIS}).

All the metallic films are patterned in a single electron-beam lithography step and fabricated by three-angle shadow-mask evaporation with \textit{in situ} oxidation, as described in Sec.~\ref{sec:requirements}. Four additional Al/AlO$_x$/Al tunnel junctions (green) are coupled to $\SAl$ and operated as Joule heaters; the heaters are powered by generators electrically floating with respect to the thermoelectric circuit. 
Since the bipolar TE is only observed when the higher-gap electrode is the hottest (see Sec.~\ref{Sec: Th}), the heaters are coupled to the $\SAl$ film.
Notably, the geometry S/I/SN has been used in the experiment, with Al in direct contact with the AlO$_x$ tunnel barrier and Cu forming the outer layer. A later theoretical analysis by Hijano \textit{et al.}~\cite{hijano_bipolar_2023} showed that the inverse arrangement, in which the normal layer of the cold bilayer faces the tunnel barrier (S/I/NS rather than S/I/SN), could yield sharper subgap features and improved thermoelectric performance.

\subsubsection{\textbf{Equilibrium characterization}}
The measurement of the IV curves at thermal equilibrium ($\tAl=\tAlCu=T$) is a crucial preliminary analysis in characterizing the TE.
The QP current flowing through the Al/AlO$_x$/Al-Cu junction is exponentially suppressed (for $T\ll\tcAl,\tcAlCu$) below the gap-sum threshold $eV=\dAl+\dAlCu$ and rises sharply above it (cf.\ Fig.~\ref{fig:IvOriginal}a). Thermally activated QPs also contribute a finite subgap current, with a characteristic matching peak at $eV=|\dAl-\dAlCu|$ whose value increases with temperature, at least for $T<\tcAlCu$. 
Fits of the full $I(V)$ trace to the tunneling expression of Eq.~\eqref{eq:IVandQ}, with Dynes broadening to account for residual subgap states~\cite{DynesPRL41,PekolaDynes}, can be used to extract the zero-temperature gaps, the tunnel conductance ($G_T$), and, possibly, also the electronic temperatures of the two electrodes, after appropriate data analysis (see Supplementary information of Ref.~\cite{germanese_bipolar_2022}).

\begin{figure}[t]
    \centering
    \includegraphics[width=1\linewidth]{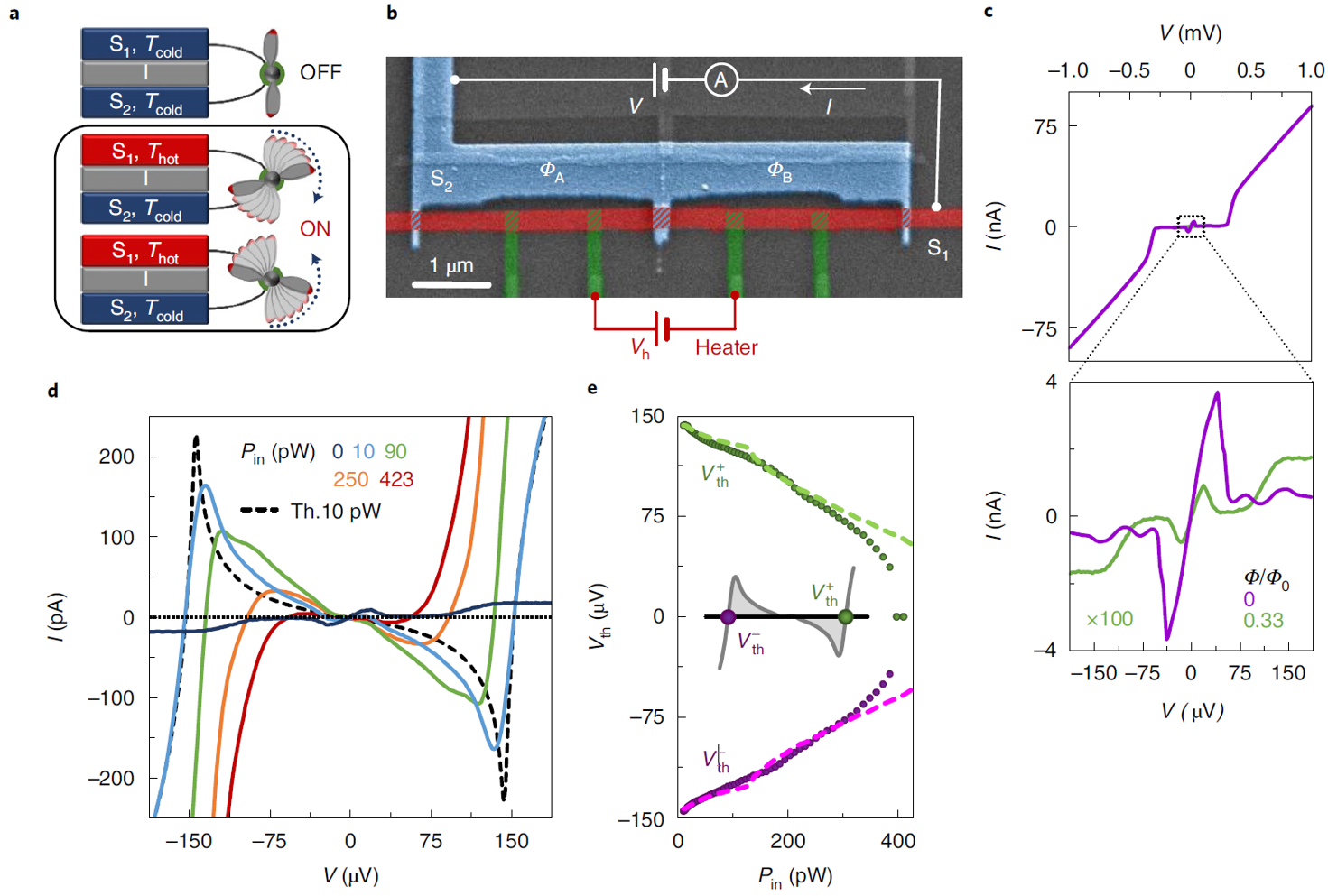}
   \caption{Bipolar Thermoelectric Josephson Engine (BTJE). 
(a) Operating principle: heat injected into one electrode of a SIS$'$ tunnel junction is converted into electrical work. (b) False-color scanning electron micrograph of a representative interferometer. $\SAl$ (labeled $S_1$ in the original figure) is shown in red, $\SAlCu$ ($S_2$) in light blue, and the heaters in green. 
(c) IV characteristic measured at base temperature and zero magnetic flux (purple). The inset shows a zoom around zero bias, highlighting the suppression of the Josephson current. 
(d) Thermoelectric IV characteristics under thermal bias for increasing injected power $P_{\mathrm{in}}$. 
(e) Seebeck voltage $V_S$ (labeled $V_\mathrm{th}$ in the original figure) 
versus $P_{\mathrm{in}}$. Adapted from~\cite{germanese_bipolar_2022}.}
    \label{fig:BTJE1}
\end{figure}

The IV characteristic of the BTJE measured at base temperature and zero magnetic flux is shown in Fig.~\ref{fig:BTJE1}(c). The three Al/AlO$_x$/Al-Cu junctions are connected in parallel, and the total QP current is the sum of the individual contributions. The measured trace also displays a DC Josephson contribution at low bias, better visualized in the inset of Fig.~\ref{fig:BTJE1}(c). Tuning the flux to the operating point $\Phi=0.33\,\Phi_0$, at the destructive-interference condition of the double-SQUID, suppresses the supercurrent to a few parts per thousand ($1.57$\textperthousand), enabling a better identification of the QP current. 

\subsubsection{\textbf{Bipolar thermoelectric response}}
Injecting a power $P_\mathrm{in}$ as Joule heating into the $\SAl$ film through the on-chip heaters establishes a temperature difference across the junction and consequently affects the IV characteristic mainly in the subgap region $eV<\dAl,\dAlCu$. At the operating flux $\Phi=0.33\,\Phi_0$, where the Josephson contribution is minimized, the device displays ANC for $|V|\leq 150~\mu$eV, as shown in Fig.~\ref{fig:BTJE1}(d); current flowing against the applied voltage, $IV<0$, provides the direct signature of thermoelectric energy conversion. In agreement with the modeling of Sec.~\ref{Sec: Th}, the maximum negative current occurs when the BCS singularities in the QP DoS of the two electrodes align, defining a matching peak at
\begin{equation}
V_p = \frac{\dAl(\tAl)-\dAlCu(\tAlCu)}{e}.
\label{eq:matching}
\end{equation}
As $P_\mathrm{in}$ increases, $\tAl$ also increases and $V_p$ shifts toward lower bias, reflecting the suppression of $\dAl(\tAl)$ with $\tAl$. The ANC appears symmetrically in $V$ at a fixed temperature difference due to the current reciprocity $I(-V)=-I(V)$ dictated by the EIS of the transport in each junction (Sec.~\ref{Sec: Th}). This feature manifestly differentiates the bipolar response with respect to the (conventional) unipolar thermoelectrics.

The Seebeck voltage $V_S$, or thermovoltage, [labeled $V_\mathrm{th}$ in Fig.~\ref{fig:BTJE1}(e)], is plotted as a function of $P_\mathrm{in}$. At low injected power ($P_\mathrm{in}\simeq 10\,$pW), $V_S$ reaches $\pm 150\,\mu$V; it decreases monotonically with $P_\mathrm{in}$ and vanishes for $P_\mathrm{in}\gtrsim 400\,$pW, where the gap asymmetry is suppressed by the increase in $\tAl$. 
A Seebeck voltage that decreases with increasing temperature difference provides additional evidence of the nonlinear nature of the effect, as anticipated before. 

The nonlinear Seebeck coefficient $\mathcal{S}=V_S/\Delta T$, with $\Delta T$ extracted by fitting the out-of-equilibrium IV characteristics, reaches $\pm 300\,\mu$V/K in the low-power regime. This value exceeds the Mott--Jones estimate for normal-state aluminum~\cite{mott1936theory} by roughly five orders of magnitude, and is comparable to that of spin-split superconductor/ferromagnetic-insulator tunnel junctions~\cite{kolenda2016observation,kolenda2017thermoelectric} and to quantum-dot heat engines operating at cryogenic temperatures~\cite{thierschmann_three-terminal_2015,josefsson_quantum-dot_2018,
jaliel_experimental_2019}.

\subsubsection{\textbf{Bipolar thermoelectric current-voltage hysteresis}}\label{Sec. BTJE}
\begin{figure}[t]
    \centering
\includegraphics[width=1\linewidth]{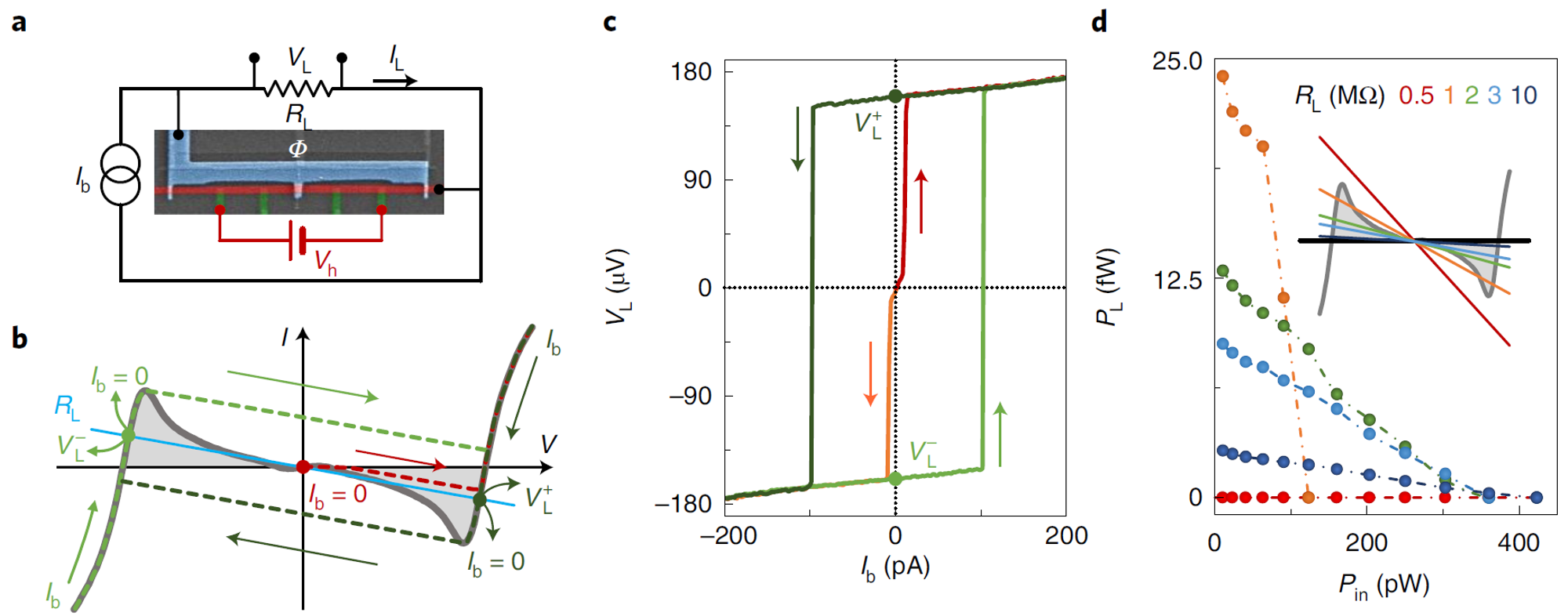}
    \caption{Operation of the bipolar thermoelectric Josephson engine (BTJE).
(a) Electrical circuit used to operate the interferometer as a thermoelectric engine, where the device acts as an active element connected in parallel to a load resistor $R_L$ and biased by a DC current $I_b$. (b) Load-line representation of the stationary solutions obtained from the intersections between the nonlinear $I$--$V$ characteristic of the BTJE and the load line defined by $R_L$. (c) Measured voltage across the load $V_L$ as a function of the bias current $I_b$, showing the hysteretic ignition of the engine. (d) Extracted power $P_L$ as a function of the injected heating power $P_{\mathrm{in}}$ for different load resistances $R_L$. The inset shows the load-line construction and the loss of intersections with the device $I$--$V$ characteristic for small $R_L$. Adapted from~\cite{germanese_bipolar_2022}.}
    \label{fig:BTJE_engine}
\end{figure}

The thermoelectric element operates as a heat engine when connected in 
parallel with a load resistor $R_L$.  
The direction of the thermoelectric current in the circuit can be controlled with
a DC current generator, as sketched in Fig.~\ref{fig:BTJE_engine}(a). The stationary operating points correspond to the intersections of the  $I(V)$ characteristic with the load line of slope $-1/R_L$ and current axis-intercept selected by the bias current $I_b$ [Fig.~\ref{fig:BTJE_engine}(b)]\footnote{In this setup, the current bias rigidly shifts the load line along the current axis.}. When the device is purely dissipative [$I(V)V\geq 0$], $V=0$ is the unique intercept for $I_b=0$; once the BTJE develops ANC in the subgap region, two additional stable states $\pm V_L$ appear in the region $IV<0$ for small values of $I_b$, where electrical power is delivered to the load~\footnote{The electrically stable solutions are identified by the condition $dI/dV|_{V_L}>0$, as discussed in Sec.~\ref{SubSec:symmetryBreaking}.}.
We remark that every thermoelectric generator is an active element [$IV<0$] that can deliver power to a passive load~\cite{goldsmid2010introduction}; the distinctive feature of the BTJE is that two thermoelectric states with finite-voltage values $\pm V_L$ are equivalently possible.
These two states correspond to opposite branches of the spontaneously broken particle--hole symmetry~\cite{marchegiani_nonlinear_2020}, where the bias current $I_b$ selects the state in which the system relaxes when the bias is removed ($I_b=0)$.
In other words, the bias current plays the role of an external magnetic field in selecting one of two degenerate magnetization states in a ferromagnet; this analogy is evidenced by the hysteresis cycle $V_L(I_b)$ shown in Fig.~\ref{fig:BTJE_engine}(c). The ignition current $I_b$ is in general required because the residual Josephson coupling makes the zero-voltage state metastable. 

More precisely, starting from the symmetric state ($V=0$) at $I_b=0$, a finite current bias $I_b$ shifts the load line toward one of the two thermoactive branches; the sign of $I_b$ selects the branch, with positive (negative) bias selecting $V_L^+$ ($V_L^-$). 
Afterward, the finite-voltage state persists even when the current bias is removed. The current flowing in the circuit, and consequently the power dissipated by the load, are sustained by the power generated in the junction.

To switch the system into the opposite thermoelectric state, $I_b$ is increased in absolute value (with opposite sign compared to the one initially chosen) until the shifted load line becomes tangent to the local minimum (or maximum, depending on the initial sign of $I_b$) of the $I(V)$ characteristic (see Fig.~\ref{fig:BTJE_engine}b); since no thermoelectric stationary state is available, a further increase in $|I_b|$ determines a switch in the sign of the voltage drop across the junction.
The resulting bistability directly implements a volatile current-controlled superconducting thermoelectric 
memory~\cite{Giazotto2025ThermoelectricMemoryUS} \footnote{The memory is volatile because the finite-voltage state is sustained by the temperature difference and is lost when the heating source is removed.}.

In Fig.~\ref{fig:BTJE_engine}(d), we display the electrical power delivered to the load,
\begin{equation}
P_L = \frac{(V_L^\pm)^2}{R_L},
\label{eq:PL}
\end{equation}
as a function of $P_\mathrm{in}$ for a few values of $R_L$. Output powers up to $P_L\sim 20\,$fW are measured, comparable to those reported for quantum-dot heat engines operating at similar cryogenic temperatures~\cite{josefsson_quantum-dot_2018,jaliel_experimental_2019}. The specific output power depends on the tunneling resistance for both quantum dots and the BTJE. However, in the BTJE, the power can be increased, for instance, by decreasing the junction tunnel resistance with wider-area junctions. 
For quantum dots, there are more restrictive requirements, since the barrier tunneling resistance should remain larger than the resistance quantum to satisfy the Coulomb blockade conditions~\cite{Kouwenhoven1997}.  

At sufficiently small values of $R_L$, the load line no longer intersects the thermoactive branches at $I_b=0$ and the engine cannot generate power in the junction [cf.\ inset of Fig.~\ref{fig:BTJE_engine}(d)]. Since the voltage drop in the load is 
bounded by
$|V_p|\leq|V_L|\leq|V_S|$, the junction cannot drive enough current to generate a voltage drop larger than $V_S$ and, moreover, the maximum power is approximately given by $|I(V_p)V_p|$ which can be sustained for a minimum load resistance of $R_L\sim V_p/I(V_p)$. $P_L$ decreases monotonically with $P_\mathrm{in}$, consistently with the reduction of the thermoelectric response discussed above. The experiment reports that the effect persists up to bath temperatures of about $200\,$mK, above which it is progressively suppressed (not shown).

\subsubsection{\textbf{Josephson contribution to thermoelectric transport}}
\label{Sec:Phase_Exp}
For a finite DC voltage bias, the Josephson current of an ideal tunnel junction 
oscillates in time with zero average (AC Josephson effect)~\cite{barone1982physics,tinkham_introduction_2004}. In a realistic device, however, this AC generation radiates into the surrounding electromagnetic environment, including the resonant modes of the junction itself, and feeds back on the junction phase, generating net finite DC contributions that modify the IV characteristic in the subgap region~\cite{werthamer1966nonlinear,likharev1986dynamics,ingold_charge_1992}. These self-coupling effects can hinder the observation of the bipolar thermoelectric signal, especially at low biases, where the QP response is usually suppressed by the presence of the superconducting gap~\cite{marchegiani_phase-tunable_2020,germanese_phase_2023}.

\begin{figure}[t]
    \centering
    \includegraphics[width=0.9\linewidth]{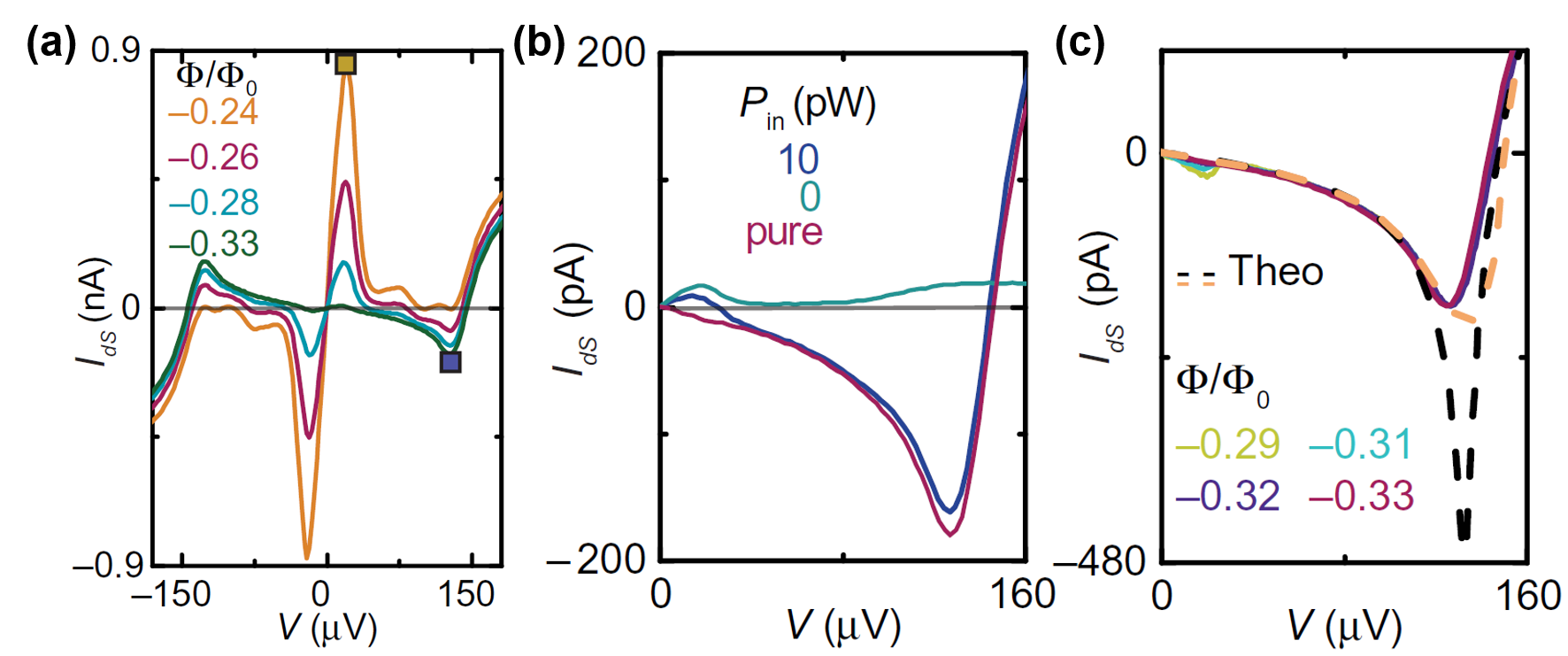}
  \caption{Impact of the Josephson contribution to the thermoelectric transport. 
(a) $I_{\mathrm{dS}}$–$V$ characteristics for $P_{in}=10$ pW for different magnetic flux values at $T_b=30$ mK. (b) Total current vs bias for $P_{in}=0$ (green) and $P_{in}=10$ pW (blue), and extracted QP current $I_{qp}$ (red, see text). 
(c) $I_{qp}$ vs voltage bias for different flux values, showing only a very weak dependence on $\Phi$. The dashed curves are theoretical approximations for different models of the Al/Cu bilayer. Adapted from~\cite{germanese_phase_2023}.}
    \label{fig:JJ_contribution}
\end{figure}

The role of the Josephson coupling in the BTJE can be investigated
by measuring the charge current for different values of  the magnetic flux $\Phi$ threading the double-loop interferometer, as shown in Fig.~\ref{fig:JJ_contribution}(a). As the supercurrent increases away from the optimal point $\Phi/\Phi_0\sim -0.33$ (equivalent by symmetry to the $+0.33$ operating point discussed above), the current near-zero bias is enhanced while the thermoelectric current around the matching peak $-I(V_p)$ is progressively suppressed,  eventually vanishing. The Seebeck voltage $V_S$ is, by contrast, less sensitive to the flux (while still vanishing if $I_c$ gets too large), since it mainly depends on the gap difference $\dAl-\dAlCu$.

To discriminate the QP contribution from the Josephson-related current, a semi-phenomenological decomposition of the measured current can be introduced,
\begin{equation}
I_{dS}(V,\Phi) = I_J(V,\Phi) + I_\mathrm{qp}(V),
\label{eq:decomp}
\end{equation}
where $I_J(V,\Phi)$ accounts for the residual DC Josephson contribution at finite bias and $I_\mathrm{qp}(V)$ for the QP current. 

In the subgap regime at thermal equilibrium ($P_\mathrm{in}=0$), the QP term is exponentially suppressed and the measured trace is dominated by the Josephson component, so that $I_J(V,\Phi)\simeq I_{dS}(V,\Phi)|_{P_\mathrm{in}=0}$. Under finite thermal bias, the QP current is then obtained by subtraction, $I_\mathrm{qp}(V,\Phi)\simeq I_{dS}(V,\Phi)-I_{dS}(V,\Phi)|_{P_\mathrm{in}=0}$, assuming that the temperature difference does not significantly affect the Josephson component. The 
resulting curve [violet line in Fig.~\ref{fig:JJ_contribution}(b)] is given by the difference between the current measured at $P_{in}=10$ pW (blue) and $P_{in}=0$ (green). This difference displays a clear thermoelectric signal with ANC also around $V=0$, i.e., the linear-in-bias bipolar TE (see Sec.~\ref{SubSec:linear-in-bias}). The extracted $I_\mathrm{qp}$ is then essentially independent of $\Phi$ [Fig.~\ref{fig:JJ_contribution}(c)], confirming that the flux primarily modulates the Josephson channel while leaving the thermoelectric response intact, and hence validating the decomposition in Eq.~\eqref{eq:decomp}.

\begin{figure}[t]
    \centering
    \includegraphics[width=0.7\linewidth]{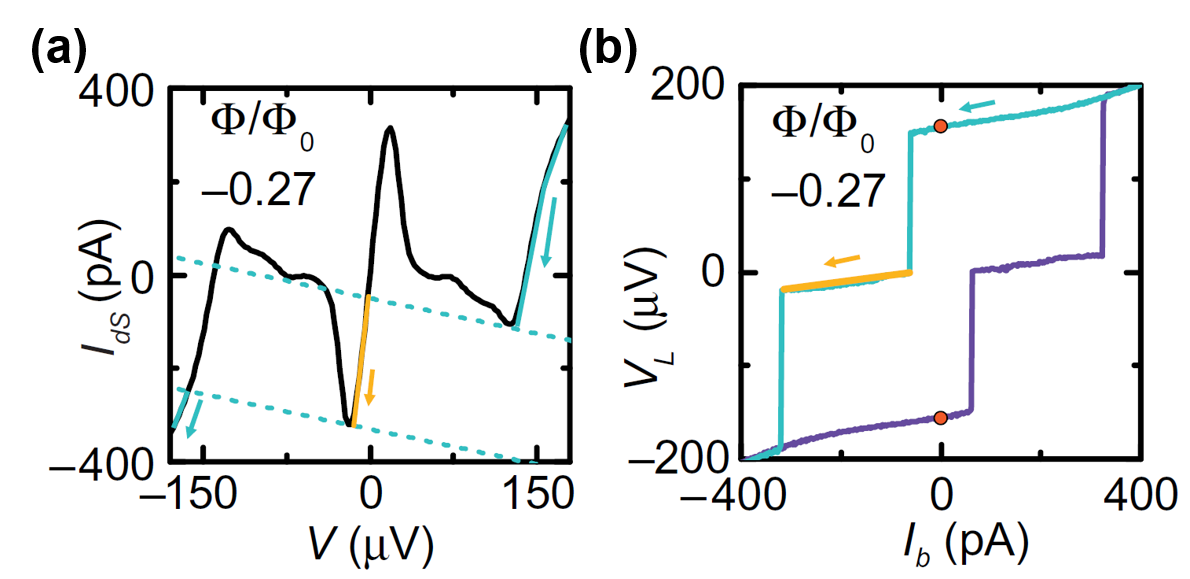}
\caption{Effect of the Josephson contribution on the engine operation. 
(a) Load-line construction. Dashed blue lines show the load line at $I_b=0$ and $I_b\neq0$, with additional intersections near $V_L \simeq 0$. 
(b) Engine characteristics with additional branches. Adapted from~\cite{germanese_phase_2023}.
}
    \label{fig:JJ_engine_Ph}
\end{figure}
The residual supercurrent also affects the engine operation, as illustrated in 
Fig.~\ref{fig:JJ_engine_Ph}. The load-line construction acquires additional 
electrically stable solutions close to the zero-voltage branch [Fig.~\ref{fig:JJ_engine_Ph}(a)], and the hysteretic cycle $V_L(I_b)$ develops 
an intermediate state around $V_L\simeq 0$ [Fig.~\ref{fig:JJ_engine_Ph}(b)]. 
As $I_b$ is reduced from the thermoactive branch, the system first relaxes to 
this intermediate metastable state before switching to the opposite thermoactive
branch at even lower bias. Despite the additional metastable states, the engine 
still operates as a bipolar thermoelectric generator, sustaining two stationary voltage states $\pm V_L$ at $I_b=0$. Then, tuning the Josephson coupling can control these extra states, thereby increasing the number of states in the bipolar thermoelectric memory.

\newpage
\section{Applications of BTJEs}\label{Sec: Applications}
Superconducting tunnel junctions with a bipolar thermoelectric response turn a temperature difference into an electrical output. This heat-powered operation makes BTJEs attractive as on-chip cryogenic elements and suitable platforms for the study of thermodynamics in superconducting electronic circuits~\cite{pekola2015towards}.

Possible applications are radiation detection, thermal management, and circuit elements driven by the effect itself, such as memories, oscillators, and amplifiers. Below, we review proposals and implementations in each of these directions.

\subsection{\textbf{Bipolar Thermoelectric Single-Photon Detection}}
\label{SubSec:singlePhoton}
Superconducting devices are among the leading technologies for cryogenic radiation sensing~\cite{AlexeiDSemenov_2002,Morozov2022SuperconductingPhotonDetectors}, particularly in the microwave and sub-terahertz range, where the small energy carried by individual photons makes their detection challenging~\cite{inomata2016single}. 
Superconducting detectors are based on various equilibrium and nonequilibrium effects: transition-edge sensors exploit the sharp superconducting-to-resistive transition~\cite{irwin1995application,IrwinHilton2005TES}, kinetic inductance detectors the temperature-dependent kinetic inductance~\cite{day2003broadband,Mazin_MKID_2009}, and superconducting nanowire detectors the current-induced breakdown of superconductivity in a biased strip~\cite{gol2001picosecond,Natarajan_2012}.

Thermoelectric-based detectors provide an additional route:
a photon absorbed at one electrode raises its electronic temperature, and the resulting temperature difference across the barrier generates a thermovoltage at open circuit.

Radiation detection exploiting TEs has been proposed in superconductor/ferromagnet junctions~\cite{heikkila2018thermoelectric,
chakraborty_thermally_2018}, in graphene--insulator--superconductor and 
two-dimensional semiconductor--superconductor junctions~\cite{lucchesi2026graphene,lucchesi2026superconductor}, 
and in gap-asymmetric SIS$'$ junctions~\cite{paolucci_highly_2023}. The latter example exploits the bipolar TE
~\cite{paolucci_highly_2023}, which we review here.
Since the detector requires no external electrical bias, the thermoelectric detector operates passively, which is advantageous in large detector arrays 
where the overcrowding of bias lines would limit the achievable pixel count and introduce spurious heat injection~\cite{heikkila2018thermoelectric}.
\begin{figure}[t]
    \centering
    \includegraphics[width=\linewidth]{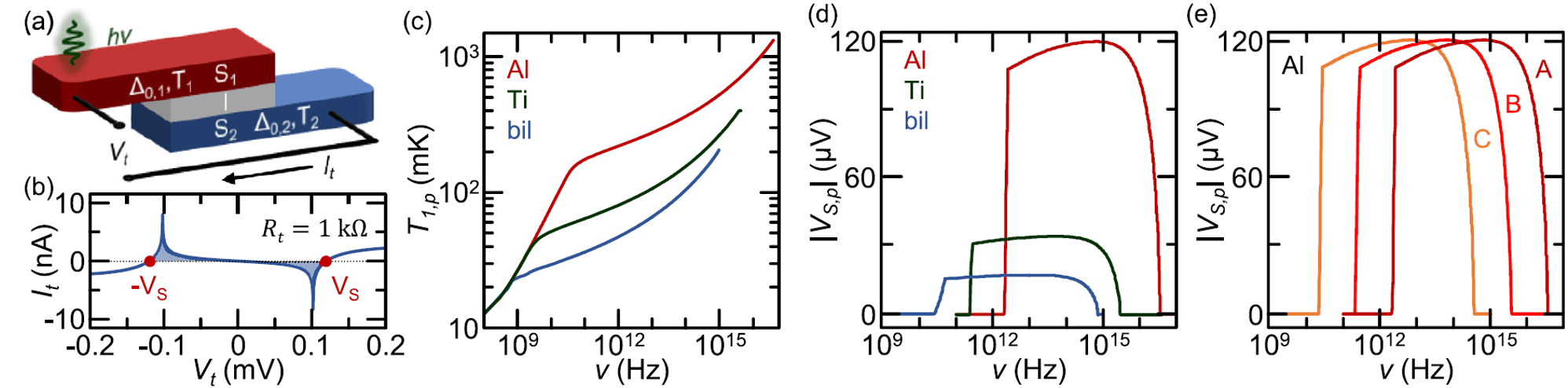}
    \includegraphics[width=0.4\linewidth]{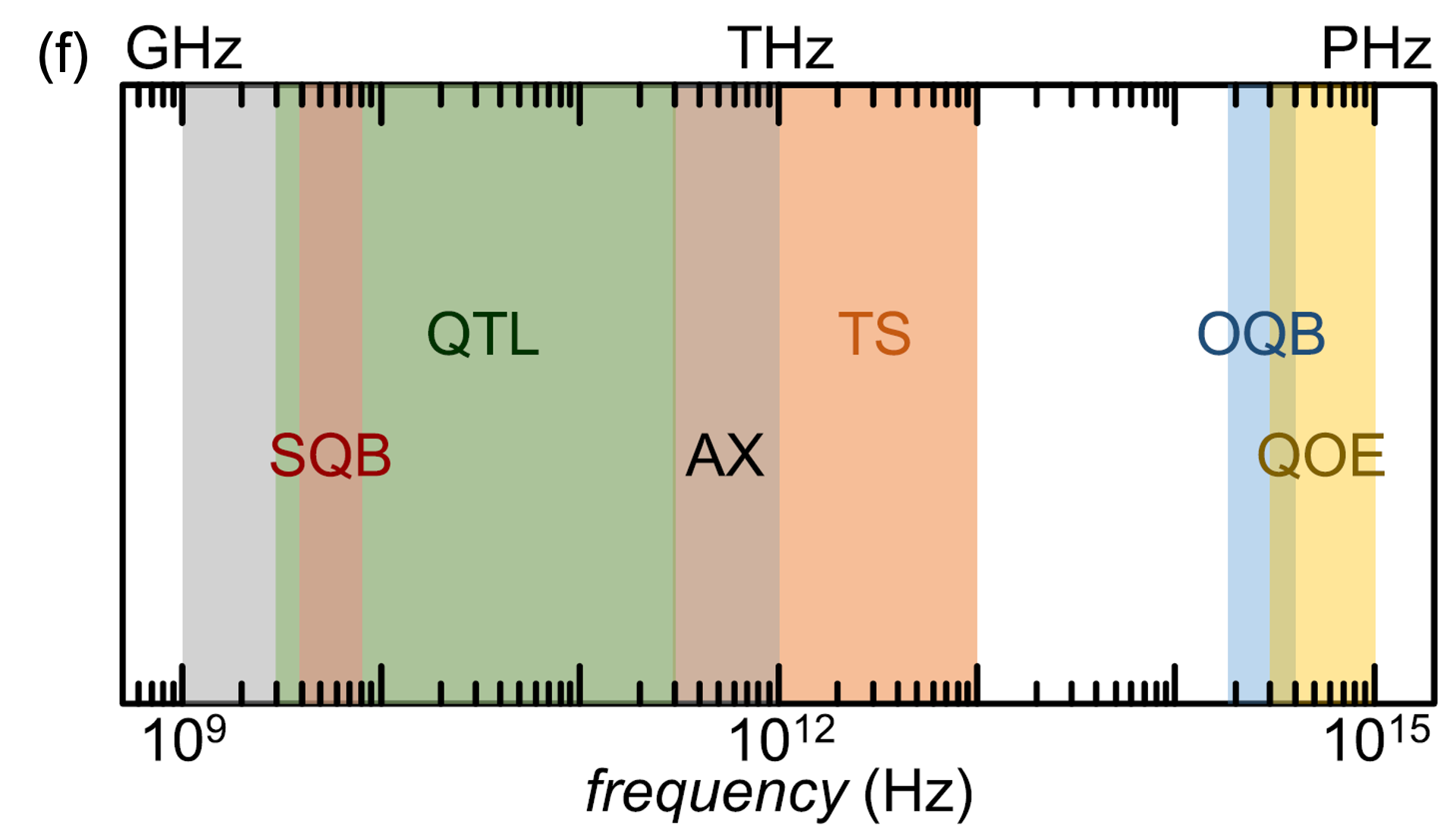}
    \caption{Superconducting thermoelectric detector. (a) Sketch of the S$_1$IS$_2$ tunnel junction ($\Delta_{0,1}>\Delta_{0,2}$): absorption of a photon of energy $h\nu$ raises the electronic temperature of $S_1$ ($T_1>T_2$), driving an open-circuit Seebeck voltage. (b) Thermoelectric $I$--$V$ characteristic for $\Delta_{0,1}=210\,\mu$eV, gap ratio $r=0.5$, $C_i=10^{-4}\Delta_{0,i}$, $R_T=1\,$k$\Omega$, at $T_1=840\,$mK and $T_2=10\,$mK.(c) Peak electronic temperature $T_{1,p}$ vs.\ photon frequency $\nu$, for an absorber of volume $V_1=0.25\,\mu$m$^3$ made of Al, Ti, or an Al/Cu bilayer, at bath temperature $T_b=10\,$mK. (d) Corresponding open-circuit thermovoltage $|V_{S,p}|$ vs.\ $\nu$, for the same systems as in (c), with $r=0.5$ and $T_2=10\,$mK. (e) $|V_{S,p}|$ vs.\ $\nu$ for an Al detector with absorber volumes $V_{1,A}=0.25\,\mu$m$^3$, $V_{1,B}=2.5\times10^{-2}\,\mu$m$^3$, and $V_{1,C}=2.5\times10^{-3}\,\mu$m$^3$, showing how a smaller volume shifts the detection window to lower frequencies. (f) Potential applications of broadband superconducting thermoelectric single-photon detectors, spanning axion dark-matter searches to quantum optoelectronics and qubit read-out. Adapted from ~\cite{paolucci_highly_2023}.}
    \label{fig:TED1}
\end{figure}
In Fig.~\ref{fig:TED1}(a), we sketch the thermoelectric detector, featuring a superconductor $S_1$ (absorber element) tunnel coupled to a lower gap superconductor $S_2$. 
Upon absorption of a photon of energy $h\nu$, the electronic temperature in the $S_1$ electrode is raised to the value $T_{1,p}$~\footnote{This description is valid in the quasiequilibrium regime, when electron–electron scattering is much faster than electron–phonon scattering and tunneling through the junction. This regime is expected to hold for sufficiently small absorber volumes~\cite{paolucci_highly_2023,giazotto_opportunities_2006}.}, which can be computed using
\begin{equation}\label{eq:Tpeak}
  \int_{T_b}^{T_{1,p}(\nu)} C_e(T)\,dT = h\nu ,
\end{equation}
assuming that the photon energy is fully absorbed by the electrons in $S_1$. In Eq.~\eqref{eq:Tpeak}, $T_b$ is the QP temperature of $S_1$ (assumed equal to the bath temperature) before the photon absorption, 
$C_e(T)=T\,dS_1/dT|_{V_1}$ is the electronic heat capacity of the absorber, 
and $S_1(T)$ is its entropy. Within the mean-field BCS treatment, the entropy of a free gas of Bogoliubov QPs reads~\cite{tinkham_introduction_2004},
\begin{equation}\label{eq:entropy}
  S_1(T) = -2 V_1 \nu_F k_B \int_{-\infty}^{\infty} d\varepsilon\,
  N_1(\varepsilon,T)\, f(\varepsilon,T)\, \ln f(\varepsilon,T) ,
\end{equation}
with $V_1$ the absorber volume, $\nu_F$ the normal-state DoS at the Fermi level, $N_1(\e,T)$ the BCS QP DoS, and $f(\e,T)$ the Fermi function at electron temperature $T$.

Once $T_{1,p}$ is sufficiently large, the junction develops the nonlinear thermoelectric response discussed in the previous section, as illustrated by the $I$--$V$ characteristic of Fig.~\ref{fig:TED1}(b): the curve develops a branch of ANC, and its open-circuit point ($I=0$) sets the detector's working point. In practice, the measurement circuit closely approximates this open-circuit condition, so the readout signal is well approximated by the Seebeck voltage $V_{S}$, which encodes the photon energy and is the quantity used to detect the event.

The detector's operating window depends on the absorber's electronic temperature as a function of photon frequency $\nu$. Figure~\ref{fig:TED1}(c) displays $T_{1,p}$ as a function of $\nu$ for several absorber materials. As expected, it rises monotonically with $\nu$. At fixed $\nu$, a larger superconducting gap yields a higher peak temperature owing to the correspondingly smaller heat capacity $C_e(T)$. Indeed, with a larger gap fewer QPs are thermally excited at a given temperature; these QPs determine the electronic heat capacity. 
Since $C_e(T)$ is smaller for a higher gap superconductor, the temperature $T_{1,p}$ becomes correspondingly higher to satisfy Eq.~\eqref{eq:Tpeak} at fixed $\nu$.
For a representative gap ratio $\Delta_{0,2}=\Delta_{0,1}/2$, panel~(d) shows the corresponding Seebeck voltage $V_{S,p}$ generated by this temperature difference. Above threshold, $V_{S,p}$ depends only weakly on $\nu$: the detector essentially behaves digitally, 
with the absorber material setting the threshold and the signal size.
The absorber volume $V_1$ also tunes the operating window through the heat capacity: a smaller $V_1$ moves the detection range to lower photon energies while leaving the output voltage nearly unchanged [Fig.~\ref{fig:TED1}(e)].

Material choice and absorber design together give broad spectral coverage, with a single-photon range spanning roughly $15\,$GHz to $150\,$PHz. This reach suits the many areas of quantum science and technology that rely on single-photon detection [Fig.~\ref{fig:TED1}(f)]: searches for axion dark matter~\cite{lamoreaux2013analysis,dixit2021searching}, terahertz spectroscopy of materials and molecules~\cite{tonouchi2007cutting}, and quantum telecommunications, optoelectronics, and qubit read-out~\cite{hadfield2009single}.

\subsection{\textbf{Thermoelectric heat management: heat pipes and diodes}}\label{SubSec: Pipe}
Heat management is a crucial requirement in cryogenic and quantum circuits, where parasitic heat loads spoil device performance and stability~\cite{pekola2021colloquium}. Controlling how heat flows on the chip is therefore as important as minimizing it. One strategy is thermal rectification: in a thermal diode, heat flows predominantly in the forward direction rather than the backward direction~\cite{li2012colloquium}. This heat-management element has been theoretically investigated in various contexts, including nonlinear lattice models~\cite{terraneo2002controlling,li2004thermal}, spin-boson nanojunction models~\cite{segal2005spin}, and near-field photonic radiation~\cite{Otey_Thermal_2010,BenAbdallah_Phase_2013}. This effect has also been studied (see, e.g., Ref.~\cite{Fornieri_AIPAdv}) and observed in superconducting junctions~\cite{martinez2015rectification}. More generally, superconducting nanostructures can be assembled into thermal circuits that guide, redistribute, or suppress heat flow~\cite{fornieri_towards_2017}.
As we show in Ref.~\cite{antola_tunable_2024}, the bipolar TE enriches this toolbox in two ways: it strengthens heat rectification, increasing the rectification coefficient and widening the parameter range over which the junction acts as a thermal diode, and it enables a heat pipe that carries excess heat away from a local hot spot toward a colder stage. Both functionalities are inherently passive: the only input is a temperature difference across the junction.
 
First, we discuss how the bipolar TE enhances non-reciprocal heat transport in gap-asymmetric SIS$'$ junctions. This enhancement is related to a key asymmetry in the bipolar TE: thermoelectricity occurs only when the higher-gap material is at the higher temperature.
In this configuration, a finite thermoelectric voltage $\pm\bar{V}$ develops across the junction; reversing the temperature difference, by contrast, produces no thermoelectric voltage, i.e., $V=0$. In the thermoelectric state, the heat exchange is strongly enhanced when $\bar{V}$ approaches $V_p$.

\begin{figure}[t]
    \centering
    \includegraphics[width=\linewidth]{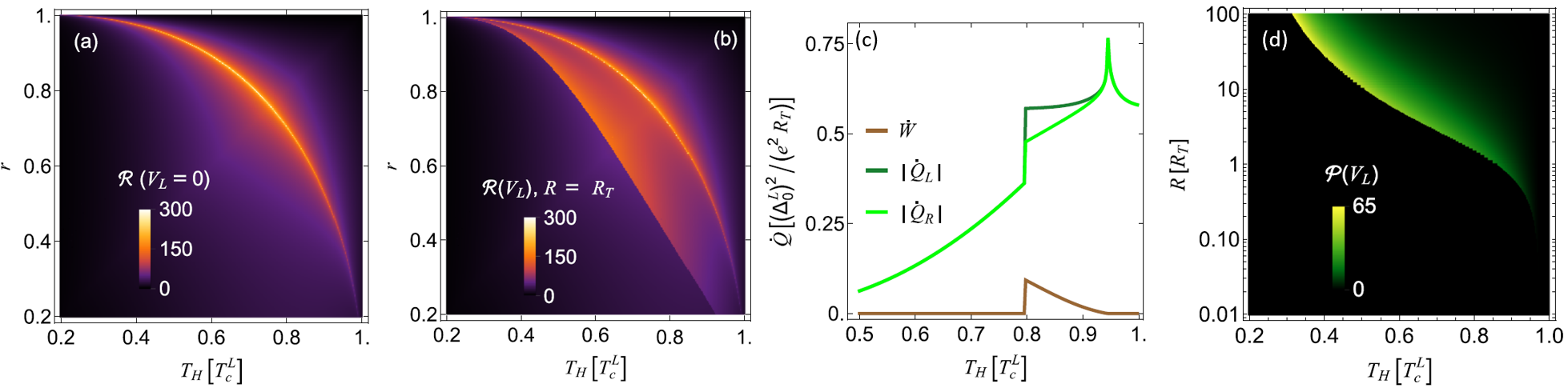}
\caption{Thermoelectric thermal management: diode and pipe.
(a) Heat rectification coefficient $\mathcal{R}$ at $V=0$ as a function of the gap ratio $r$ and the hot-electrode temperature $T_H$, for $T_C=0.01\,\tcL$ and $\gamma=10^{-4}\,\dLo$. (b) $\mathcal{R}(\bar{V})$ in the presence of a load resistor $R_L=R_T$, for the same parameters as in panel (a). (c) Heat currents and thermoelectric power as a function of $T_H$ for $r=0.4$ and $R_L=R_T$. (d) Heat-piping efficiency $\mathcal{P}$ as a function of $R_L$ and $T_H$ for $r=0.4$. Adapted from~\cite{antola_tunable_2024}.}
    \label{fig:PIPE}
\end{figure}

The asymmetry between the forward and backward heat flow can be quantified using the rectification coefficient
\begin{equation}\label{eq:rect}
  \mathcal{R}(\bar V) = \frac{\dot Q_+(\bar V) - \dot Q_-(0)}{\dot Q_-(0)}\times 100 ,
\end{equation}
where $T_H$ and $T_C$ are the hot and cold temperatures, respectively, $\dot Q_+(\bar V)=\dot Q_R(\tL=T_H,\tR=T_C,\bar V)$ is the heat flowing out of the higher-gap electrode when it is hotter, and $\dot Q_-(0)=\dot Q_L(\tL=T_C,\tR=T_H,V=0)$ is the heat current in the reversed configuration.
We remark that a finite rectification occurs even when the thermoelectric symmetry breaking is ignored (setting $V=0$ in both configurations), due to the gap asymmetry~\cite{martinez2013efficient}. Figure~\ref{fig:PIPE}(a) shows the density plot of $\mathcal{R}$ as a function of the gap ratio $r$ and the hot-electrode temperature $T_H$ at $V=0$. Values up to about $260\%$ are reached, but only within a narrow region in the parameter space ($T_H$ and $r$).

For a junction coupled in parallel with a load resistor, the picture changes: the bipolar thermoelectric junction can relax into a finite-voltage thermoelectric state, and the rectified region widens considerably [Fig.~\ref{fig:PIPE}(b)], as long as the load line meets the nonlinear branch and yields an electrically stable finite-voltage solution (see Secs.~\ref{SubSec:symmetryBreaking} and~\ref{Sec. BTJE}). The junction then uses the temperature difference to generate the bias that brings it near the matching-peak condition, where heat exchange is enhanced. The thermoelectric response thus reinforces the gap-induced asymmetry into a more effective diode. This voltage-enhanced rectification remains formally passive, since the bias is supplied by the junction's own thermoelectric response rather than by an external source~\cite{Khomchenko_PRB106}.

The same circuit element also operates as a heat pipe. In a finite-voltage thermoelectric state, the junction delivers a power $\dot W = I_L V_L$ to the load. Since the load can be physically placed far from the junction, the fraction $\dot W$ of heat current flowing out of the hot electrode ($\dot{Q}_L$) is dissipated remotely, and only the remaining heat current $\dot{Q}_R=\dot{Q}_L-\dot W$ reaches the cold electrode. In this way, heat can be partially evacuated from the cryogenic region and transferred elsewhere; this feature is particularly desirable in superconducting elements, since the heat removal via electron-phonon coupling is strongly suppressed at low temperatures.  Figure~\ref{fig:PIPE}(c) displays the heat currents and the thermoelectric power as a function of the hot-electrode temperature; the heat-pipe regime sets in once $T_H$ is large enough for the load line to cross the nonlinear $I$--$V$ characteristic, so that the junction sustains the TE across the load (see Sec.~\ref{Sec. BTJE}). 
The efficiency of the heat pipe can be defined as
\begin{equation}\label{eq:pipe}
  \mathcal{P} = \frac{\dot W}{\dot Q_R}.
\end{equation}
This quantity differs from the efficiency of the thermoelectric heat engine, as it refers to a different figure of merit. Figure~\ref{fig:PIPE}(d) shows how the heat-piping efficiency $\mathcal{P}$ depends on the load resistance $R_L$ and the hot-electrode temperature $T_H$. A larger $R_L$ extends the operating regime toward lower $T_H$, where the thermodynamic efficiency of the bipolar conversion is highest (Sec.~\ref{SubSec:Efficiency}), and $\mathcal{P}$ is also optimized, approaching about $65\%$.

\subsection{\textbf{Thermoelectric circuit applications}}
\label{subsec:TEcircuit}
Bistable switches, relaxation oscillators, and amplifiers can be realized with a single nonlinear element possessing a region of negative differential conductance, the standard example being the tunnel diode, which has been known since the late 1950s~\cite{esaki1958new}. In a tunnel diode, the differential conductance is negative, but the absolute conductance remains positive ($IV>0$): the element is dissipative and requires a current bias to operate. A bipolar thermoelectric junction differs in this respect: under an applied temperature difference, it displays both absolute negative ($IV<0$, cf.\ Sec.~\ref{Sec: Th}) and differential negative conductance at subgap voltages. The junction is therefore an active element~\footnote{The bipolar thermoelectric junction was called ``passive'' previously and ``active'' here. We remark that the terms refer to different contexts: ``passive'' denotes operation without any external electrical bias, with the driving power coming from heat, whereas ``active'' denotes the ability to deliver power ($IV<0$), unlike a dissipative element such as a tunnel diode.}, generating power from the heat current, and enabling circuit operations at zero bias with no static dissipation. Below, we discuss the bipolar-thermoelectric switch and memory, the relaxation oscillator, and the voltage amplifier, together with their integration into arrays.

\subsubsection{\textbf{Thermoelectric memory}}
Compact, low-dissipation memory is a key enabling component of cryogenic and superconducting electronic systems, which remains challenging to realize~\cite{holmes2013energy,Alam2023}. The bipolar thermoelectric SIS$'$ junction can tackle this need, powered by heat alone~\cite{marchegiani_superconducting_2020}. 

We recall that this circuit element can develop a finite-voltage state of either polarity, $\pm V_L$, for a suitable temperature difference,  if connected in parallel with a load resistor $R_L$ (see discussion in Sec.~\ref{SubSec:symmetryBreaking} and Fig.~\ref{fig:PRLEH}). 
This feature can be exploited to store aclassical bit~\cite{Giazotto2025ThermoelectricMemoryUS}. Under a fixed temperature difference, the two states $\pm V_L$ are metastable and encode two logic levels. A current bias $I_b$ injected into the parallel branch writes and erases the cell, with the current direction selecting the state. The stored value is then read as the voltage across the load. A residual Josephson contribution can add a further state near $V\simeq 0$, modifying the switching dynamics and shifting the current needed to move between states (see Sec.~\ref{Sec. BTJE}). The stored state survives after the writing current is removed, as long as the temperature difference is sustained. The memory is therefore volatile: the bit is lost when the thermal bias is switched off, and the junction returns to the zero-voltage state. The capacitance in parallel with the junction fixes both stability and speed. Small capacitances leave the cell sensitive to fluctuations, which can trigger spontaneous switching (see Sec.~\ref{Sec: Noise}). Larger ones improve stability but increase the write and erase time, which, for realistic parameters, operate in the GHz range~\cite{marchegiani_noise_2020,Giazotto2025ThermoelectricMemoryUS}.
\begin{figure}[t]
    \centering
    \includegraphics[width=\linewidth]{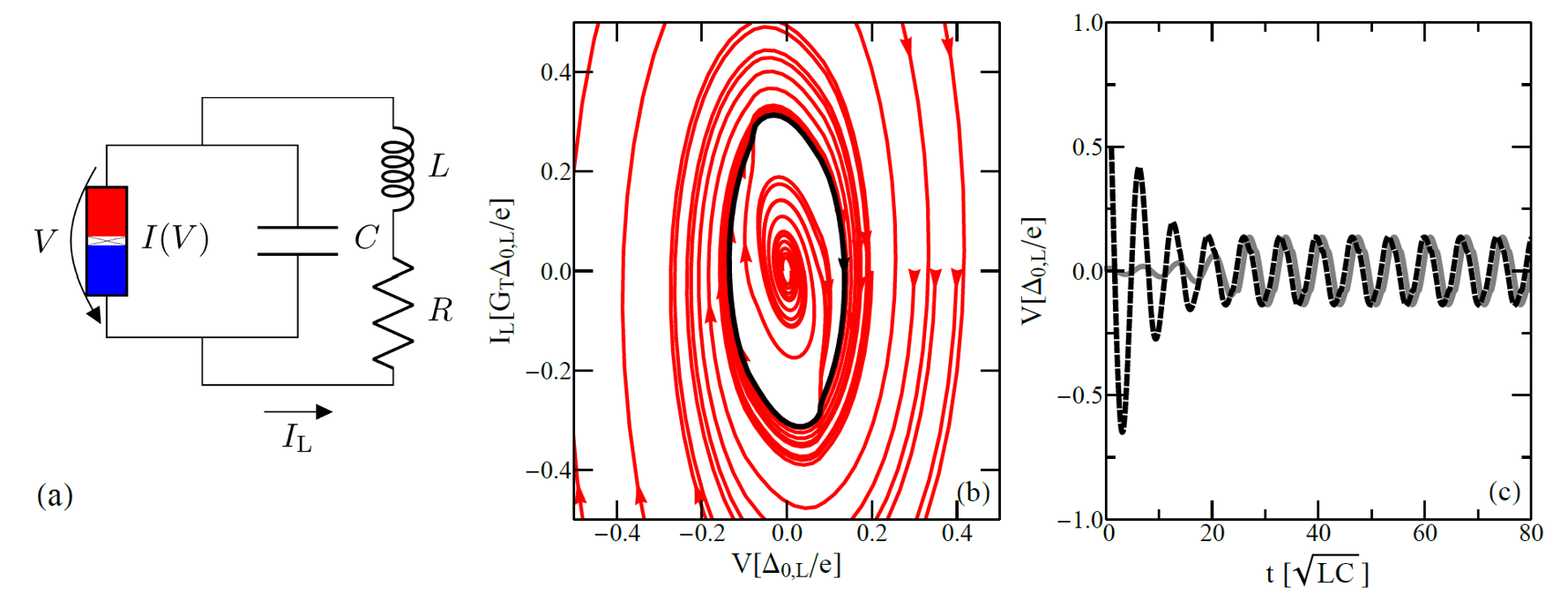}
    \caption{Thermoelectric relaxation oscillator. (a) Circuit schematic including the thermoelectric junction, the load resistor, and the reactive elements $L$ and $C$. (b) Phase portrait for $R=0.1\,G_T^{-1}$, showing the convergence of all trajectories to a limit cycle (black curve). (c) Time evolution of the voltage for selected initial conditions, $(V,I)=(\Delta_{0,L}/e,0)$ (dashed black) and $(V,I)=(0.01\Delta_{0,L}/e,0)$ (solid gray), illustrating the approach to the same periodic steady state. Parameters: $G_T^{-1}=100\,\Omega$, $L=100\,\mathrm{pH}$, $C=50\,\mathrm{fF}$, $T_L=0.7\,T_{c,L}$, $T_R=0.01\,T_{c,L}$, and $r=0.75$. Adapted from~\cite{marchegiani_nonlinear_2020}.}
    \label{fig:oscillator}
\end{figure}

\subsubsection{\bf Relaxation oscillator}
The applications described in the previous subsection are based on stationary DC thermoelectric states.
A different regime appears when the $I$--$V$ characteristic displays thermoelectricity, but the load resistance is too small, $R<\min\{G_0^{-1},\, I(V_p)/V_p\}$, leaving $V=0$ as the only static solution. In such a case, thermoelectricity can sustain oscillatory dynamics. This possibility can be studied in the setup of Fig.~\ref{fig:oscillator}(a), including a series inductance $L$.
The charge dynamic is governed by
\begin{equation}\label{eq:osc}
  \begin{cases} I_L = C\dot V + I(V),\\ V = -L\dot I_L - R\,I_L, \end{cases}
\end{equation}
for the load current $I_L$ and junction voltage $V$.
Linearizing the system of Eq.~\eqref{eq:osc} around $V=0$ shows that the stability depends on the zero-bias differential conductance $G_0 = dI/dV|_{V=0}$: for $G_0>0$ the $V=0$ state is always stable, while for $G_0<0$ it is stable only if
\begin{equation}\label{eq:osc-stability}
  |G_0|\sqrt{L/C} < 1 .
\end{equation}
If this inequality is violated, the static solution $V=0$ becomes unstable. With no additional fixed points, the trajectory winds onto a closed orbit in phase space, a limit cycle [see, e.g., Fig.~\ref{fig:oscillator}(b)]. Both the voltage and current then evolve periodically in time (with zero average) at the steady state, with a period and waveform set by the circuit parameters and the shape of the nonlinear $I(V)$ [Fig.~\ref{fig:oscillator}(c) shows how different initial conditions lead to the same steady-state dynamics, up to a phase shift]. This is the relaxation oscillator familiar from nonlinear dynamics~\cite{van1926lxxxviii}, with a long experimental history in superconducting circuits: hysteretic Josephson junctions driven into relaxation oscillations~\cite{vernon1968relaxation}, and the relaxation-oscillation SQUID used for sensitive flux readout~\cite{adelerhof1994double}. 
Those devices are current-biased and switch between the zero-voltage and finite-voltage states; the bipolar junction differs in that its oscillation is sustained at zero average bias by the temperature difference alone, with no DC power source. 

\subsubsection{\textbf{Zero-bias thermoelectric voltage amplifier}}
Superconducting amplifiers based on flux-sensitive architectures are well established, from DC SQUID  amplifiers~\cite{clarke2004squid} to Josephson
parametric~\cite{aumentado2020superconducting} and traveling-wave parametric amplifiers~\cite{macklin2015near}. Direct voltage amplification at low frequencies, from near DC to about $100$~MHz, is less developed, yet that is the band needed for multiplexed detector readout~\cite{IrwinHilton2005TES}, transition-edge sensors~\cite{IrwinHilton2005TES,lucia_transition_2024}, superconducting single-photon detectors~\cite{mccaughan2018readout}, and nanomechanical resonators~\cite{ekinci2005nanoelectromechanical}. In this frequency range, the amplification is usually handled by cryogenic semiconductor amplifiers, whose dissipation keepsthem at the $4$~K stage rather than at millikelvin temperatures~\cite{ivanov2011cryogenic}. A superconducting amplifier with voltage gain, nanowatt dissipation, and zero-bias operation would be beneficial to enable lower-temperature operation. Bipolar thermoelectricity offers one route, through asymmetric superconducting tunnel junctions~\cite{trupiano2026zerobias}.

The minimal circuit features an asymmetric-gap SIS$^\prime$ junction in series with a load resistor $R_L$, driven by a purely AC input $V_{\rm in}$, with the output $V_{\rm out}$ read across the junction. It works as a voltage divider set by the nonlinear thermoelectric response. Where the junction shows negative differential resistance, the load resistor establishes a load line, and a small AC input shifts the operating point along it, yielding an amplified swing in the junction voltage.
Unlike conventional negative-resistance amplifiers, which bias into the active region and dissipate DC   power~\cite{esaki1958new}, the operating point here sits
at $V=0$: no DC power is drawn, and the energy for amplification comes from the heat current from the hot element. Because the $I$--$V$ characteristic is odd, this zero-bias point lies at the center of the negative-resistance region, which suppresses even-order
distortion and improves the trade-off between linearity, gain, and dynamic range.

The negative-resistance region exists only in a finite thermal window. At low $T_H$, the thermal broadening is too small to form it; at high $T_H$, the gap of the hot electrode collapses, and the effect is quenched. A relatively large Dynes broadening is assumed, $\Gamma_{1,2}=0.075\,\Delta_{1,2}(0)$, smoothing the QP characteristic so that sharp features do not create multiple load-line crossings, which keeps a single operating point and a stable response. This feature may be effectively implemented by replacing a pure superconductor with superconductor--normal-metal bilayers, where broadening of this magnitude is commonly observed (see, e.g., Ref.~\cite{germanese_bipolar_2022}). The gain follows from small-signal voltage division between $R_L$ and the junction differential resistance $r_d = (dI/dV)^{-1}$. At zero bias, where $r_d<0$,
\begin{equation}\label{eq:gain}
  G_V = \frac{|r_d|}{|r_d| - R_L} .
\end{equation}
Gain above unity needs $|r_d| > R_L$ and grows as $R_L$ approaches $|r_d|$. The gain holds from near DC to a cutoff near $100$~MHz, set by the junction $RC$ time, and falls as the bath temperature rises, tracking the weakening of the bipolar TE and the growth of $|r_d|$. 

\subsubsection{\bf Arrays and circuit integration}
\label{Subsec:arrayIntegration}
\begin{figure}[t]
    \centering
    \includegraphics[width=0.7\linewidth]{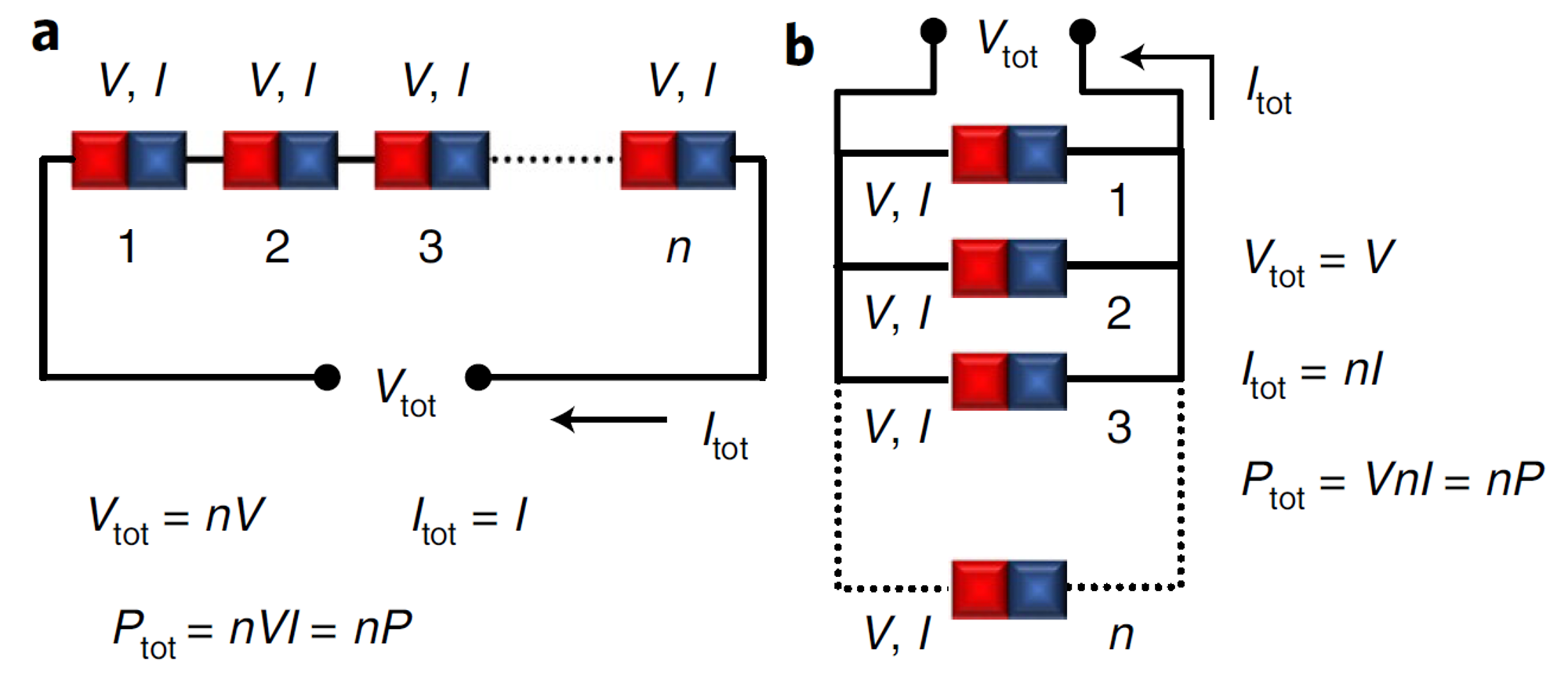}
    \caption{Series and parallel integration of thermoelectric junctions. (a) Series architecture, where the output voltage adds across multiple elements, yielding $V_{\mathrm{tot}}=nV$. (b) Parallel architecture, where the output current scales with the number of elements, yielding $I_{\mathrm{tot}}=nI$. Adapted from~\cite{germanese_bipolar_2022}.}
    \label{fig:arrays}
\end{figure}
Series and parallel integration of individual elements is the standard way to scale a thermoelectric generator: conventional modules wire many alternating $n$- and $p$-type legs of opposite Seebeck sign together, so their thermovoltages add along the series string, building up voltage and current~\cite{goldsmid2010introduction}. The same 
idea scales the bipolar junctions, with one key advantage: a bipolar junction provides thermopower of either sign on its own, so the array can be built from a single type of element. The same advantage underlies electron cooling in 
SINIS refrigerators~\cite{leivo1996efficient}.

In the series connection [Fig.~\ref{fig:arrays}(a)] the voltages add, so $V_{\rm tot}=nV$ and $P_{\rm tot}=nP$. This choice enables compact layouts in which neighboring junctions share an electrode, as in an SIS$'$IS stack. A shared hot island improves efficiency, since the same injected heat drives both elements. In the parallel connection [Fig.~\ref{fig:arrays}(b)] the currents add, $I_{\rm tot}=nI$, again giving $P_{\rm tot}=nP$. Experiments on superconducting bipolar thermoelectric devices already use parallel combinations of junctions~\cite{germanese_bipolar_2022}, which suggests scalable implementations.
Together, series and parallel architectures provide a natural route toward integrated superconducting thermal circuits that are powered by local heating.
\newpage

\section{Bipolar thermoelectricity in hybrid and gated platforms}\label{Sec: hybrids}
The bipolar TE was originally investigated in tunnel junctions between two superconductors, but it is not exclusive to this configuration.
Hybrid structures in which superconductors are combined with normal metals or semiconductors, as well as gated devices, 
enable enhanced control over the TE by exploiting the versatility of mesoscopic physics. 

This section reviews four representative examples.
In single-electron transistors, charging effects enrich the bipolar thermoelectric response and make it gate-tunable. Bilayer graphene junctions enable electrostatic control of the spectral gap, extending bipolar thermoelectricity to non-superconducting electrodes and to higher operating temperatures. Proximity interferometers achieve a similar spectral-gap modulation through an applied magnetic field, enabling phase-coherent tuning of the spectrum. Finally, spin-split superconducting junctions couple thermal and spin transport, producing a thermospin response while preserving the 
reciprocity of the $I$--$V$ characteristic.

\subsection{\textbf{Thermoelectric single-electron transistors}}
Charging effects enable the electrostatic control of bipolar thermoelectricity in superconducting junctions. The standard setup is the single-electron transistor, comprising a small conducting island tunnel-coupled to two leads and gated through a capacitor. The small size of the island determines a large charging energy $E_C$, the electrostatic energy cost of adding a single electron, giving rise to Coulomb blockade, which stabilizes discrete charge states when $E_C \gg k_BT$ and tunnel resistances are large enough, $R_T\gg R_K=h/e^2$~\cite{AverinLikharev1986,fulton1987observation}. The thermopower of such a gated island is itself an established effect: as the gate sweeps the island through successive charge states, it oscillates in a sawtooth pattern~\cite{beenakker1992theory,staring1993coulomb}, reshaped at low temperatures by cotunneling~\cite{turek2002cotunneling} and by the superconductivity of the island~\cite{turek2005thermopower}. 
In all these cases, the thermopower arises from Coulomb-blockade energy filtering through a gate-induced, energy-asymmetric transmission, in contrast to the bipolar mechanism of Sec.~\ref{Sec: Th}, which operates in junctions also satisfying EIS. For our purposes, the Coulomb blockade primarily serves as a gate-controlled knob for the TE.

\begin{figure}[t]
    \centering
    \includegraphics[width=0.7\linewidth]{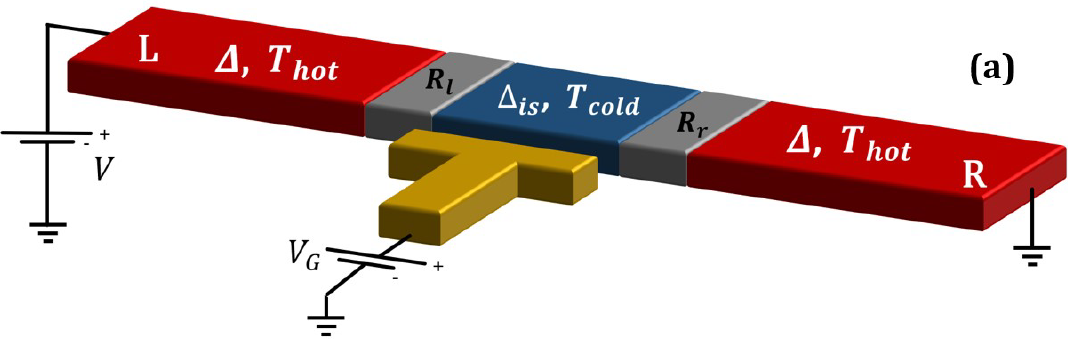}
    \caption{\textbf{Bipolar thermoelectric SET device.} Superconducting island (blue) tunnel-coupled to two superconducting leads (L, R, in red) and capacitively coupled to a gate electrode (yellow). The gate voltage $V_g$ controls the offset charge $N_G = C_g V_g/e$ in the island. Adapted from~\cite{battisti2024bipolar}.}
    \label{fig:SET1}
\end{figure}

The device proposed in Ref.~\cite{battisti2024bipolar} consists of a superconducting island tunnel-coupled to two superconducting leads, left (L) and right (R), forming an SIS$'$IS structure [Fig.~\ref{fig:SET1}]. A gate electrode is capacitively coupled to the island through $C_g$, while the two tunnel barriers have capacitances $C_L$ and $C_R$, so that the total capacitance is $C_{\mathrm{tot}} = C_g + C_L + C_R$ and the charging energy is $E_C = e^2/2C_{\mathrm{tot}}$. For $E_C \gg k_B T_l$, with $l=\{\mathrm{is},L,R\}$, and for $R_T\gg R_K$~\cite{devoret1992introduction}, the charge transport in the island enters the Coulomb blockade regime. This setup has been thoroughly investigated experimentally, in relation to the charge periodicity of the transistor response~\cite{tuominen1992experimental,joyez1994observation}, providing the sensitivity to measure single-electron charges~\cite{Yoo1997} and one of the first demonstrations of QP poisoning~\cite{Aumentado_QP_poisoning}. 
The tunnel barriers are assumed to be resistive enough to suppress the Josephson coupling ($E_C\gg E_J$). In this regime, the current and the output power are consequently small, as expected for a single-electron transistor.
We remark that the thermal configuration considered here is nonstandard. The two leads are held at the same temperature, which is higher than that of the island, whereas in all the references cited above, the temperature bias is between the left and right leads. For linear (unipolar) TEs, the thermoelectric current would vanish in this symmetric setup. By contrast, the bipolar TE operates and provides a finite response in this configuration~\cite{battisti2024bipolar} (see also Sec.~\ref{Subsec:arrayIntegration}).

In the sequential-tunneling approximation, the current follows from a standard master equation for the populations $P_n$ of the island charge states $n$, $\dot P_n = \sum_{n'} W_{nn'}P_{n'}$~\cite{NazarovCounting}. The kernel is built from the golden-rule tunneling rates of Sec.~\ref{Sec: Th}, evaluated at the electrostatic energy difference $\delta U_{n,n\pm1}$ associated with each transition $n\to n\pm1$; the latter is set by the charging energy $U(n)=E_C(n-q_{\rm is}/e)^2/2$ of Coulomb-blockade theory~\cite{AverinLikharev1986}, where $q_{\rm is}=C_gV_g+\sum_j C_jV_j$ is the island offset charge. Denoting these rates $\Gamma_j^{n,n\pm1}$ ($j=L,R$), the stationary current in the right lead reads
\begin{equation}\label{eq:setcurrent}
  I = I_R = -I_L = e\sum_n \left[\Gamma_R^{n,n+1}-\Gamma_R^{n,n-1}\right]P_n^0,
\end{equation}
with $P_n^0$ the stationary probability of island charge state $n$.

\begin{figure}[t]
    \centering
    \includegraphics[width=\linewidth]{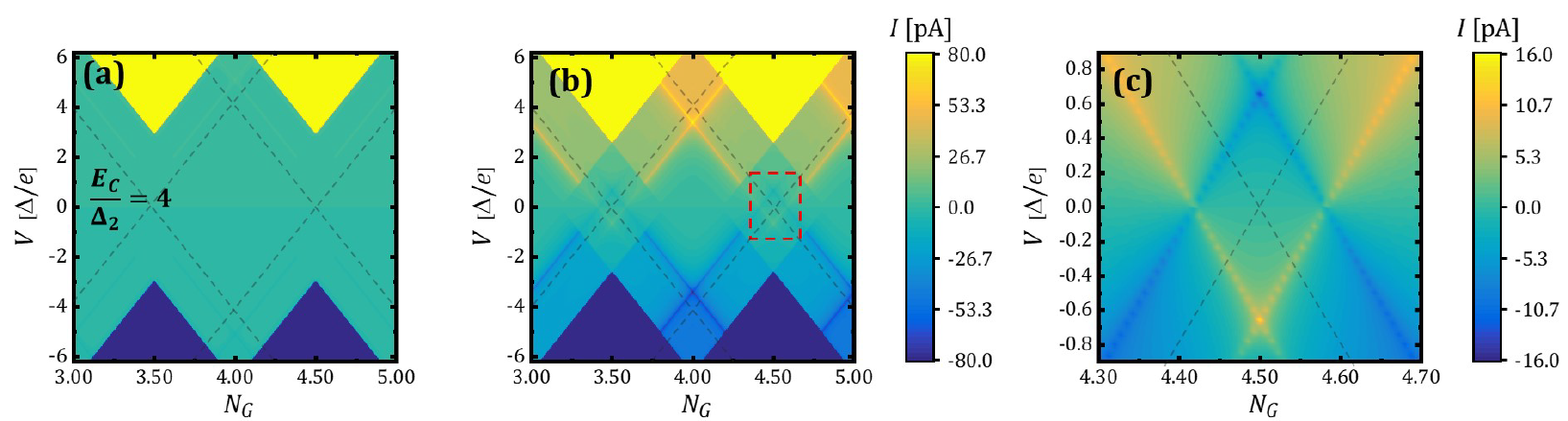}
    \caption{\textbf{Bipolar thermoelectric response in a SET.} 
    (a) Stability diagram at thermal equilibrium, showing the current $I$ as a function of bias voltage $V$ and gate-induced offset charge $N_G = C_g V_g/e$, with characteristic Coulomb blockade diamonds and superconductivity enlarging the blockade region.(b) Same as (a) under a finite thermal bias, $T_{\mathrm{hot}} > T_{\mathrm{cold}}$, where additional subgap transport channels emerge within the diamonds. (c) Zoom highlighting the thermoelectric regime near half-integer $N_G$, where the current reverses sign with respect to the applied bias ($IV < 0$), indicating a bipolar thermoelectric response tunable via the gate. Adapted from~\cite{battisti2024bipolar}.}
    \label{fig:SET2}
\end{figure}

Figure~\ref{fig:SET2}(a) shows the resulting stability diagram (current output as a function of gate voltage and voltage bias) at equilibrium and low temperature, $k_B T_l\ll\Delta,\Delta_{\rm is}$ for all elements $l$, with the characteristic Coulomb diamonds periodic in the gate-induced offset charge $N_G = C_gV_g/e$~\cite{nazarov_quantum_2009}. In this regime, QP tunneling additionally requires overcoming the superconducting gaps of the island, thereby suppressing transport for any bias below the sum of the gaps,
\begin{equation}\label{eq:setgap}
    e|V| < 2(\Delta+\Delta_{\rm is}).
\end{equation}
However, with superconductivity the effective threshold for transport combines the gap sum of Eq.~\eqref{eq:setgap} with the charging energy, so that the blockade region is enlarged relative to the normal-metal case. As a result, the transport stays suppressed over a finite bias window even at the charge-degeneracy points (half-integer $N_G$), where the normal-state diamonds would otherwise touch. Under a finite thermal bias, $T_{\mathrm{hot}} > T_{\mathrm{cold}}$, with $T_{\rm hot}\equiv T_L=T_R$ the common lead temperature and $T_{\rm cold}\equiv T_{\rm is}$ the island temperature, additional subgap channels appear inside the diamonds, producing the richer structure visible in Fig.~\ref{fig:SET2}(b). The key regime lies near the charge-degeneracy points: for half-integer $N_G$ and subgap bias, the current flows against the applied voltage ($IV<0$) [Fig.~\ref{fig:SET2}(c)], signaling a thermoelectric response. By contrast, near integer values of $N_G$, the current is suppressed, and the system is dissipative at sufficiently high biases. Crucially, the response does not change sign across adjacent degeneracy points, reflecting its bipolar nature, contrasting the usual sawtooth behavior of the standard unipolar TE in quantum dots. Consequently, the Seebeck voltage and the thermoelectric current depend strongly on $N_G$, thereby enabling electrostatic control of the TE through the gate. For Al leads and an Al/Cu bilayer island with $\Delta_{\mathrm{is,0}} = \dAlo/2$ and $E_C = 4\dAlo$~\cite{battisti2024bipolar}, the nonlinear Seebeck coefficient can reach values of the order of $3\,\mathrm{mV/K}$ at sub-Kelvin temperatures.
\subsection{\bf Gate-tunable thermoelectricity in graphene-based junctions}
\label{subsec:grapheneSuper}
Graphene and low-dimensional conductors have long been studied as possible thermoelectric platforms~\cite{hicks1993effect}. In graphene, the Seebeck coefficient essentially follows the Mott relation and changes sign across the charge neutrality point, as the majority carrier switches between electron and hole~\cite{zuev2009thermoelectric}, with the sign of the thermopower depending on the gate. The bipolar effect discussed here is intrinsically different (see Sec.~\ref{Sec: Th}), with both thermovoltage polarities arising for the same configuration. Nonetheless, bilayer graphene (BLG) has been theoretically investigated for bipolar thermoelectricity~\cite{bernazzani_bipolar_2023}, when the gap in the BLG spectrum is suitably tuned electrostatically and tunnel-coupled to a superconductor.

Differently from the SET discussed above, where the gate 
tunes the charging energy, here the gate acts directly on the spectral gap. A perpendicular electric field (oriented normal to the graphene sheet, e.g.\ via top and bottom gates) opens a bandgap in BLG that can be tuned 
electrostatically~\cite{castro2007biased,zhang2009direct}; for suitable 
gate-voltage values the bandgap can be comparable to the superconducting gaps of low-temperature superconductors (tenths to a few meV). This tunability makes BLG-superconductor tunnel junctions attractive for bipolar thermoelectricity. Since only one of the electrodes in the junction is superconducting, no Josephson coupling is established in this setup, and so no suppression scheme is needed.

\begin{figure}[t]
    \centering
    \includegraphics[
    ]{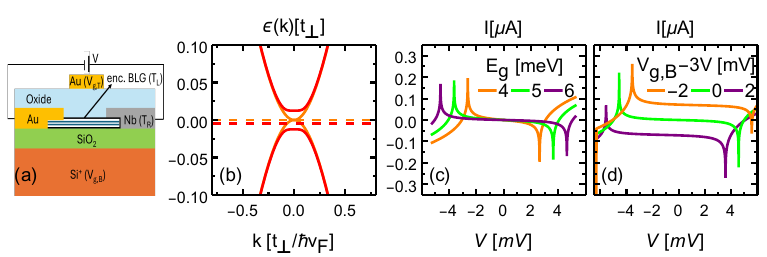}
    \caption{Gate control of bipolar thermoelectricity in a BLG-based tunnel junction. 
(a) Device scheme of the BLG--I--S structure. Bilayer graphene is controlled by top and bottom gates and contacted by a normal metal, so that the bias drops across the tunnel barrier. 
(b) BLG band structure with gate-induced gap $E_g$. The gates tune both the gap and the Fermi energy $E_F^{\mathrm{BLG}}$; dashed lines show how $E_F^{\mathrm{BLG}}$ shifts under asymmetric gating. Fermi velocity $v_F\approx10^6\,\mathrm{m/s}$.
(c) Thermoelectric response for different gate-induced gaps along the particle-hole-symmetric line ($E_F^{\mathrm{BLG}}=0$), with matching peaks at $|eV_p| \simeq E_g(U)-\Delta$, at $T_L=50\,$K. 
(d) Response for finite $E_F^{\mathrm{BLG}}\neq0$ at $T_L=50\,$K and gate voltage $V_{g,T}=-3\,$V (orange, violet), compared to the particle-hole-symmetric case (green): the $I$--$V$ characteristics lose their exact antisymmetry, and a conventional thermoelectric contribution appears alongside the bipolar one, signaled by $P=-IV>0$ near the peaks. Adapted from \cite{bernazzani_bipolar_2023}.}
    \label{fig:BLG}
\end{figure}

In Fig.~\ref{fig:BLG}(a), we schematize a concrete setup for BLG--I--S tunnel junctions. A thin hBN layer acts as the insulating tunnel 
barrier between the BLG and the superconductor. The BLG is coupled to independent top and bottom gates, which open a tunable bandgap in the BLG spectrum [Fig.~\ref{fig:BLG}(b)]. A normal-metal contact fixes the BLG potential, so that the applied bias drops across the tunnel barrier. 
In this way, the gate controls the BLG DoS close to the gap edge and, in turn, the thermoelectric response.

The gate-induced gap depends on the potential difference $U$ between the two graphene layers~\cite{mccann2006asymmetry,mccann2013electronic},
$2E_g = |U|\,t_\perp/\sqrt{U^2+t_\perp^2},$ where $t_\perp \simeq 0.4\,\mathrm{eV}$ is the interlayer hopping parameter and $E_g$ is the band edge.  For $|E|\ll t$, the 
BLG DoS can be approximated as~\cite{mccann2013electronic}
\begin{equation}
\label{eq:DoSBLG}
N_{\mathrm{BLG}}(E)=\frac{t_\perp}{\sqrt{3}\,\pi t^2}\frac{|E|}{\sqrt{E^2-E_g^2}}
\,\Theta(|E|-E_g),
\end{equation}
with $t\simeq 3\,\mathrm{eV}$ the in-plane hopping parameter. The current across the junction follows by inserting Eq.~\eqref{eq:DoSBLG}, together with the BCS DoS (cf.\ Eq.~\eqref{eq:BCSdos}), into the tunneling expression of Eq.~\eqref{eq:RateTunneling}.

In Fig.~\ref{fig:BLG}(c), we display selected thermoelectric IV characteristics for a BLG-I-S junction in the presence of a suitable temperature difference. 
The gate mainly shifts the matching peak, located at $|V_p|\simeq [E_g(U)-\Delta]/e$, by tuning $E_g(U)$, while the peak amplitude is set primarily by the BLG temperature. Since we use the BLG as the hot electrode and its gap is temperature-independent, the peak position remains unaffected by heating. This behavior is starkly different from SIS$'$ junctions, where heating the hot electrode suppresses its gap monotonically. As a result, the output power increases monotonically with the BLG temperature over a broad range~\cite{bernazzani_bipolar_2023} and the nonlinear Seebeck coefficient is estimated to reach values of the order of $1\,\mathrm{mV/K}$ with Nb superconductors in the cold terminal. 

In the setup of Fig.~\ref{fig:BLG}(a), controlled deviations from the particle-hole symmetric configuration are possible. The two gates play a double role: their differential action sets the interlayer asymmetry and opens the BLG gap, while their common-mode component controls the BLG Fermi energy, inducing a finite $E_F^{\mathrm{BLG}}\neq 0$, where $E_F^{\mathrm{BLG}}=0$ indicates that the Fermi energy is at the center of the gap [Fig.~\ref{fig:BLG}(b)]. The $I$--$V$ characteristics are then no longer antisymmetric in the voltage bias [Fig.~\ref{fig:BLG}(d)], and a conventional thermoelectric contribution appears alongside the bipolar one. For small deviations from the symmetric configuration, the maximum thermoelectric response still occurs at the matching peak, i.e., bipolar thermoelectricity dominates the response to the thermal bias; when the Fermi-energy shift is sufficiently large, the unipolar component becomes dominant.

However, other two-dimensional platforms may present nonlinear TE, which should not be confused with the bipolar TE, since the IV reciprocity is lost. Monolayer graphene--insulator--superconductor junctions have also been investigated~\cite{bianco2024coexistence}, but do not meet the canonical conditions discussed for the  bipolar thermoelectric response discussed above: in particular the graphene is gapless (cf.\ Sec.~\ref{Sec: Th}). A nonlinear TE response nonetheless appears, as the graphene side breaks the energy symmetry away from charge neutrality. While some nonlinear features can accompany the resulting conventional thermoelectric signal, their contribution remains, in general, marginal. Another example is a junction between a superconductor and a two-dimensional electron gas, which replaces the gapped BLG spectrum discussed above with a one-sided band edge; the resulting thermoelectricity is strongly nonlinear, with thermovoltages of several $\Delta/e$ and a heat-engine efficiency close to the Carnot limit, but still unipolar, with the $I$--$V$ curve not antisymmetric in bias~\cite{lucchesi2026strong}. 

\subsection{\bf Phase-tunable bipolar thermoelectricity in SQUIPTs}\label{Subsec: SQUIPT}
\begin{figure}[t]
    \centering
    \includegraphics[width=0.5\linewidth]{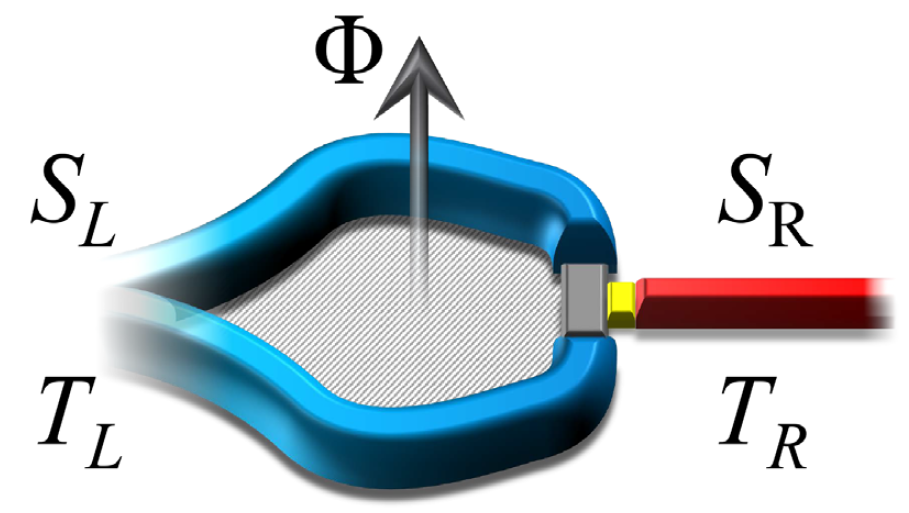}
    \caption{Bipolar thermoelectric SQUIPT. A normal metal wire (gray) embedded in a superconducting loop (blue) and tunnel-coupled through an insulating barrier (yellow) to a superconducting probe (red). The proximitized region and the loop are kept at temperature $T_L$, while the probe is at $T_R$. The loop is threaded by a magnetic flux $\Phi$, which controls the proximity-induced minigap. Adapted from \cite{guarcello_bipolar_2023}.}
    \label{fig:BTSQUIPT1}
\end{figure}

Magnetic flux control offers an alternative route for tuning the gap in the DoS, providing a magnetostatic analog of the electrostatic gap control discussed in Sec.~\ref{subsec:grapheneSuper}. A possible circuit element to reach this goal is the superconducting quantum interference proximity transistor (SQUIPT)~\cite{giazotto2010superconducting}, comprising a normal-metal wire embedded in a superconducting loop and coupled through an insulating barrier to a superconducting probe (see Fig.~\ref{fig:BTSQUIPT1}). The flux threading 
the loop sets the phase difference across the wire (due to fluxoid quantization) and, consequently, the proximity-induced minigap in the normal metal wire. 
 
The SQUIPT can be exploited for bipolar thermoelectricity, as theoretically investigated in
Ref.~\cite{guarcello_bipolar_2023}. The proximitized wire and the loop are assumed to be kept at a lower temperature compared to the probe~\footnote{The cold side is assumed to be thermalized to the phonons temperature, thanks to the comparatively large volume of the superconducting loop.}.
In the dirty limit, i.e., when the mean free path is short compared to the wire length, the spectral properties of the normal-metal wire can be described by the one-dimensional Usadel equation~\cite{usadel1970generalized}. For a narrow probe contact ($w\ll L$) and in the short junction limit, the DoS at the wire center reads~\cite{GiazottoTaddei_PRB84}
\begin{equation}\label{Eq. SQUIPT}
N_N(\varepsilon, T, \Phi) = \Re \left[
\sqrt{\frac{(\varepsilon + i\Gamma)^2}{(\varepsilon + i\Gamma)^2 - 
\Delta_L^2(T)\cos^2\left(\pi \Phi/\Phi_0\right)}}
\right],
\end{equation}
which has a BCS-like form (smeared by a Dynes factor) with an effective flux-dependent gap $\Delta_L(T)|\cos(\pi\Phi/\Phi_0)|$: the gap is maximized at $\Phi=0$, where it is equal to the gap in the loop $\Delta_L(T)$, and closes at $\Phi=\Phi_0/2$. The tunnel barrier with the probe is taken in the high-resistance limit, where the Josephson contribution is suppressed. In this regime, the maximum output power is also reduced.

\begin{figure}[t]
    \centering
    \includegraphics[width=\linewidth]{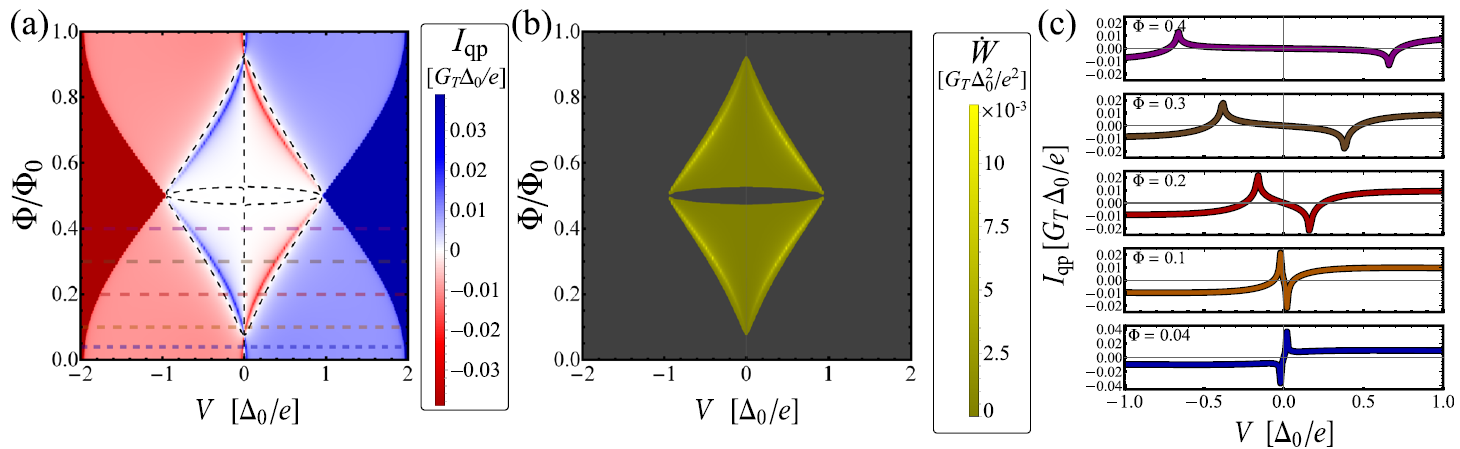}
    \caption{Thermoelectric response of a SQUIPT.
(a) Color plot of the QP current as a function of bias voltage and magnetic flux for a narrow probe contact, computed with $T_R=0.4\,T_C$, $T_L=0.01\,T_C$, and Dynes parameter $\Gamma=10^{-4}\Delta_0$. (b) Extracted power, highlighting the thermoelectric region in the $(V,\Phi)$ plane. 
(c) Representative $I(V)$ characteristics at different flux values, showing the flux-dependent shift of the matching peaks. Adapted from \cite{guarcello_bipolar_2023}.}
    \label{fig:BTSQUIPT2}
\end{figure}

Figure~\ref{fig:BTSQUIPT2}(a) shows the density plot of the QP current as a function of bias voltage and magnetic flux. The current is odd in the bias voltage, reflecting the IV reciprocity enforced by the particle-hole symmetry of the DoS. The flux dependence is periodic with period $\Phi_0$ and symmetric around $\Phi=\Phi_0/2$, as expected from Eq.~\eqref{Eq. SQUIPT}. At large bias, transport is purely dissipative, and the current rises sharply once the voltage bias exceeds the sum of the probe gap and the flux-dependent minigap, ranging from $\Delta_R(T_R)+\Delta_L(T_L)$ at $\Phi=0$ to $\Delta_R(T_R)$ when the minigap closes at $\Phi=\Phi_0/2$. The thermoelectric regime ($IV<0$) appears at lower bias, and is better visualized in the extracted power map of Fig.~\ref{fig:BTSQUIPT2}(b). The flux controls the matching peak position as follows 
\begin{equation}
eV_p=\Delta_R(T_R)-\Delta_L(T_L,\Phi),
\end{equation}
so that $|V_p|$ increases as the flux reduces the minigap [see Fig.~\ref{fig:BTSQUIPT2}(c)].

Flux control also sets the device's operating window. For $\Phi \approx n\Phi_0$, the gap asymmetry is too weak to sustain bipolar thermoelectricity. For flux values too close to $\Phi_0(n+1/2)$, the minigap becomes too small, and the thermoelectric response is similarly suppressed, as expected for an NIS junction (see Sec.~\ref{SubSec:linear-in-bias}). The bipolar TE mainly appear in the intermediate flux range, where the minigap and the superconducting loop gap are of comparable size.

\subsection{Thermospin response in spin-split superconducting junctions}
Tunnel junctions comprising spin-split superconductors can host a rich set of nonequilibrium charge, spin, and energy transport 
phenomena~\cite{bergeret2018colloquium}. The key element is a thin superconducting film whose spin-components of the QP DoS are shifted with respect to each other, i.e., $N_{S_m,\uparrow(\downarrow)}(E) = N(E\pm h, \Delta_{S_m}, \gamma)$, by the effective spin-splitting field $h$, originating from exchange interaction with the spin of an adjacent ferromagnetic insulator or by an in-plane magnetic field~\cite{hao1991thin}. Each spin branch is individually energy-asymmetric, yet the two are related by $N_{S_m,\uparrow}(E) = N_{S_m,\downarrow}(-E)$, so that the total DoS remains energy-symmetric and satisfies Eq.~\eqref{eq:PHSdos}. When superconductors with spin-split DoS are combined with spin-filtering elements, such as ferromagnetic insulator barriers, the particle-hole asymmetry in the individual spin component produces a linear thermoelectric response, as predicted in ferromagnet-superconductor junctions~\cite{Ozaeta} and observed in high-field superconductor-ferromagnet tunnel junctions~\cite{kolenda2016observation}. 

\begin{figure}[t]
    \centering
    \includegraphics[width=\linewidth]{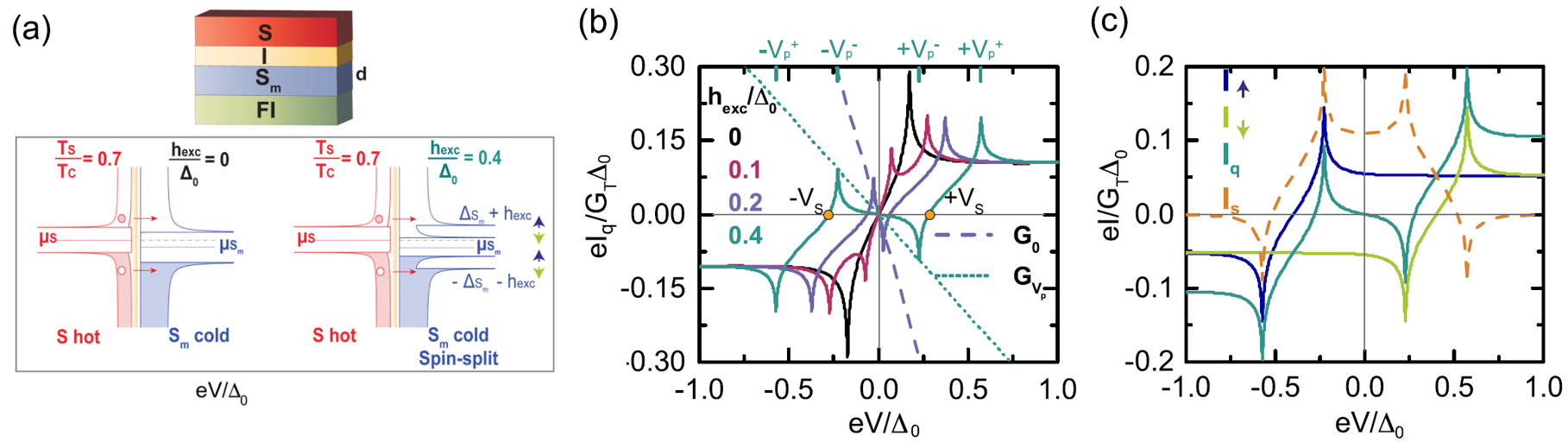}
\caption{Thermospin response in spin-split superconducting junctions. 
    (a) Sketch of the spin-split superconductor $S_m$ tunnel-coupled to a conventional superconductor $S$, and the resulting spin-resolved DoS. Computed at $T_S=0.7\,T_C$, $T_{S_m}=0.01\,T_C$, and Dynes parameter $\Gamma=10^{-3}\Delta_0$. (b) QP current-voltage characteristics $I_q(V)$ for different values of $h_{\mathrm{exc}}/\Delta_0$, at $T_S=0.7\,T_C$ and $T_{S_m}=0.01\,T_C$. In the linear regime, the curves are well approximated by $I_q=G_0V$ (dashed violet line), while in the nonlinear regime, the peak conductance $G_{\max}$ is evaluated at the thermoelectric peaks (dotted aquamarine line). The Seebeck voltages $\pm V_S$ (orange circles) and the matching-peak positions $\pm V_p^{\pm}$ are reported for $h_{\mathrm{exc}}=0.4\Delta_0$.(c) The charge current $I_q$ is separated into its spin-up ($I_\uparrow$) and spin-down ($I_\downarrow$) components at $h_{\mathrm{exc}}=0.4\Delta_0$, $T_S=0.7\,T_C$, $T_{S_m}=0.01\,T_C$. Only one spin component generates a thermoactive peak for a given bias sign ($I_\downarrow$ for $V>0$, $I_\uparrow$ for $V<0$); the resulting spin current $I_s$ is shown as the dashed orange line. Adapted from ~\cite{germanese2021spontaneous}.}
    \label{fig:TSpin1}
\end{figure}

Spin-split superconductors can also be used for bipolar thermoelectricity, as proposed in Ref.~\cite{germanese2021spontaneous}. In this reference, the spin-split superconductor is tunnel-coupled to a conventional one through a nonmagnetic barrier [Fig.~\ref{fig:TSpin1}(a)]; since there is no spin filtering, the transport satisfies the EIS, enforcing a reciprocal charge characteristic, and no linear TE occurs. However, in the nonlinear regime, a finite temperature difference can generate a bipolar thermoelectric response that is tunable via the exchange field $h_{exc}$. Due to the spin-splitting, in this structure, a spin current can also be generated by a thermal bias
~\cite{germanese2021spontaneous}. 
Denoting $I_{\uparrow}$ and $I_{\downarrow}$ the two spin components of the tunneling current, we can express the charge current as 
$I_q=I_{\uparrow}+I_{\downarrow}$ and the spin current as $I_s=I_{\uparrow}-I_{\downarrow}$. The EIS implies that the charge current is reciprocal, $I_q(V)=-I_q(-V)$, while the spin current is even, $I_s(V)=I_s(-V)$, as follows by the identity $N_{S_m,\uparrow}(E)=N_{S_m,\downarrow}(-E)$ quoted above.

When the $S$ electrode is hot and the spin-split electrode cold, the $I$--$V$ characteristic develops multiple subgap features due to the matching of the spin-resolved coherence peaks. In contrast to the $h_{\mathrm{exc}}=0$ case, the thermoelectric peaks are split and appear at
\begin{equation}
eV_p^\sigma = \pm|\Delta_S(T_S)-\Delta_{S_m}(T_{S_m})+\sigma h_{\mathrm{exc}}|,
\end{equation}
as shown in Fig.~\ref{fig:TSpin1}(b). For sufficiently large $h_{\mathrm{exc}}$, one of the two spin channels becomes thermoactive, so that the total charge current flows against the applied bias in the subgap regime ($IV<0$).

The junction is also spin-active, generating a pure spin current driven by a temperature difference in the absence of a spin-polarized barrier. In Fig.~\ref{fig:TSpin1}(c) we illustrate the thermospin response by plotting  
$I_{\uparrow}$, $I_{\downarrow}$, $I_q$, and $I_s$ as a function of the voltage bias for $h_{\mathrm{exc}} = 0.4\Delta_0$. At $V=0$, the two spin contributions are finite and exactly opposite, $I_{\uparrow}(0)=-I_{\downarrow}(0)$, so that $I_q(0)=0$ while $I_s(0)\neq 0$: a pure spin current flows with no net charge transport. At finite biases, the two spin channels play different roles, with one thermoactive and the other dissipative. The Seebeck voltage identifies a second 
compensation point, $I_{\uparrow}(V_S)=-I_{\downarrow}(V_S)$, where again $I_q(V_S)=0$ and a pure spin current is sustained. The spin current reaches its maximum at the internal matching peaks, where the imbalance between the two spin channels is largest.

These results place spin-split superconducting junctions within superconducting spin caloritronics~\cite{bauer2012spin,linder2015superconducting}, demonstrating that thermal gradients alone can generate and control pure spin currents in the absence of any magnetic barrier.

\newpage
\section{Quantum Bipolar Thermoelectricity}\label{Sec: Env}
The bipolar TE discussed in the previous sections is driven by a temperature difference across the junction. The thermal bias determines spontaneous breaking of the EIS when the two electrodes of the junction have suitable DoS, thus producing a nonlinear thermoelectric response (Sec.~\ref{Sec: Th}), as measured in superconducting tunnel junctions (Sec.~\ref{Sec: Exp}) and proposed for several hybrid platforms (Sec.~\ref{Sec: hybrids}).

Here we discuss a distinct mechanism in which the two junction electrodes are kept at the same temperature~\cite{antola2026quantum}: the EIS is not broken by a temperature difference across the junction but rather by coupling to a colder electromagnetic environment. The effect is intrinsically quantum; it relies on the imbalance between emission and absorption of energy quanta $\hbar\omega$ with a bosonic environment. This imbalance becomes significant when the environment is sufficiently cold, $k_B T_e \lesssim \hbar\omega$, and negligible in the classical limit $k_B T_e \gg \hbar\omega$, where the rates of emission and absorption are asymptotically equal.

The asymmetry between emission and absorption has been measured directly, both in the spontaneous emission of double quantum dots coupled to a bosonic bath~\cite{fujisawa_spontaneous_1998,aguado_double_2000} and in the high-frequency quantum noise of superconducting devices~\cite{deblock_detection_2003,billangeon_emission_2006,basset_emission_2010,basset_high-frequency_2012}. More broadly, the coupling to the electromagnetic environment can be used as a resource for heat management, a few examples being Brownian refrigerators~\cite{pekola2007normal,Peltonen_Brownian_2011}, photon-assisted Cooper-pair tunneling heat engines~\cite{hofer_quantum_2016}, and quantum-circuit refrigerators~\cite{tan_quantum-circuit_2017,Silveri_Theory_2017}. Finally the dynamical Coulomb blockade can also modify conventional thermoelectric transport~\cite{mecklenburg_thermopower_2017,rossello_dynamical_2017}. In the literature, quantum effects in TEs have been 
exploited in nanostructures~\cite{Hicks_Thermoelectric_1993,Dresselhaus_New_2007} through energy selectivity: quantum dots, for instance, use their discrete levels as optimal energy filters~\cite{sanchez_optimal_2011,jordan_powerful_2013,thierschmann_three-terminal_2015,josefsson_quantum-dot_2018,jaliel_experimental_2019}. We instead inquire whether the quantum nature of the environment spectrum can be used to develop thermoelectricity.

The environment affects the junction transport by exchanging energy with the tunneling electrons through the dynamical Coulomb blockade mechanism~\cite{devoret_effect_1990}. Notably, the mechanism may appear quite similar to photon-assisted tunneling~\cite{tien1963multiphoton}, but, as we will discuss, it presents important differences.
As anticipated before, the emission/absorption asymmetry in the quantum regime of the environment can be exploited using a system with a resonant response at a specific energy~\cite{aguado_double_2000}.
In the SIS$'$ junction, this energy selection is 
provided by the sharp, diverging BCS DoSs at the gap edge: photon emission/absorption processes are strongly enhanced at photon energies near resonance with the gap difference $\hbar\omega \approx \Delta - \Delta'$ due to the diverging DoS in the initial and final QP states.
However, when the environment temperature $T_e$ satisfies $k_BT_e\ll \Delta - \Delta'$, the environment cannot emit quanta at the matching energy, while it can absorb those emitted by the junction. Then the environment emission-absorption asymmetry, in combination with the SIS$'$ spectrum selectivity, produces a strong violation of detailed balance that underlies the bipolar TE (Sec.~\ref{Sec: Th}). Finally, due to the EIS in the SIS$'$ junction, the response preserves the signatures of the bipolar TE: zero current at zero voltage bias, reciprocal $I$--$V$ characteristic, and opposite-sign Seebeck voltages $\pm V_S$ at a given temperature configuration.

Section~\ref{subsec:PE_description} introduces the $P(E)$ framework necessary to incorporate the probability that a tunneling event exchanges an energy $E$ with the surrounding circuit~\cite{ingold_charge_1992,nazarov_quantum_2009}, and Sec.~\ref {SubSec:QBT_mechanism} applies it to quantum bipolar thermoelectricity.

\subsection{$P(E)$ Description of Environment-Assisted Tunneling}
\label{subsec:PE_description}
To include the electromagnetic environment in the tunneling framework of Sec.~\ref{Sec: Th}, the initial ($|i\rangle$) and final ($|f\rangle$) states in the golden rule treatment must carry both electronic ($|E\rangle$) and electromagnetic environmental ($|R\rangle$) degrees of freedom,
\[
|i\rangle = |E\rangle \otimes |R\rangle, \qquad |f\rangle = |E'\rangle \otimes |R'\rangle ,
\]
where the two subsystems are assumed uncorrelated both before and after the tunneling event~\cite{ingold_charge_1992}. The tunneling Hamiltonian keeps the structure of Sec.~\ref{Sec: Th}, but its phase factor, which couples the 
junction's leads to the environment, is now treated as a quantum operator. The phase $\hat\varphi$ and the charge $\hat Q$ stored on the capacitance in parallel with the junction are canonically conjugate, $[\hat\varphi,\hat Q]=ie$, so that the operator $e^{-i\hat\varphi}$ [cf.\ Eq.~\eqref{eq:Htun}] acts as a charge-translation operator,
\begin{equation}
e^{i\hat\varphi}\hat Q e^{-i\hat\varphi} = \hat Q - e.
\end{equation}
This confirms that each tunneling event shifts the environmental charge by one elementary charge $e$, and, consequently, a transition $|R\rangle \to |R'\rangle$ of the environment. This is the central idea of the $P(E)$ description.
\begin{figure}[t]
    \centering
    \includegraphics[width=0.5\linewidth]{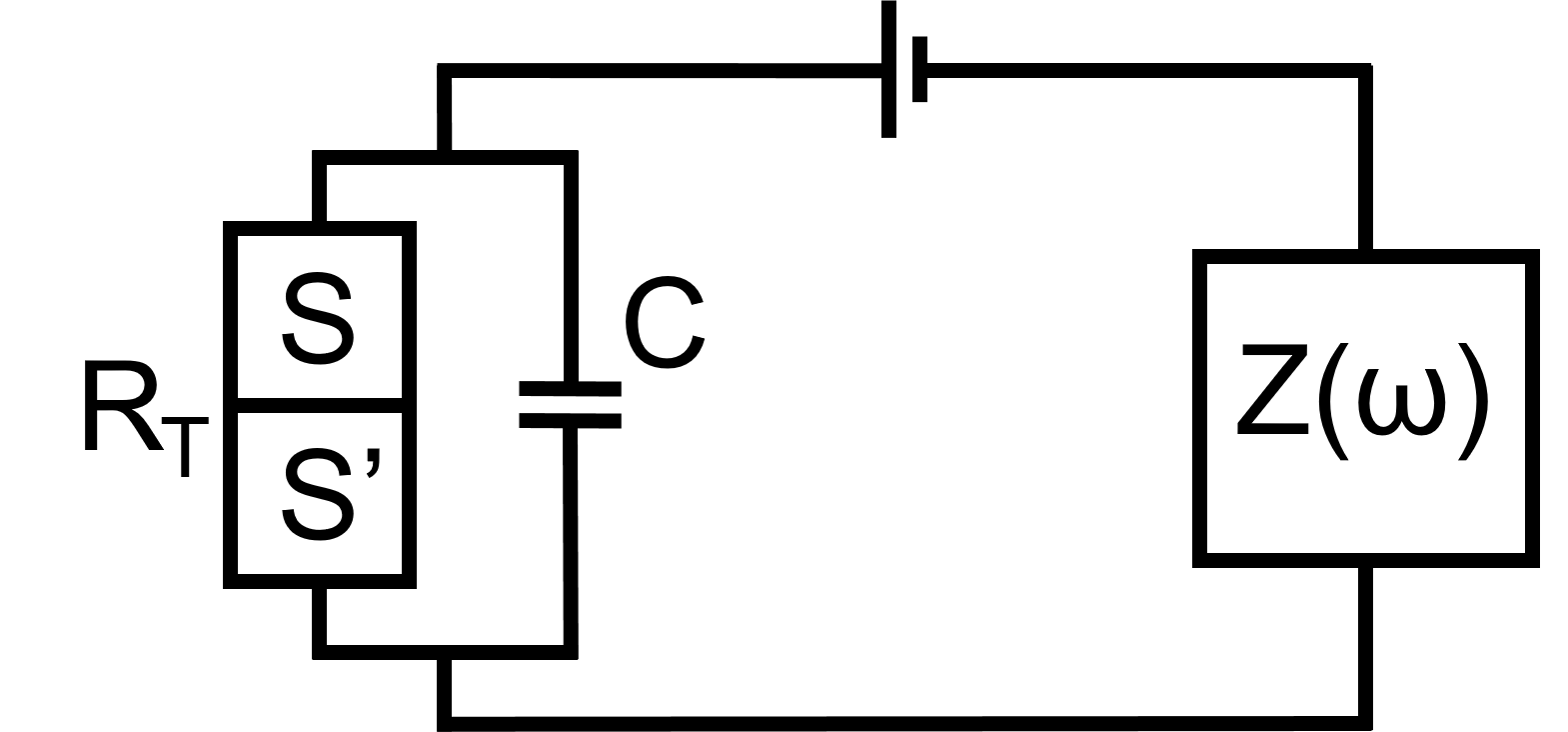}
    \caption{Circuit scheme for the \(P(E)\) description of environment-assisted tunneling. An ultrasmall tunnel junction with capacitance \(C\) and tunnel resistance \(R_T\) is biased by a DC voltage source and coupled to an external impedance \(Z(\omega)\), which models the electromagnetic environment.}
    \label{fig:P(E)_Scheme}
\end{figure}
In Fig.~\ref{fig:P(E)_Scheme}, we display the minimal circuit scheme of the system: a SIS$^\prime$ superconducting tunnel junction of capacitance $C$ and normal-state tunnel resistance $R_T$, in series with an external impedance $Z(\omega)$ and biased by a DC voltage source~\cite{ingold_charge_1992}. The electromagnetic environment seen by the junction is modeled by the effective impedance
\begin{equation}
Z_t(\omega)=\frac{1}{i\omega C+Z^{-1}(\omega)} ,
\end{equation}
i.e., the parallel combination of the junction capacitance and the external circuit impedance $Z(\omega)$.  
To be consistent with the notation of Sec.~\ref{Sec: Th}, we denote the two leads as $L$ and $R$ and the forward rate as $\Gamma_{LR}$, and, for our goals, we consider the electronic temperatures of the two leads to be in equilibrium, $T_L=T_R\equiv T_j$. Under a DC bias $V$, the forward tunneling rate of Eq.~\eqref{eq:RateTunneling} is then generalized to the inelastic case by a double energy integral expression~\cite{ingold_charge_1992},
\begin{equation}\label{eq:rate_PE}
\Gamma_{LR}(V)=\frac{1}{e^{2}R_T}\int_{-\infty}^{+\infty} d\e\, 
\int_{-\infty}^{+\infty}d\e'\, N_L(\e)\,N_R(\e'+eV)\, 
f_L(\e)\,[1-f_R(\e'+eV)]\,P(\e-\e'),
\end{equation}
which accounts for the energy $\e-\e'$ exchanged with the environment during 
each tunneling event: positive $\e-\e'$ corresponds to energy emitted into the environment, negative $\e-\e'$ to energy absorbed from it.
The Fermi functions $f_L$ and $f_R$ are both evaluated at the junction temperature $T_j$, while $P(\e)$ is set by the environment temperature $T_e$. This temperature configuration is the key difference from standard $P(E)$ treatments, where a single temperature is assumed, $T_e=T_L=T_R$~\cite{ingold_charge_1992}; here, instead, $T_j\neq T_e$ is the relevant regime for environment-induced thermoelectricity~\cite{antola2026quantum} or, eventually, for quantum-circuit refrigerators~\cite{tan_quantum-circuit_2017}.
The emission/absorption probability density, also resulting from the golden rule calculation, is given as
\begin{equation}\label{eq:PE_def}
P(E)=\frac{1}{2\pi\hbar}\int_{-\infty}^{+\infty} dt\, 
\exp\!\left[J(t)+\frac{iEt}{\hbar}\right],
\end{equation}
where $J(t)=\langle [\tilde{\varphi}(t)-\tilde{\varphi}(0)]\,\tilde{\varphi}(0)
\rangle$ is the correlation function of the fluctuating phase operator 
$\tilde\varphi(t)=\frac{e}{\hbar}\left(\int_{-\infty}^{t}dt'\,\frac{\hat Q(t')}
{C}-Vt\right)$ taken in the rotating frame~\cite{ingold_charge_1992}, with the 
average taken over the environment at thermal equilibrium at $T_e$:
\begin{equation}\label{eq:JofTPE}
J(t)=2\int_{0}^{\infty}\frac{d\omega}{\omega}\,\frac{\mathrm{Re}[Z_t(\omega)]}
{R_K}\left\{\coth\!\left(\frac{\hbar\omega}{2k_BT_e}\right)
\left[\cos(\omega t)-1\right]-i\sin(\omega t)\right\},
\end{equation}
with $R_K=h/e^2$, the von Klitzing constant. Being a property of the environment kept at equilibrium temperature $T_e$, $P(E)$ obeys detailed balance at the bath temperature, $P(-E)=e^{-E/k_BT_e}P(E)$, so at low $T_e$ absorption ($E<0$) is suppressed relative to emission processes ($E>0$)~\cite{ingold_charge_1992}. \footnote{We assume that the environment's stationary temperature is fixed, and the phase correlator $J(t)$ is given exactly by the fluctuation-dissipation theorem, yielding Eq.~\eqref{eq:JofTPE} at equilibrium. This thermal equilibrium assumption of the electromagnetic environment is guaranteed only for sufficiently strong dissipative environments $\propto \mathrm{Re}[Z_t(\omega)]$; otherwise, the environment may be driven out of equilibrium. A self-consistent treatment of the environment beyond the $P(E)$ framework is then required~\cite{cailleaux_photonic_2025}. The present approach is the minimal consistent treatment in the equilibrium-bath limit.}

The impact of the coupling to the environment on the tunneling rate is determined by the real part $\mathrm{Re}[Z_t(\omega)]$ of the effective impedance. When this quantity is small compared with $R_K$ over the relevant frequency range, phase fluctuations are weak, $P(E)\simeq\delta(E)$, and the elastic limit of Eq.~\eqref{eq:RateTunneling} is recovered. When $\mathrm{Re}[Z_t(\omega)]$ becomes comparable to or larger than $R_K$, inelastic processes grow and the junction enters the dynamical Coulomb blockade regime. For concreteness, we consider two reference cases for the external impedance $Z_t(\omega)$, even though several more examples can be found in the literature~\cite{ingold_charge_1992}.

\textit{Ohmic environment.} The impedance is real and frequency independent, $Z(\omega)=R$, with the junction capacitance $(C)$ providing an effective high-frequency cutoff that keeps $J(t)$ finite. The coupling strength is set by the dimensionless conductance $g=R_K/R$. In the low-impedance limit $g\gg1$, the tunneling is essentially elastic, as discussed before. In the high-impedance limit $g\ll1$, the circuit back-action is strong, and charge transfer incurs additional energy costs. The distribution then reads
\begin{equation}
P(E)=\frac{1}{\sqrt{4\pi E_C k_B T_e}}
\exp\!\left[-\frac{(E-E_C)^2}{4E_C k_B T_e}\right]\,,
\end{equation}
a Gaussian centered at the charging energy $E_C$, which in the low-temperature limit $k_BT_e\ll E_C$ reduces to
\begin{equation}
P(E)\to\delta(E-E_C),
\end{equation}
so each tunneling event requires exactly the energy $E_C$ to be emitted into the electromagnetic environment. For intermediate $g$, the full expression of Eq.~\eqref{eq:JofTPE} must be evaluated numerically.

\textit{Resonant environment.} For a purely inductive environment $Z(\omega)=i\omega L$, the junction capacitance $C$ and the inductance $L$ set the resonant LC frequency $\omega_{LC}=1/\sqrt{LC}$. The environment then behaves as a single mode, and energy is exchanged only in discrete quanta of energy $E_k=k\hbar\omega_{LC}$ with integer $k$. The corresponding distribution is
\begin{equation}
P(E)=\exp\!\left[-\rho\,\coth\!\left(\frac{\beta_e\hbar\omega_{LC}}{2}\right)\right]
\sum_{k=-\infty}^{+\infty}
I_k\!\left(\frac{\rho}{\sinh(\beta_e\hbar\omega_{LC}/2)}\right)
\exp\!\left(\frac{\beta_e E_k}{2}\right)
\delta(E-E_k),
\end{equation}
with $\beta_e=1/(k_BT_e)$, $\rho=E_C/(\hbar\omega_{LC})$, and $I_k$ the modified Bessel function of the first kind. In the limit $T_e\ll\hbar\omega_{LC}$, absorption (negative $k$) is suppressed, and the small-argument form $I_k(x)\simeq(x/2)^k/k!$ gives
\begin{equation}
P(E)=\sum_{k=0}^{+\infty}p_k\,\delta(E-k\hbar\omega_{LC}),
\qquad
p_k=e^{-\rho}\frac{\rho^k}{k!}.
\end{equation}
The weights $p_k$ follow a Poisson distribution, reflecting that there is only a discrete emission of energy quanta into the resonant mode.
\subsection{Quantum Bipolar Thermoelectricity}
\label{SubSec:QBT_mechanism}
We discussed above how, in a cold electromagnetic environment, tunneling 
processes in which the junction absorbs energy are suppressed relative to those 
in which it emits energy. Here, we discuss how this emission-absorption asymmetry 
generates a thermoelectric response in an otherwise dissipative SIS$'$ junction in the absence of a thermal gradient between its leads.

\begin{figure}[t]
    \centering
    \includegraphics[width=0.8\linewidth]{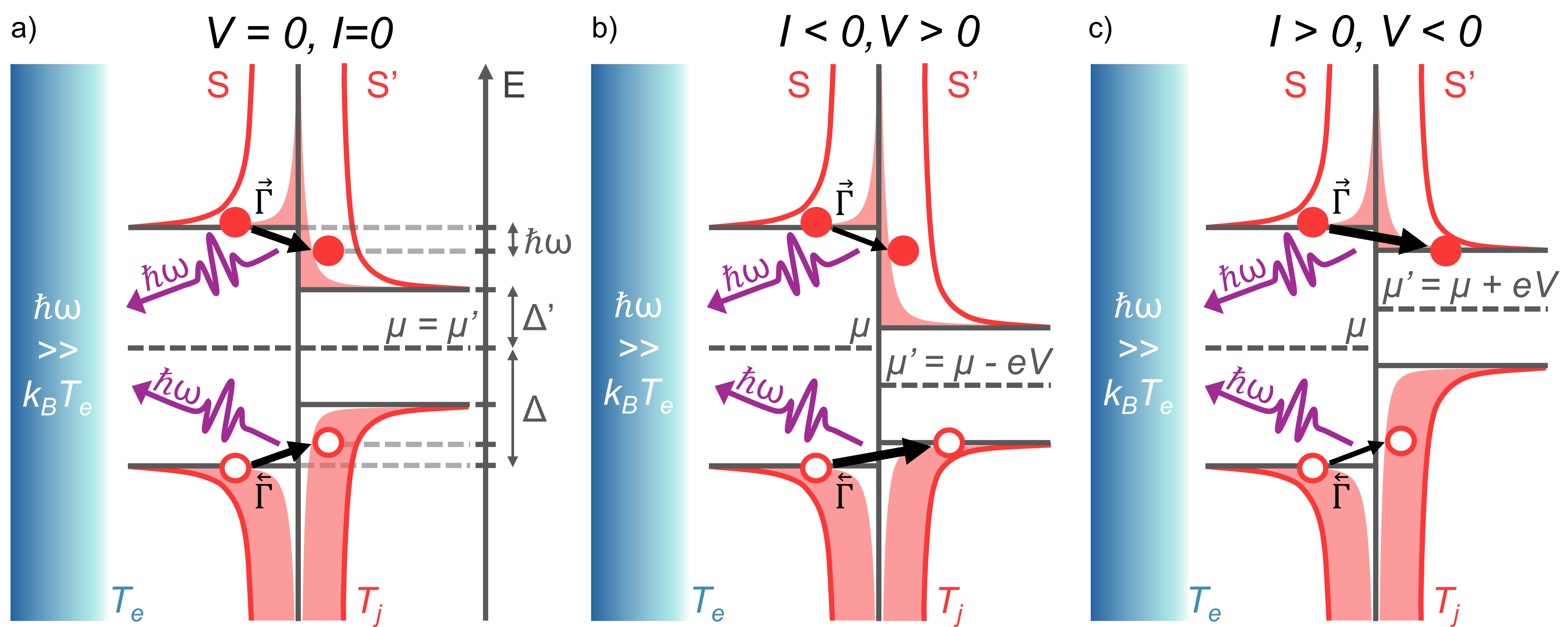}
    \caption{Mechanism of quantum bipolar thermoelectricity. Energy-band diagrams for an asymmetric S-I-S$'$ junction at electronic temperature $T_j$, coupled to a cold environment at $T_e$, for (a) zero, (b) small positive, and (c) small negative chemical potential bias. In the quantum regime $k_BT_e\lesssim\hbar\omega$, hole-assisted transitions feed the backward rate and particle-assisted transitions the forward rate, with the dominant weight set by the DoS near the gap edges. Adapted from~\cite{antola2026quantum}.}
    \label{fig:QBT1}
\end{figure}

Figure~\ref{fig:QBT1} schematically illustrates the physical mechanism. Panels (a)--(c) show the energy-band diagrams for QP excitations in an S-I-S$'$ junction whose electrodes are kept at the same electronic temperature $T_j$, coupled to an electromagnetic environment at temperature $T_e$, for different bias configurations.\footnote{A bias voltage is needed to measure the 
$I$--$V$ characteristic; when operating as a thermoelectric generator, no 
external bias is required, provided the load resistance is sufficiently high.} The relevant photon-assisted transitions are the ones connecting QP states near the gap edges, where the DoS is strongly peaked, for which the photon energy is near resonantce with the gap asymmetry $\hbar\omega\approx\Delta-\Delta'$. 
In the quantum regime $k_BT_e\lesssim\hbar\omega$, the environment is too cold to supply energy quanta to the junction, and tunneling is dominated by processes in which QPs only emit energy into the environment.
At zero bias [Fig.~\ref{fig:QBT1}(a)], forward and backward processes balance and no net current flows across the junction, as required by the reciprocity property of the IV characteristic~\cite{ingold_charge_1992}. At a small positive bias [Fig.~\ref{fig:QBT1}(b)], the relative alignment of the two superconducting DoSs shifts. Because the DoS decreases away from the gap edge, the dominant emission-assisted hole processes outweigh the corresponding particle processes, so the backward rate exceeds the forward ones, $\Gamma_{RL}(V)>\Gamma_{LR}(V)$, and a current flows against the applied voltage, $IV<0$. At the opposite polarity [Fig.~\ref{fig:QBT1}(c)], the roles of the two rates are exchanged, and the same argument makes the rate corresponding to a charge transfer against the bias dominant, again producing $IV<0$. The response is therefore bipolar as for the TE discussed in Sec.~\ref{Sec: Th}.

\begin{figure}[t]
    \centering
    \includegraphics[width=\linewidth]{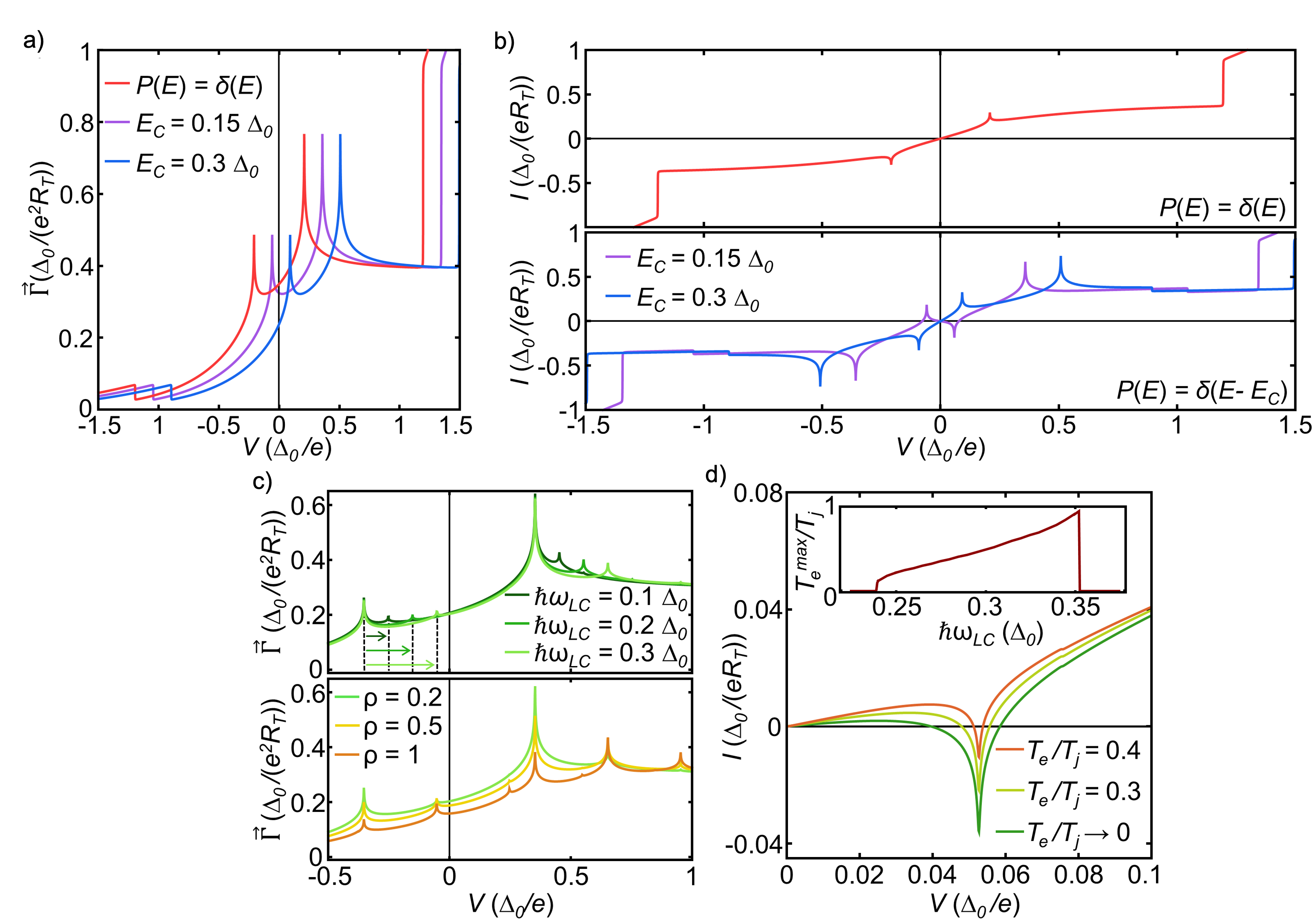}
    \caption{Quantum bipolar thermoelectricity for the two environments. (a) Forward rate $\Gamma_{LR}(eV)$ for an Ohmic environment in the low-impedance $P(E)\simeq\delta(E)$ and high-impedance regimes, and (b) the corresponding $I$--$V$, where the high-impedance case develops the thermoelectric region $I(V)V<0$. (c) Forward rate for a resonant environment, showing emission-side photon-assisted replicas at multiples of $\hbar\omega_{LC}$, and (d) the corresponding $I$--$V$. Adapted from~\cite{antola2026quantum}.}
    \label{fig:QBT_results}
\end{figure}
For an Ohmic environment, the effect is better visualized in the high-impedance, zero-temperature limit $g\to0$, $T_e\to0$, where $P(E)=\delta(E-E_C)$ (Sec.~\ref{subsec:PE_description}). Figure~\ref{fig:QBT_results}(a) shows the forward rate in this limit (violet and blue lines) alongside the low-impedance elastic case $g\to\infty$(red line): while the elastic rate  displays the standard subgap matching peak at $eV=\Delta-\Delta'$, the high-impedance rate is shifted by the charging energy $E_C=e^2/2C$ (see label for the $E_C$ values of the violet and blue lines). When $E_C\lesssim\Delta-\Delta'$, this shift reverses the balance between positive and negative energy states, which corresponds to making the backward rates dominate over the forward ones in a finite voltage window. The junction can display ANC, $I(V)V<0$, as shown from the blue line in Fig.~\ref{fig:QBT_results}(b). Electrical power is generated even if no thermal bias is present across the junction.\footnote{When $T_L\neq T_R$, the classical bipolar thermoelectric 
contribution and the quantum one can coexist and contribute in parallel.} An additional transport signature of this mechanism is the shift of the matching peak to $eV_p=\Delta-\Delta'-E_C$. 
By increasing the environmental temperature, or reducing its impedance, the effect weakens: a finite $T_e$ increases the emission processes from the environment and a
finite $g$ lowers the inelastic weight; the response survives in the quantum regime $k_BT_e\lesssim E_C$. For realistic Nb/AlO$_x$/Nb tunnel junctions, $R_T\sim100~\mathrm{k}\Omega$,$r=0.9$, $E_C\sim0.2~\mathrm{meV}$, and $R\sim2.5~\mathrm{M}\Omega$ at $T_e\sim100~\mathrm{mK}$, the generated power is of order $10~\mathrm{fW}$.

The resonant environment generalizes, to some extent, the results of the Ohmic case to multiphoton processes. In the Ohmic case, for the high-impedance and zero-temperature limit, the relevant energy exchange is given by $E_C$. Here, the junction can instead exchange energy in multiple quanta $k\hbar\omega_{LC}$ set by the resonant mode. 
The forward rate therefore develops a set of photon-assisted replicas [Fig.~\ref{fig:QBT_results}(c)], each corresponding to the emission of $k$ photons into the resonator. For $k_BT_e\ll\hbar\omega_{LC}$, the environment cannot supply photons, so only emission-assisted processes contribute, and the replicas are one-sided, appearing at positive multiples of $\hbar\omega_{LC}$. Increasing the coupling $\rho=E_C/(\hbar\omega_{LC})$ shifts spectral weight to multiphoton processes and makes higher-order peaks visible.

As in the Ohmic case, this one-sided exchange reverses the balance between forward 
and backward rates over a finite voltage range. Depending on $\omega_{LC}$ and 
$\rho$, it produces the strong detailed-balance violation that this review 
identifies as the essential requirement for thermoelectricity, and the resulting 
$I$--$V$ shows a thermoelectric region with $I(V)V<0$ 
[Fig.~\ref{fig:QBT_results}(d)].

It is interesting to contrast this latter mechanism with microwave-assisted thermoelectricity, analyzed in a previous work which connects the photon-assisted tunneling to the bipolar TE~\cite{hijano_microwave-assisted_2023}. In this reference, an SIS$'$ junction under a thermal gradient is driven by an AC voltage $V(t)=V+a\cos(\omega t)$, so that the averaged current becomes a Bessel-weighted sum of DC responses shifted by multiples of $\hbar\omega/e$~\cite{tien1963multiphoton},
\begin{equation}
\label{eq:TienGordonIV}
\bar{I}(V)=\sum_{n=-\infty}^{+\infty} 
J_n^2\!\left(\frac{ea}{\hbar\omega}\right)\,I\!\left(V-\frac{n\hbar\omega}{e}\right),
\end{equation}
with $J_n$ the Bessel function of the first kind. The drive redistributes the thermoelectric response over sidebands at $eV=eV_p+n\hbar\omega$ rather than enhancing it, widening the range in which thermoelectricity is observable, and allowing its operating point to be tuned by $\omega$. However, the essential difference with respect to the quantum case lies in the sign of $n$: in a classical drive which corresponds to an arbitrarily large number of photons, there is no asymmetry between emission and absorption processes so the sum runs symmetrically over both emission ($n>0$) and absorption ($n<0$). As a result, the drive alone cannot generate the effect: bipolar TE still requires a genuine temperature difference across the junction, as in the SIS$^\prime$ without without irradiation~\cite{hijano_microwave-assisted_2023}. A nonclassical drive instead reweights the sidebands according to the quantum statistics of the field~\cite{souquet_photon-assisted_2014}. 
This discussion shows that bipolar thermoelectricity operates in a quantum or, eventually, photon-dominated regime. Intriguingly, this research opens a perspective to the investigation of the interplay between thermoelectricity and electromagnetic or phononic modes, which are still largely unexplored but promise unexpected results~\cite{TAGANI_2013,Khedri_2017,Zhang_Enhanced_2025}.
\newpage
\section{Conclusions and perspectives}
\label{sec:conclusions}
In this review, we provided a comprehensive overview of bipolar thermoelectricity in tunnel-coupled two-terminal systems from its prediction to the most recent results. After discussing the main theoretical aspects in Sec.~\ref{Sec: Th}, we summarized in Sec.~\ref{sec:Experiments} the experimental results to date in gap-asymmetric superconducting tunnel 
junctions. Then, we discussed some proposals for specific applications in Sec.~\ref{Sec: Applications}, and highlighted the tunability opportunities offered by the superconducting and hybrid platforms (Sec.~\ref{Sec: hybrids}). Finally, in Sec.~\ref{Sec: Env} we summarized some new directions to realize environment-assisted 
thermoelectricity by coupling tunnel junctions to bosonic degrees of freedom, including quantum effects.

A common foundation underlies all of these cases. 
The bipolar TE occurs in tunnel junctions with energy-symmetric DoSs in the leads, characterized by a reciprocal $I$--$V$ characteristic, and can be expressed as a strong violation of detailed balance.
In superconductors, the DoS symmetry reflects electron pairing in the mean-field description, so the gap opens around the chemical potential; the presence of the gap and the monotonically decreasing DoS above it enable the generation of thermoelectric power in SIS$^\prime$ junctions when a temperature difference is applied across them. 
Reciprocity shapes the unique features of the effect. In the thermoelectric state, the junction displays at the same time negative differential conductance and absolute negative conductance for both voltage bias polarities; the negative differential conductance, in turn, can make the zero-voltage bias state electrically unstable so that the junction spontaneously develops a thermovoltage, selecting one of two opposite-sign values
$\pm V_S$~\cite{marchegiani_nonlinear_2020,marchegiani_superconducting_2020}. In other words, the bipolar TE can be regarded as a spontaneous nonequilibrium breaking of the energy inversion symmetry by the temperature difference, in stark contrast with standard TEs where this symmetry is explicitly broken (see, e.g.~\cite{bergeret2018colloquium}). 
Notably, this thermoelectric generation mechanism represents, to some extent, the dual of the SIS$^\prime$ cooling that was investigated over twenty years earlier~\cite{FRANK1997281,ManninenAPL74}.
In the quantum regime (c.f. Sec.~\ref{Sec: Env}), the bipolar TE is the thermoelectric reverse of the quantum-circuit refrigerator, which utilizes work to cool microwave modes via the same environment-assisted tunneling~\cite{tan_quantum-circuit_2017,sevriuk_initial_2022}. 

The SIS$'$ element enables both cooling and power generation, and this versatility, combined with its heat-driven operation, is what makes the effect attractive for thermal management in quantum technologies, where the heat load of wiring and control lines is itself a limit to processor scaling~\cite{krinner_engineering_2019}. 
The bipolar TE in SIS$'$ junctions has already been proposed
for routing and rectifying heat, through the thermal diodes and heat pipes of Sec.~\ref{SubSec: Pipe}~\cite{fornieri_towards_2017,antola_tunable_2024}, contributing to superconducting caloritronics' toolbox. The same circuit element can convert a thermal difference directly into usable electrical power, powering the circuit applications of Sec.~\ref{subsec:TEcircuit}. The bias-free operation makes radiation detection another promising direction. The passive, self-biased design of the bipolar single-photon detector of Sec.~\ref{SubSec:singlePhoton}~\cite{paolucci_highly_2023} is especially suitable for cryogenic detector arrays, where bias lines carry heat and limit scalability. Even though thermoelectric detectors can also be realized using superconductor-ferromagnet junctions~\cite{heikkila2018thermoelectric,chakraborty_thermally_2018}, the SIS$'$ bipolar TE is convenient on the material side, requiring no ferromagnetic elements. In the environment-assisted mechanism discussed in Sec.~\ref{Sec: Env}, bipolar thermoelectricity can serve an additional purpose: the sensitivity of the thermoelectric current to the environmental spectrum hints at its use as an on-chip spectrometer of the electromagnetic modes~\cite{antola2026quantum}.

Our results defy the usual premise of thermoelectricity. While traditional approaches focus on engineering an electron-hole asymmetry to obtain a thermoelectric response, we propose an alternative strategy based on symmetric transport, in which the interplay of nonequilibrium physics (i.e., temperature difference), interaction effects, and a strong energy-dependent spectrum spontaneously breaks this symmetry. 
The bipolar TE is not specific to fully superconducting platforms. Combining superconductors with normal metals, ferromagnetic films, or even monolayer or bilayer graphene extends the tunability of this response, as we discussed in Sec.~\ref{Sec: hybrids}. In particular, transferring the mechanism to different systems could also push it toward higher operating temperatures, a topic that remains largely unexplored. Possible alternative setups compatible with current reciprocity, and hence with  bipolar thermoelectricity, may include 
Kondo systems~\cite{Hewson_1993}, 2D materials~\cite{bernazzani_bipolar_2023}, and topological materials~\cite{sato_topological_2017,flensberg_engineered_2021,aslani2026enhanced}.
We leave the assessment of the generality of bipolar TE beyond the paradigmatic SIS$^\prime$ junction to future work. The discovery of the bipolar thermoelectrical effect has opened a new perspective in thermoelectric research~\cite{Balduque_Scattering_2025,Khomchenko_PRB106}. These investigations suggest that the general principle discussed in this review may be extended further. In particular, they could also clarify the role of interactions in nonequilibrium spontaneous symmetry breaking, a largely unexplored field. Finally, the recent proposal of the quantum bipolar TE could enable new tests of quantum thermodynamic bounds~\cite{Landi_Irreversible_2021} and applications in quantum detection~\cite{Tucker_Quantum_1985}.

\newpage

\section{Appendix:}
\subsection{\bf Thermoelectric Power for a few values of the Dynes parameter}
\label{app:optimalPowerGamma}
\begin{figure}[t]
    \centering
    \includegraphics[width=1\linewidth]{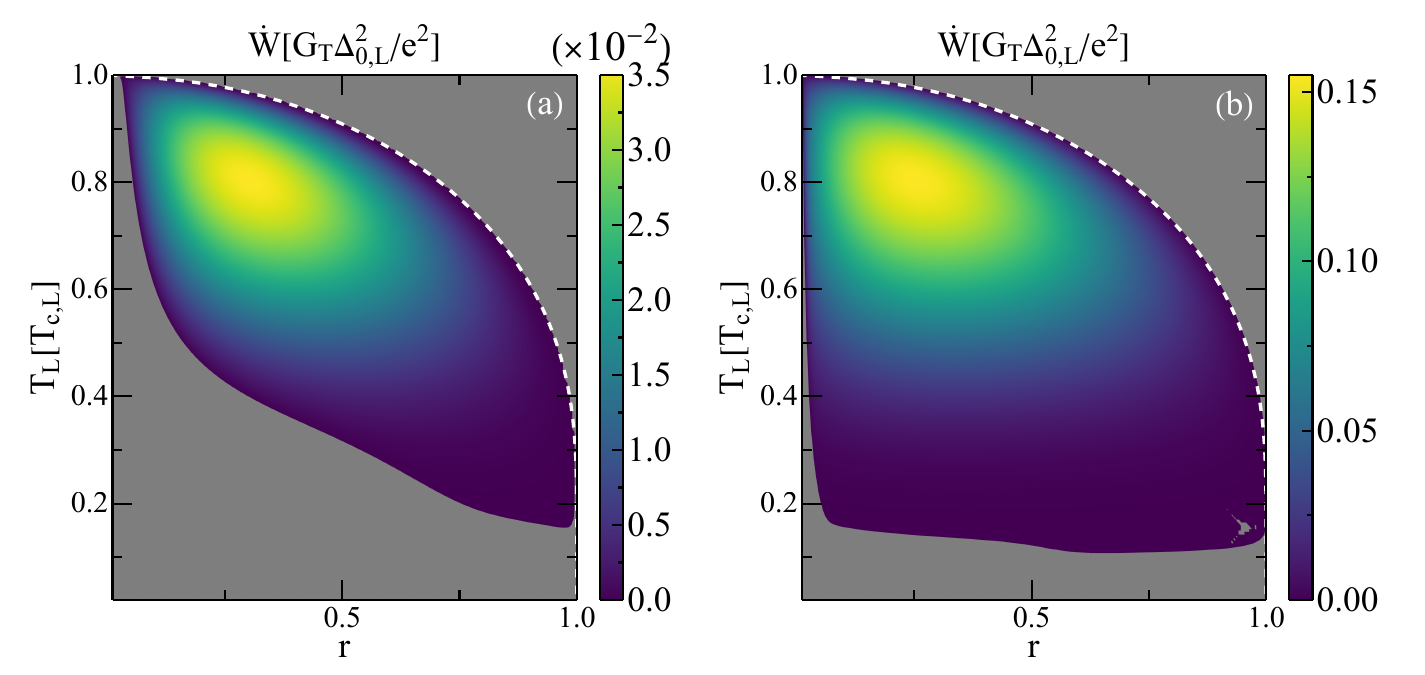}
   \caption{
   Dynes parameter and parameter dependence of the thermoelectric power. (a)-(b) Density plot of the thermoelectric power $\dot{W}$ as a function of $r$ and $\tL$ (a) $\gamma_j=10^{-2}\Delta_{0,j}$ and (b) $\gamma_j=10^{-2}\Delta_{0,j}$. The low-gap electrode temperature is fixed to $\tR = 0.01\tcL$. The dashed line gives the condition $\dL(\tL)=\dR(\tR)$, and the gray regions a dissipative behavior of the junction [$I(V_p)V_p\geq 0$].
}
    \label{fig:PRL2appendix}
\end{figure}

In Sec.~\ref{SubSec:PowerSIS}, we discussed the thermoelectric power $\dot{W}=-IV$ provided by the thermally biased SIS$'$ junction at the matching peak condition. As mentioned there, we choose a realistic value of the Dynes parameter, i.e., $\gamma_j=10^{-4}\Delta_{0,j}$ (with $j=1,2$); the thermoelectric power is otherwise divergent at the matching peak singularity for BCS DoSs. Here we show that the Dynes parameter only affects the magnitude of the thermoelectric power, but not its overall dependence on other junction parameters, such as the temperature of the high-gap superconductor and the zero-temperature gap ratio. Specifically, in Fig.~\ref{fig:PRL2appendix}, we display density plots analogous to Fig.~\ref{fig:PRL2}c for higher (panel a) and lower (panel b) values of $\gamma_j$. The plots clearly show that the optimal region for thermoelectricity depends very weakly on $\gamma_j$; as expected, the Dynes parameter mainly affects the size of the current (by less than a factor ten, decreasing it of three orders of magnitude) and the parameter region for the observation of a thermoelectric response, which is observable at smaller values $\tL$
by decreasing $\gamma_j$. However, in such a region, the thermoelectric response is typically low. 
\section*{Acknowledgments}
The authors acknowledge F. Taddei, L. Arrachea, L. Tosi, R. Whitney, L. Bernazzani, G. De Simoni, and S. Battisti for fruitful discussions.
\section*{Funding}
AB acknowledges funding from MUR-PRIN 2022 -- Grant No. 2022B9P8LN (PE3) -- Project NEThEQS ``Non-equilibrium coherent thermal effects in quantum systems'' in PNRR Mission 4 -- Component 2 -- Investment 1.1 "Fondo per il Programma Nazionale di Ricerca e Progetti di Rilevante Interesse Nazionale (PRIN)" funded by the European Union -- Next Generation EU, the project "Thermoelectric effects in solid-state quantum devices based on multiterminal Josephson junctions" of the bilateral agreement CNR/CONICET (Italy/Argentina) 2026-2027, and the CNR Project QTHERMONANO.
\section*{Competing Interests}
The authors have no relevant financial or non-financial interests to disclose.

\bibliography{Bibliography}
\end{document}